\documentclass[12pt,preprint]{aastex}

\def\hal{H$\alpha$}
\def\be{\begin{equation}}
\def\ee{\end{equation}}

\def\m{~$\mu$m}
\def\HI{\ion{H}{1}}
\def\HII{\ion{H}{2}}

\def\ISO{{\it ISO}}
\def\IRAS{{\it IRAS}}
\def\Spitzer{{\it Spitzer}}

\def\2MASS{{\it 2MASS}}

\begin {document}

\title{The \Spitzer\ Local Volume Legacy: Survey Description and Infrared Photometry}
 
\author {D.A.~Dale\altaffilmark{1}, S.A.~Cohen\altaffilmark{1}, L.C.~Johnson\altaffilmark{1,7}, M.D.~Schuster\altaffilmark{1}, D.~Calzetti\altaffilmark{2}, C.W.~Engelbracht\altaffilmark{3}, A.~Gil~de~Paz\altaffilmark{4}, R.C.~Kennicutt\altaffilmark{5,3}, J.C.~Lee\altaffilmark{6}, A.~Begum\altaffilmark{5}, M.~Block\altaffilmark{3}, J.J.~Dalcanton\altaffilmark{7}, J.G.~Funes\altaffilmark{8}, K.D.~Gordon\altaffilmark{9}, B.D.~Johnson\altaffilmark{10}, A.R.~Marble\altaffilmark{3}, 
S.~Sakai\altaffilmark{11}, E.D.~Skillman\altaffilmark{12}, L.~van~Zee\altaffilmark{13}, F.~Walter\altaffilmark{14}, D.R.~Weisz\altaffilmark{12}, B.~Williams\altaffilmark{7}, S.-Y.~Wu\altaffilmark{2}, Y.~Wu\altaffilmark{15}}
\altaffiltext{1}  {\scriptsize Department of Physics and Astronomy, University of Wyoming, Laramie, WY 82071; ddale@uwyo.edu}
\altaffiltext{2}  {\scriptsize Astronomy Department, University of Massachusetts, Amherst, MA 01003}
\altaffiltext{3}  {\scriptsize Steward Observatory, University of Arizona, Tucson, AZ 85721}
\altaffiltext{4}  {\scriptsize Departamento de Astrofisica, Universidad Complutense, Madrid, E-28040, Spain}
\altaffiltext{5}  {\scriptsize Institute of Astronomy, University of Cambridge, Cambridge CB3 0HA, United Kingdom}
\altaffiltext{6}  {\scriptsize Carnegie Observatories, 813 Santa Barbara Street, Pasadena, CA 91101}
\altaffiltext{7}  {\scriptsize Department of Astronomy, University of Washington, Seattle, WA 98195}
\altaffiltext{8}  {\scriptsize Vatican Observatory Research Group, Steward Observatory, University of Arizona, Tucson, AZ 85721}
\altaffiltext{9}  {\scriptsize Space Telescope Science Institute, 3700 San Martin Drive, Baltimore, MD 21218}
\altaffiltext{10} {\scriptsize Department of Astronomy, Columbia University, New York, NY 10027}
\altaffiltext{11} {\scriptsize Division of Astronomy and Astrophysics, University of California, Los Angeles, CA 90095}
\altaffiltext{12} {\scriptsize Astronomy Department, University of Minnesota, Minneapolis, MN 55455}
\altaffiltext{13} {\scriptsize Department of Astronomy, Indiana University, Bloomington, IN 47405}
\altaffiltext{14} {\scriptsize Max Planck Institut f\"{u}r Astronomie, K\"{o}nigstuhl 17, 69117 Heidelberg, Germany}
\altaffiltext{15} {\scriptsize Infrared Processing and Analysis Center, California Institute of Technology, MC 314-6, Pasadena, CA, 91125}

\begin {abstract}
The survey description and the near-, mid-, and far-infrared flux properties are presented for the 258 galaxies in the Local Volume Legacy (LVL).  LVL is a {\it Spitzer Space Telescope} legacy program that surveys the local universe out to 11~Mpc, built upon a foundation of ultraviolet, H$\alpha$, and HST imaging from 11HUGS (11~Mpc H$\alpha$ and Ultraviolet Galaxy Survey) and ANGST (ACS Nearby Galaxy Survey Treasury).  LVL covers an unbiased, representative, and statistically robust sample of nearby star-forming galaxies, exploiting the highest extragalactic spatial resolution achievable with {\it Spitzer}.  As a result of its approximately volume-limited nature, LVL augments previous {\it Spitzer} observations of present-day galaxies with improved sampling of the low-luminosity galaxy population.  The collection of LVL galaxies shows a large spread in mid-infrared colors, likely due to the conspicuous deficiency of 8\m\ PAH emission from low-metallicity, low-luminosity galaxies.  Conversely, the far-infrared emission tightly tracks the total infrared emission, with a dispersion in their flux ratio of only 0.1~dex.  In terms of the relation between infrared-to-ultraviolet ratio and ultraviolet spectral slope, the LVL sample shows redder colors and/or lower infrared-to-ultraviolet ratios than starburst galaxies, suggesting that reprocessing by dust is less important in the lower mass systems that dominate the LVL sample.  Comparisons with theoretical models suggest that the amplitude of deviations from the relation found for starburst galaxies correlates with the age of the stellar populations that dominate the ultraviolet/optical luminosities.
\end {abstract}
 
\keywords{surveys --- galaxies: photometry --- infrared: galaxies}
 
\section {Introduction}
Although star formation rates based on optical spectroscopy and {\it GALEX} ultraviolet and {\it Spitzer} infrared imaging have been measured for thousands of galaxies (and hundreds of thousands via the Sloan Digital Sky Survey), most currently available datasets are derived from flux-limited samples, and thus suffer from well-known biases against low-mass, low surface brightness systems.  Multi-wavelength datasets that do include such systems generally only provide representative samples of this low-mass galaxy population (e.g., SINGS; Kennicutt et al.\ 2003), and are thus not suitable for studies that seek to probe the low metallicity dwarf galaxy regime and that require datasets which are true to the statistics rendered by volume-limited sampling.  The goal of the {\it Spitzer} Local Volume Legacy (LVL) survey\footnote{http://www.ast.cam.ac.uk/IoA/research/lvls} is to fill a vital niche in existing multi-wavelength surveys of present-day galaxies with a statistically robust, approximately volume-complete study of our nearest 
neighbors.  

With LVL, we have directly addressed this issue by performing the most complete census to date of dust and star formation within the Local Volume.  LVL consolidates and builds upon recent Local Volume galaxy surveys that have acquired ground-based narrowband H$\alpha$ (Kennicutt et al.\ 2008), {\it GALEX} ultraviolet (Lee et al.\ 2009a) and {\it HST} resolved stellar population imaging (Dalcanton et al.\ 2009), by collecting {\it Spitzer} IRAC and MIPS infrared imaging for a complete sample of 258 galaxies derived from these programs.  The resultant LVL multi-wavelength dataset provides information on each galaxy's ({\it i}) current star formation rate, as traced by H$\alpha$ emission, which is produced by the recombination of gas ionized by massive, short-lived OB stars ($<$20~Myr; Meynet \& Maeder 2000), ({\it ii}) star formation rate averaged over a longer $\sim$100~Myr timescale, as traced by the non-ionizing ultraviolet continuum which originates in the photospheres of OB stars, ({\it iii}) overall stellar mass, from 3.6 and 4.5\m\ luminosities which are generally dominated by the light from old stellar populations, and ({\it iv}) dust content, from both the strength and the shape of the infrared emission, which represents the stellar light that has been absorbed and re-radiated by dust.  Temporally resolved star formation histories derived from the modeling of stellar population color-magnitude diagrams from {\it HST} resolved stellar photometry are also available for 69 of the closer galaxies in the sample (e.g., Williams et al.\ 2009).  The collection of these observations enable a wealth of spatially-resolved and spatially-integrated studies probing present-day star formation, chemical abundance, stellar structure, and dust properties as well as galaxy evolution, particularly for metal-poor, low-mass galaxies which dominate the LVL sample by number.  As part of the LVL program, we are providing homogeneously processed H$\alpha$, {\it GALEX} ultraviolet and {\it Spitzer} IRAC and MIPS infrared images to the community.  Public data releases have begun through the NASA/IPAC Infrared Science Archive\footnote{http://ssc.spitzer.caltech.edu/legacy/lvlhistory.html} (IRSA).

Principal science issues to be addressed by LVL include: constraining the physical mechanisms underlying dust heating and understanding correlations between infrared emission, dust content, and global galaxy properties; establishing the primary factors which influence polycyclic aromatic hydrocarbon (PAH) emission and evaluating the robustness of PAH emission as a star formation rate indicator, particularly at low metallicities and high specific star formation rates; and probing the temporal variation of star formation as a function of global properties, with special focus on the dwarf galaxy population (e.g., Lee et al.\ 2009b).  Specific forthcoming papers concentrate on the development of an accurate photometric technique for gauging PAH emission (Marble et al.\ 2009), the impact of relatively young, luminous AGB stars on stellar masses derived from the near-infrared (Johnson et al.\ 2009), utilizing ultraviolet-infrared colors to investigate the ``inside-out'' growth of galaxies (Gil de Paz et al.\ 2009), a characterization of the population of heavily-obscured star-forming regions (Staudaher et al.\ 2009), using integrated fluxes to model the spectral energy distributions with the aim of quantifying parameters such as dust mass and temperature, radiation field strength, dust-to-gas ratio, and PAH mass fraction (Gordon et al.\ 2009; Draine et al.\ 2009), and combining \hal\ and infrared data to formulate optimal star formation rate indicators.  Efforts are also being made to collect new and existing optical {\it UBVRI} imaging and spectroscopy for the sample, which enable, for example, work updating the local mass/luminosity-metallicity relationship from the $B$ to 4.5\m\ bands (Lee et al.\ 2009c), and stellar energy distribution fitting using stellar population synthesis model grids to provide constraints on star formation histories and present-day stellar masses.

In this paper, we provide a requisite component for much of this work by presenting the {\it Spitzer} observations, data reduction and IRAC and MIPS infrared flux densities for the LVL sample.  Near-infrared photometry measured from 2MASS data within the same apertures that are used on the {\it Spitzer} imaging are also provided.  Section~\ref{sec:sample} describes the sample, \S~\ref{sec:observations} reviews the observational and data processing programs, \S~\ref{sec:data} covers details of the integrated aperture photometry, \S~\ref{sec:results} presents initial results based on the photometry, and \S~\ref{sec:summary} summarizes our work.

\section{The Sample}
\label{sec:sample}
The Local Volume Legacy public dataset consists of {\it GALEX} ultraviolet, H$\alpha$ and {\it Spitzer} IRAC and MIPS imaging for a tiered sample of 258 galaxies that have been drawn from two existing volume-limited surveys.  The inner tier of LVL mainly consists of galaxies targeted by ANGST, the ACS Nearby Galaxy Survey Treasury (Dalcanton et al.\ 2009).  This includes all known galaxies within 3.5~Mpc which lie outside the Local Group and the Galactic plane ($|b|>20^{\circ}$), as well as galaxies in the M81 group and Sculptor filament.  ANGST provides {\it GALEX} ultraviolet imaging and has augmented existing deep {\it HST} imaging with new observations to provide uniform stellar photometry with homogeneous depth for these galaxies.  The outer tier of LVL is derived from the larger 11~Mpc narrowband H$\alpha$ imaging survey of Kennicutt et al.\ (2008), and {\it GALEX} ultraviolet follow-up observations of a sub-sample which avoids the Galactic plane ($|b|>30^{\circ}$) (Lee et al.\ 2009a).  The H$\alpha$ survey and the {\it GALEX} ultraviolet component taken together has been referred to as 11HUGS, the 11~Mpc H$\alpha$ and Ultraviolet Galaxy Survey.  The sample used by 11HUGS is given in Kennicutt et al.\ (2008) and was compiled as follows.  A primary subset of the sample aims to be as complete as possible in its inclusion of galaxies that meet a combined criteria of $D\leq 11$~Mpc, $|b|>20^{\circ}$, $m_B<15$~mag and RC3 type $T\geq0$ (i.e., galaxies with spiral and irregular morphologies later than S0a).  A secondary subset is comprised of galaxies within 11~Mpc for which H$\alpha$ data are available, but fall outside one of the limits in magnitude, Galactic latitude and morphological type, and have available H$\alpha$ data (i.e., galaxies that were either observed as available telescope time allowed, or had existing H$\alpha$ measurements in the literature).  

The sum of the primary and secondary subsets compiled in Kennicutt et al.\ (2008) encompasses the majority of the ANGST galaxies, as well as Local Group galaxies not targeted by ANGST.  To build the outer tier of LVL, we have targeted those galaxies in the primary subset, but with more stringent limits on Galactic latitude ($|b|>30^{\circ}$, consistent with that applied for {\it GALEX} follow-up\footnote{A Galactic latitude limit of $|b|>30^{\circ}$ for the ultraviolet observations avoids objects with excess Galactic extinction and high foreground star density which would violate {\it GALEX}'s bright object safety limits.}) and a slightly relaxed brightness limit ($m_B<15.5$~mag).  Beyond these bounds the original surveys that have provided the bulk of our knowledge about the Local Volume galaxy population are known to become severely incomplete (e.g., Tully 1988).  Within the bounds, statistical tests and comparison with blind all-sky \HI\ surveys confirm that the sample completeness is excellent ($>$95\%; Lee et al.\ 2009b). More details on the sample selection and properties of the precursor surveys are given in Dalcanton et al.\ (2009; ANGST sample definition, {\it HST} observations and reduction), Kennicutt et al.\ (2008; 11~Mpc sample compilation, H$\alpha$ observations, and integrated flux and equivalent width catalog), Lee et al.\ (2007; Local Volume star formation demographics as traced by the H$\alpha$ equivalent width), Lee et al.\ (2009b; 11~Mpc sample completeness properties) and Lee et al.\ (2009a; {\it GALEX} observations and integrated ultraviolet photometry catalog).  A schematic illustration of LVL's tiered volume coverage is shown in Figure~\ref{fig:volume_coverage}, and the final LVL sample of 258 galaxies is given in Table~\ref{tab:sample}.

The two tiers of LVL are highly complementary.  Figure~\ref{fig:sample} presents distributions in the primary selection criteria (morphology, apparent $B$ magnitude, Galactic latitude, and distance) for LVL, where the ANGST sub-sample has been marked separately.  Data on these basic properties are taken from the compilations given in Kennicutt et al.\ (2008) and Dalcanton et al.\ (2009).  ANGST provides complete coverage within an inner volume, and includes both early (dwarf spheroidals, ellipticals, lenticulars) and late (spiral and irregular) morphological types, and many of the lowest mass galaxies.  11HUGS covers a 30 times larger volume, and therefore offers better statistical sampling of the star-forming galaxy population as a whole.  As would be expected for an approximately volume-limited sample, the sample population is dominated by low-luminosity, dwarf galaxies.  Approximately 61\% are irregulars, 31\% have spiral morphology, 5\% are dwarf spheroidals, and 2\% are early-type galaxies.  For comparison, the SINGS sample is dominated more by luminous spiral galaxies (63\%), with 17\% irregular, 12\% S0, and 8\% elliptical morphologies.  While the 11HUGS-based portion of the sample goes as faint as $m_B=15.5$~mag, as explained above, Figure~\ref{fig:sample} shows that the faintest systems in the ANGST inner-tier approach $m_B=19$~mag (e.g., M81~Dwarf~A and BK3N).  In terms of their absolute $B$ magnitudes, 81\% of the galaxies in LVL are intrinsically fainter than the LMC ($M_B=-17.9$~mag).

The distances in the LVL sample range from 50-60~kpc for the Magellanic Clouds out to 11~Mpc at the outer edge of the survey.  Kennicutt et al.\ (2008) describe in detail the origins of the adopted distances displayed in Figure~\ref{fig:sample}.  Many of the galaxies within $\sim$5~Mpc have distance determinations based on standard candles, whereas estimates based on secondary indicators or flow-corrected velocities (assuming $H_{\circ}$=75~km~s$^{-1}$~Mpc$^{-1}$) are adopted for the more distant systems.  About half of the sample galaxies have reliable distances from measurements of the tip of the red giant branch ($\sim$40\% of the sample) and Cepheid variables (6\% of the sample).  It is important to note that an inherent difficulty with efforts to construct a volume-limited sample is that its membership will necessarily be fluid until accurate distance and photometric measurements are available for all of the galaxies that are within the volume and around its periphery.  Since the inception of the LVL {\it Spitzer} program, four galaxies included in the sample (and in Table~\ref{tab:sample}) have updated distances which place them outside of 11~Mpc.  In addition, the flow model initially applied was updated in Kennicutt et al.\ (2008) to provide consistency with one used by NED.\footnote{The NASA/IPAC Extragalactic Database (NED) is operated by the Jet Propulsion Laboratory, California Institution of Technology, under contract with NASA.}  As a result 30 galaxies with $|b|>30^{\circ}$ and $m_B<15.5$~mag in the parent sample (Kennicutt et al. 2008) are not included in LVL.  The galaxies are generally between 10 and 11~Mpc, where flow distance uncertainties ($\pm15$\%) would most likely scatter objects in and out of the volume.  Such uncertainties however, should not have a significant impact on studies which use the sample to statistically characterize the physical properties of local galaxies.  Further discussion of such issues is provided in \S~2 of Kennicutt et al.\ (2008).

Overall, the LVL sample covers a diverse cross-section of morphologies and star formation properties, and spans a factor of $10^4$ in optical luminosity, a factor of $10^5$ in star formation rate, and the full range of metallicities found locally ($\sim$1.5~dex).  The nature of the sample allows LVL to more robustly sample infrared properties associated with metal-poor, dwarf galaxies than previous surveys.  For example, the plots in Figure~\ref{fig:sample_phase_space} show LVL's coverage of parameter spaces defined by integrated infrared and infrared-to-optical galaxy properties.  Also shown are the distributions for SINGS, which was designed to broadly, but not statistically, sample the range of properties in nearby galaxies.  While the SINGS and LVL surveys are fairly similar in their distributions of far-infrared colors and infrared-to-optical ratios, the two surveys differ dramatically in their distributions of total infrared luminosity.  As would be expected, LVL is far more effective at filling in the distribution at faint infrared luminosities, whereas SINGS includes more infrared-bright and dusty systems.  Preliminary comparisons with the {\it IRAS} 1.2~Jy survey (Fisher et al.\ 1995) suggest that LVL and SINGS respectively sample preferentially the faint and bright envelopes of a broader distribution at lower infrared-to-optical ratios.

\section {Observational Strategy and Data Processing}
\label{sec:observations}
LVL {\it Spitzer} observations build upon IRAC and MIPS archival data which were already available for about a quarter of the sample when the program began.  In this section we describe the observational strategy employed for the new IRAC and MIPS infrared data obtained to complete {\it Spitzer} coverage of the LVL sample, and the archival data that have been reprocessed for inclusion in our dataset.

\subsection {New {\it Spitzer} IRAC 3.6, 4.5, 5.8, and 8.0\m\ Data}
\label{sec:irac_observations}

New {\it Spitzer} IRAC (Fazio et al.\ 2004) observations were obtained for 180 LVL galaxies.  The IRAC observing strategy follows that of SINGS, which shows that stellar and small grain dust emission is typically detected out to the optical radius at a surface brightness level of $\sim$0.01--0.1~MJy~sr$^{-1}$ (Regan et al.\ 2006; Dale et al.\ 2000).  For galaxies smaller than the IRAC field of view ($D_{25}\leq300$\arcsec) the Astronomical Observing Requests (AORs) were constructed using four dithered 30~s integrations.  For larger galaxies a mosaicking strategy with $\sim$half-array spatial offsets was used, with the sizes of the mosaic `cores' tailored to the optical size of each galaxy.  Two sets of IRAC maps were obtained for each source to enable asteroid removal and to enhance map sensitivity and redundancy.  Combining all eight 30~s frames thus results in a net integration per pixel of 240~s (and 120~s around the $\sim$2\farcm5-wide mosaic peripheries).  Since each source was observed in all IRAC channels, ample sky coverage is automatically provided by the non-overlapping nature of the two IRAC fields-of-view.

The basic calibrated data (BCD) used for post-pipeline processing are from the S18.0 and S18.5 versions of the IRAC pipeline.  These versions differ from their predecessors by including improved corrections for muxbleed and the first-frame effect, among other corrections.  The multi-epoch, multiple-pointing IRAC observations for each galaxy are combined into one single mosaic for each band using the MOPEX mosaicking software.  Additional post-BCD processing includes: distortion corrections, rotation of the individual frames (for multi-epoch observations), bias structure and bias drift corrections, image offset determinations via pointing refinements from the SSC pipeline (MOPEX's default), detector artifact removal, constant-level background subtraction, and image resampling to 0\farcs75 pixels using drizzling techniques.  The drizzling slightly improves the final PSF over the native one; the full-width half maxima are $\sim$1\farcs6 in the shorter wavelength channels and $\sim$1\farcs9 at 8\m.  The final images are in units of MJy~sr$^{-1}$ and have the average sky level removed; sky values are estimated via several ``blank'' regions located near but beyond the target galaxy emission.  Though the LVL IRAC data processing is built upon MOPEX while the SINGS project developed its own IRAC data processing package, the nature of the final data products in the two surveys is essentially the same.

In cases where exceptionally bright target sources saturated or entered the non-linear regime of the detector during the 30~s exposure, additional 1.2~s images are used to allow for recovery of this information.  Pixels affected by these issues, typically in the 5.8 and 8.0\m\ frames, are flagged during processing.  The correction begins by creating a mosaic of the 1.2~s exposures interpolated onto the same pixel grid as the original mosaic.  A difference image is then created from the two mosaics and any residual, systematic difference in the background sky levels is removed.  Pixels in the difference image valued at 1~MJy~sr$^{-1}$ or higher are flagged (routinely regions of $\sim$400 contiguous pixels) and these pixels in the long integration mosaic are replaced by their short integration counterparts.  Immediately outside of these saturated areas, the photometry of the 1.2~s-based mosaics is consistent with that from the far-deeper mosaics described above, and further away in the fainter surface brightness regions the deeper mosaics obviously more effectively detect emission.  The nuclear regions for the following galaxies were corrected for saturation: NGC~0253, NGC~2903, NGC~3031, NGC~3034 (at all IRAC wavelengths), NGC~3351, NGC~3593, NGC~3627, NGC~4258, NGC~5195, and NGC~5253.

\subsection {New {\it Spitzer} MIPS 24, 70, and 160\m\ Data}
\label{sec:mips_observations}

New {\it Spitzer} MIPS (Rieke et al.\ 2004) observations were obtained for 201 LVL galaxies.  Galaxies were imaged in all three MIPS bands centered at 24, 70, and 160\m, using the highly successful scan mapping strategy employed in the SINGS project.  The scan mode was used even on galaxies small enough to fit within the array field of view, because achieving adequate background measurements for extended targets in the photometry mode is less efficient than in the scan mode.  Each map was executed at the medium scan rate, and includes multiple scan legs tailored to the size of the galaxy and half-array offsets between scan legs.  Each galaxy was mapped twice, with the maps separated by 10--40 days to allow time for the field-of-view to rotate and for asteroids to move out of the field.  This second map was performed in the reverse direction (the ``backward mapping'' mode), with offsets in the cross-scan and in-scan directions.  Taken together, these mapping strategies ensure that each point on the galaxy is scanned over in two different directions, which aids reduction of array artifacts on both Si:As and Ge:Ga arrays.  The in-scan offset ensures that Ge:Ga stimflashes do not occur at the same point in both maps and thereby improves the calibration.  The integration time per point is 160, 80, and 16~s at 24, 70, and 160\m, respectively.

The MIPS images are processed with the MIPS Data Analysis Tool (DAT; Gordon et al.\ 2005), supplemented by custom scripts for the specific data reduction and mosaicking of extended sources.  The latter include at 24\m: readout offset correction, array-averaged background subtraction, and exclusion of the first five images in each scan leg due to boost frame transients.  At 70 and 160\m, the custom scripts include a pixel-dependent background subtraction for each map to remove residual detector drifts and background cirrus and zodiacal emission.  This method of reduction was used for all the SINGS galaxies as well as very large galaxies (M31, M33, M101, SMC, LMC, etc.).  The resulting PSFs have full-width half maxima of $\sim$6, 18, and 40\arcsec\ at 24, 70, 160\m, respectively.  The pixel scales of the MIPS mosaics are 1\farcs5, 4\farcs5, and 9\farcs0 at 24, 70, and 160\m, respectively.  

Finally, a correction for 70\m\ non-linearity effects is included in the data processing.  A correction of the form
\be
f^{\rm 70{\scriptsize \micron}}_{\rm true} = 0.502 (f^{\rm 70{\scriptsize \micron}}_{\rm measured})^{1.182},
\label{eq:nonlinear}
\ee
derived from data presented by Gordon et al.\ (2009) and slightly different than the form first presented in Dale et al.\ (2007) for SINGS galaxies, is applied to pixel values above a threshold of $\sim$44~MJy~sr$^{-1}$.  The uncertainties on the parameters in Equation~\ref{eq:nonlinear} are $\sim$10\%.  The correction to the global 70\m\ flux density is $\leq$1.01 for 83\% of the sample, $\leq$1.05 for 90\% of the sample, and $\leq$1.29 for all but two sources.  The correction for NGC~0253, a galaxy with a well-known super star cluster, is 1.59.  The correction for the starburst galaxy NGC~3034 (M82) is 1.83.

\subsection {Archival {\it Spitzer} Data}
Archival IRAC and MIPS data, with spatial coverage and sensitivity similar to or greater than that described in \S~\ref{sec:irac_observations} and \S~\ref{sec:mips_observations}, are utilized for 78 (IRAC) and 57 (MIPS) galaxies.  No new IRAC or MIPS observations were obtained for these subsets of the LVL sample.  The data processing procedures for the archival data are the same as those followed for the new observations described above (including the use of the S18 IRAC data processing pipeline), except for the asteroid rejection in the few cases where only one epoch was measured.  Table~\ref{tab:data} indicates for which galaxies we use archival {\it Spitzer} data.

\section {Aperture Photometry}
\label{sec:data}
This section describes the infrared flux densities measured for the LVL program.  For a given galaxy, in most cases the same aperture was used for extracting all infrared flux densities.  Elliptical apertures were based on capturing all the galaxy emission visible for all infrared images.  Typically this means that the 3.6\m\ image was used to create the aperture, since 3.6\m\ is the bandpass within which {\it Spitzer} is most sensitive and stars are brightest.  Occasionally the emission at 160\m\ shows the greatest spatial extent, resulting in part from the smearing involved with the $\sim$40\arcsec\ of the PSF at this wavelength.  In addition, for a subset of $\sim$40 LVL galaxies, the infrared-based apertures were slightly enlarged to capture extended ultraviolet emission.  The aperture centers, semi-major and semi-minor axes $a$ and $b$, and the position angles are provided in Table~\ref{tab:sample}.  The median semi-major axis is 1.13 times $R_{25}$, and 7\% of the semi-major axes are smaller than $R_{25}$.

Table~\ref{tab:data} presents the global flux densities for the entire LVL sample, for wavelengths spanning the near- to far-infrared.  The compact table entry format T.UV$\pm$W.XYEZ implies (T.UV$\pm$W.XY)$\times10^{\rm Z}$.  The data are corrected for Galactic extinction (Schlegel, Finkbeiner, \& Davis 1998) assuming $A_V/E(B-V)\approx3.1$ and the reddening curve of Li \& Draine (2001).  The effect of airmass has been removed from the ground-based near-infrared fluxes.  No color corrections have been applied to the flux densities.  Additional issues such as sky removal, aperture corrections, and upper limits are covered in detail below.  

The uncertainties provided in Table~\ref{tab:data} include both calibration and statistical uncertainties.  Including the uncertainties in aperture corrections described below, the IRAC calibration uncertainties are, conservatively, 5-10\% for 3.6 and 4.5\m\ data, and 10-15\% for 5.8 and 8.0\m\ data (Reach et al.\ 2005; Farihi et al.\ 2008; T. Jarrett, private communication); 10\% IRAC calibration uncertainties are used in Table~\ref{tab:data}.  MIPS calibration uncertainties are 4\%, 5\%, and 12\% respectively at 24, 70, and 160\m\ (Engelbracht et al.\ 2007, Gordon et al.\ 2007, and Stansberry et al.\ 2007).  A floor to the 2MASS uncertainties is fixed by setting the calibration errors to 5\%.

\subsection {2MASS Near-Infrared JHK$_{\rm s}$ Photometry}
The Two Micron All Sky Survey (2MASS) obtained data for the entire sky at 1.25, 1.65, and 2.17\m\ using two automated, ground-based 1.3~m telescopes (Skrutskie et al.\ 2006).  Galaxy photometry is available from the 2MASS Extended Source Catalog for over a million galaxies and from the 2MASS Large Galaxy Atlas for several hundred galaxies larger than 1\arcmin\ (Jarrett et al.\ 2003).  Integrated fluxes for several LVL galaxies were adopted from the Large Galaxy Atlas, and these are generally consistent with expectations based on IRAC 3.6 and 4.5\m\ fluxes and simple stellar model extrapolations to 2MASS wavelengths.  However, most LVL galaxies do not appear in the Large Galaxy Atlas, and for these relatively faint systems many of the fluxes from the Extended Source Catalog are 0.5--2~mag low based on similar extrapolations from IRAC 3.6 and 4.5\m\ data.  We find that when Extended Source Catalog fluxes appear unexpectedly faint, it is typically due to the comparatively small apertures used in the automated 2MASS extraction (see, for example, the fairly extreme case of UGC~08245 in Figure~\ref{fig:u8245}).  Hence we have independently extracted 2MASS fluxes for the vast majority of the LVL sample using the same apertures and foreground star removals used to determine IRAC and MIPS fluxes, as discussed in the following section.  Figure~\ref{fig:2mass} displays the ratios of our near-infrared extractions with those provided in the 2MASS Extended Source Catalog.  Included in the figure are results from Kirby et al.\ (2008) based on deep $H$ band imaging of nearby galaxies with the 3.9~m Anglo-Australian Telescope; Kirby et al.\ (2008) likewise find that the fainter sources in the Extended Source Catalog have their global fluxes underestimated.  The correction factors in Figure~\ref{fig:2mass} rise steeply with decreasing flux densities below 0.1~Jy ($\sim$10~mag).  The secureness of the detections below this level also drops quickly, down to the 2--3$\sigma$ level for $f_\nu \lesssim 0.01$~Jy.

\subsection {{\it Spitzer} 3.6, 4.5, 5.8, 8.0, 24, 70, and 160\m\ Photometry}
\label{sec:aperture}

\subsubsection {Foreground Star and Background Galaxy Removal}
The presence of foreground stars and background galaxies can significantly affect the global infrared fluxes for some galaxies, particularly the fainter dwarfs and galaxies at low Galactic latitudes.  Once identified, the foreground stars and background galaxies are removed through a simple interpolation of the local sky from the images using the IRAF task {\tt IMEDIT}.  Our procedure for distinguishing between target galaxy and foreground/background sources relies on a multi-wavelength analysis (3.6, 8.0, 24\m, and \hal), looking for objects that are \hal-rich (target galaxy) or especially blue (foreground stars; $f_\nu(3.6)/f_\nu(8.0)>8$), or extended red systems with smooth morphologies (background galaxies).  Archival {\it Hubble Space Telescope} imaging was also inspected for obvious background galaxy or foreground stellar identifications, when available.  When uncertain about the identification of a particular source, we opted to err on the conservative side and allow such sources to remain in the global flux extraction.  However, these sources of uncertain origin are typically very faint and have negligible impact on global flux extractions.  The median ratios of corrected-to-stellar contaminated fluxes is [0.854,~0.846,~0.939,~0.971,~0.980,~1.00,~1.00] at [3.6,~4.5,~5.8,~8.0,~24,~70,~160]\m; very few significant corrections are made at 24, 70, and 160\m.

\subsubsection {Aperture Corrections}
Since the IRAC flux calibration is based on point source photometry for a 12\arcsec\ radius aperture, the fluxes for all extended sources and aperture radii $\neq$ 12\arcsec\ need to have an additional correction applied.  These corrections account for the ``extended'' emission due to the wings of the PSF and also for the scattering of the diffuse emission across the IRAC focal plane.  This photometric correction is different than merely subtracting off the sky value (\S~\ref{sec:irac_observations}).  As described in Dale et al.\ (2007), the IRAC extended source correction has been derived for a variety of source morphologies and extents.  For an effective aperture radius $r=\sqrt{ab}$ in arcseconds derived from the semi-major $a$ and semi-minor $b$ ellipse axes provided in Table~\ref{tab:sample}, the IRAC extended source aperture correction is
\be
f^{\rm IRAC}_{\rm true} / f^{\rm IRAC}_{\rm measured} = A {\rm e}^{-r^B} + C,
\label{eqn:irac}
\ee
where $A$, $B,$ and $C$ are listed in Table~3\footnote{See http://ssc.spitzer.caltech.edu/irac/calib/extcal/}.
The median IRAC extended source aperture corrections are [0.914,~0.941,~0.826,~0.756] at [3.6,~4.5,~5.8,~8.0]\m.

In contrast to the IRAC aperture corrections, the main reason MIPS aperture corrections are needed is the smearing of light according to the PSF profile; the measured MIPS fluxes need to be slightly boosted to account for light diffracted beyond the extent of the chosen apertures.  MIPS aperture corrections are empirically determined from a comparison of fluxes from smoothed and unsmoothed 3.6\m\ imaging, an approximate proxy for tracing the MIPS galaxy morphologies.  The aperture correction for a given MIPS flux is the ratio of the fluxes from the unsmoothed 3.6\m\ image to the flux from the 3.6\m\ image smoothed to the same PSF as the MIPS image in question.  The median MIPS aperture corrections are [1.01,~1.01,~1.03] at [24,~70,~160]\m, and the most significant corrections are 
[1.07,~1.20,~1.68] for UGC~05923.  

\subsection {Upper Limits}
Many of the optically-faint galaxies in the sample are frequently undetected in the infrared, particularly at wavelengths of 5.8\m\ and longer.  Upper limits are included in Table~\ref{tab:data} for sources undetected by infrared imaging.  In all cases ``undetected'' implies that the measured flux density is below the 5$\sigma$ upper limit.  The 5$\sigma$ upper limits for {\it Spitzer} imaging are derived assuming a galaxy spans all $N_{\rm pix}$ pixels in the aperture,
\be
f_\nu(5\sigma~{\rm upper~limit})_{\rm Spitzer} ~ = ~ 5 ~ \sigma_{\rm sky} \Omega_{\rm pix} ~ \sqrt{N_{\rm pix}+{N_{\rm pix}}^2/N_{\rm sky}} ~ \approx ~ 5 ~ \sigma_{\rm sky} \Omega_{\rm pix} ~ \sqrt{2 N_{\rm pix}}
\ee
where $\sigma_{\rm sky}$ is the sky surface brightness fluctuation per pixel (MJy~sr$^{-1}$), $\Omega_{\rm pix}$ the solid angle subtended per pixel, and $N_{\rm sky}$ ($\approx N_{\rm pix}$) the total number of pixels in the sky apertures.  The parameter $\sigma_{\rm sky}$ is approximately 0.02, 0.03, 0.11, 0.12, 0.2, 0.9, and 1.7~MJy~sr$^{-1}$ at 3.6, 4.5, 5.8, 8.0, 24, 70, and 160\m, respectively, though somewhat larger values are employed for situations where the sky fluctuations are notably larger due to flatfielding errors, scattered light, cirrus, etc.  A similar computation for 2MASS near-infrared upper limits is carried out after converting that survey's mean 10$\sigma$ point source sensitivities ($\sim$ 16.4, 15.5, and 14.8~mag for $J$, $H$, and $K_{\rm s}$, respectively; Skrutskie et al.\ 2006) to 5$\sigma$ values and accounting for the difference in the sizes of the 2MASS point source aperture ($\pi r_{\rm 2MASS}^2$; $r_{\rm 2MASS}=$4\arcsec) and the LVL apertures ($\pi a b$).  In other words,
\be
f_\nu(5\sigma~{\rm upper~limit})_{\rm 2MASS} ~ = {5 \over 10} f_\nu({\rm 2MASS}~10\sigma~{\rm point~source~limit}) ~ \sqrt{{\pi a b \over \pi r_{\rm 2MASS}^2}}.
\ee

\section{Results}
\label{sec:results}

\subsection{Detection Rate}
The lower panels of Figure~\ref{fig:detection_rate} display the detection rates for the different \Spitzer\ imaging channels as a function of $B$ band apparent and absolute magnitudes.  Nearly all galaxies are detected at all \Spitzer\ wavelengths down to $m_B\approx14$~mag and $M_B\approx -13$~mag.  Consistent with our pre-survey expectations, the $m_B\sim15.5$~mag cut-off for the outer tier of the sample that extends to 11~Mpc (see \S~\ref{sec:sample}) proved to be a useful sample selection criterion, because very few galaxies fainter than $m_B\sim15.5$~mag were detected in MIPS.  The inner tier/ANGST portion of the sample extends the sample to much fainter levels, as faint as $m_B\approx19$~mag in the cases of BK03N and M81~Dwarf~A.  As expected for the optically-faint galaxies, the highest detection rates are found for the stellar-dominated 3.6 and 4.5\m\ channels, while the 70 and 160\m\ imaging proved to be far more challenging to convincingly detect cold dust emission.  A stacking analysis (e.g., Dole et al.\ 2006) will be employed to obtain a better statistical understanding of the fainter galaxy population at long wavelengths, in particular with respect to the \HI\ emission.

\subsection{Comparison with Data from {\it IRAS}}

Secure flux measurements are available at all {\it IRAS} and MIPS wavelengths for a subset of 70 LVL galaxies.  The {\it IRAS} data are compiled from Rice et al.\ (1988), Moshir et al.\ (1990), Sanders et al.\ (2003), Lisenfeld et al.\ (2007), and our own archival extractions.  Figure~\ref{fig:iras} provides a comparison of MIPS 24\m\ and {\it IRAS} 25\m\ data.  The agreement between 24 and 25\m\ fluxes is excellent: $\nu f_\nu(24\micron)/\nu f_\nu(25\micron)=1.01$ with a dispersion of 25\%.  

The aggregate emission from all dust grains is a fundamental metric of any galaxy.  Figure~\ref{fig:iras} provides a comparison of the 3--1100\m\ total infrared ({\it TIR}) for the LVL sample as measured by MIPS and {\it IRAS}.  The MIPS-based total infrared is estimated from a linear combination of 24, 70, and 160\m\ fluxes,
\begin{equation}
f(TIR)_{\rm MIPS}=1.559 \nu f_\nu(24\mu{\rm m}) + 0.7686 \nu f_\nu(70\mu{\rm m}) + 1.347 \nu f_\nu(160\mu{\rm m}),
\label{eq:tir}
\end{equation}
and the {\it IRAS}-based total infrared is similarly computed from a linear combination of the 25, 60, and 100\m\ fluxes,
\begin{equation}
f(TIR)_{\rm IRAS}=2.403 \nu f_\nu(25\mu{\rm m}) - 0.2454 \nu f_\nu(60\mu{\rm m}) + 1.6381 \nu f_\nu(100\mu{\rm m}),
\end{equation}
which are Equations~4 and 5, respectively, in Dale \& Helou (2002; see Equation~22 in Draine \& Li 2007 for a variation of Equation~\ref{eq:tir} above that includes the IRAC~8.0\m\ flux).  The coefficients in the above two equations stem from fits to a suite of spectral templates applicable to a wide range of normal star-forming galaxies at redshift zero, where ``normal'' implies the exclusion of AGN and ultraluminous infrared galaxies (see \S~5.3 and Figure~5 in Dale \& Helou 2002 for a representative sampling of the suite of templates).  The uncertainty in using these prescriptions to compute the total infrared for normal star-forming galaxies is estimated to be of order 25\% (Draine \& Li 2007).

The MIPS-based version should be more accurate since the infrared wavelength baseline spanned by MIPS is longer than the baseline covered by {\it IRAS}, and more importantly, the {\it IRAS} detectors do not sample the bulk of the dust in the coldest, most quiescent galaxies.  To determine if these differences in wavelength coverage between {\it IRAS} and MIPS result in different estimates of the total infrared, the two righthand panels in Figure~\ref{fig:iras} compare $f(TIR)_{\rm MIPS}$ and $f(TIR)_{\rm IRAS}$.  The ratio of MIPS- and {\it IRAS}-based total infrared measures has a scatter (21\%) similar to that in the 24-to-25\m\ comparison, but the average ratio is 1.15.  These findings are similar to those by Kennicutt et al.\ (2009) for a sample of 205 nearby galaxies with both {\it IRAS} and MIPS data.  The righthand panel in Figure~\ref{fig:iras} includes semi-empirical predictions from models of infrared spectral energy distributions.  As alluded to above, part of the systematic offset in {\it TIR}$_{\rm MIPS}$/{\it TIR}$_{IRAS}$ can be attributed to the relative inability of {\it IRAS} to accurately measure the total infrared for cold galaxies.  The infrared emission for the coldest galaxies, galaxies with the lowest $f_\nu(60\micron)/f_\nu(100\micron)$ ratios, peaks beyond {\it IRAS}'s 100\m\ detector, and thus the total infrared as measured by {\it IRAS} is systematically low for the coldest galaxies.

\subsection{Multi-Wavelength Spectral Energy Distributions}
\label{sec:seds}
Figures~\ref{fig:mosaicA} and \ref{fig:mosaicB} show ultraviolet-\hal-infrared mosaics of NGC~5236 and UGC~05829, spanning wavelengths where the emission is dominated by young stars (0.15\m), \HII\ regions (\hal), old stars (3.6\m), PAHs (8.0\m), very small grains (24\m) and large grains (70\m).  The galaxies and wavelengths displayed in these two figures highlight the broad range of environments and galaxies sampled by the LVL survey (see \S~\ref{sec:sample}).  Figure~\ref{fig:seds01} provides the panchromatic ultraviolet-infrared broadband spectral energy distributions for all 258 galaxies.\footnote{The far- and near-ultraviolet data are from images acquired as part of the {\it GALEX} Nearby Galaxy Survey, Medium Imaging Survey, and All-Sky Imaging Survey along with several Guest Investigator programs including the 11HUGS Cycle~1 and Cycle~4 and ANGST Cycle~3 proposals.}  The solid curve is the sum of a dust (dashed) and a stellar (dotted) model.  The dust curve is a Dale \& Helou (2002) model (least squares) fitted to ratios of the observed 24, 70, and 160\m\ fluxes, and then scaled to match the overall infrared brightness.  The $\alpha_{\rm SED}$ listed within each panel parameterizes the distribution of dust mass as a function of heating intensity $U$ in units of the local ultraviolet interstellar radiation field, as described in Dale \& Helou (2002): 
\begin{eqnarray}
\label{eq:dMdU}
             dM_{\rm dust}(U) &\propto& U^{-\alpha_{\rm SED}} \; dU,  \; \; \; \; \; \; 0.3 \leq U \leq 10^5.
\end{eqnarray}
To quantify the uncertainty on $\alpha_{\rm SED}$ displayed within each panel of Figure~\ref{fig:seds01}, 1,000 Monte Carlo simulations of the fit to each galaxy's far-infrared fluxes were performed, utilizing the tabulated flux uncertainties to add a random (Gaussian deviate) flux offset at each MIPS wavelength.  The $\alpha_{\rm SED}$ uncertainties reflect the standard deviations in the simulations.  The stellar curve is a 1~Gyr continuous star formation, solar metallicity curve from Vazquez \& Leitherer (2005) fitted to the 2MASS data.  The initial mass function for this curve utilizes a double power law form, with $\alpha_{\rm 1,IMF}=1.3$ for $0.1<m/M_\odot<0.5$ and $\alpha_{\rm 2,IMF}=2.3$ for $0.5<m/M_\odot<100$ (e.g., Kroupa 2002).  Though this stellar curve is not adjusted for internal extinction and may not be applicable to many galaxies in the sample, it is included as a fiducial reference against which deviations in the ultraviolet can be compared from galaxy to galaxy.  

The spectral energy distributions for the LVL sample range widely.  There are stellar-dominated (NGC~0404, UGC~05373, UGCA~0193) and comparatively dusty (IC~5256, NGC~6503) systems; for sources detected by MIPS, the infrared-to-far-ultraviolet ratio in the sample spans more than three orders of magnitude, from $\lesssim$~0.1 to over 100 (\S~\ref{sec:beta}).  There are galaxies with far-infrared spectral energy distributions indicative of warm (UGCA~0281) and cold dust grains (NGC~5055).  Compared to what would be expected based on their stellar and far-infrared emission, many galaxies show a dearth of emission from PAHs in the 8.0\m\ band (e.g., ESO~245-G005, UGC~01249, UGC~05272).  The variations in global spectral energy distributions are discussed in more detail below.

\subsection{Infrared Colors}
\label{sec:colors}
The IRAC--MIPS infrared colors for the LVL sample are displayed in Panel~a of Figure~\ref{fig:global_colors}.  The $f_\nu(70\micron)/f_\nu(160\micron)$ ratio typically traces the temperature of large interstellar grains, while the $f_\nu(8.0\micron)/f_\nu(24\micron)$ ratio has several influences.  The flux at 24\m\ mostly represents emission from very small grains (grains with effective radii of 15--40\AA; Draine \& Li 2007), and the flux at 8.0\m\ can have contributions from stars, hot dust, PAHs, and AGN.  Perhaps due to the diversity of emission mechanisms responsible for 8.0 and 24\m\ flux levels, the $f_\nu(8.0\micron)/f_\nu(24\micron)$ ratio spans nearly two orders of magnitude compared to the single factor of $\sim$10 stretch in the $f_\nu(70\micron)/f_\nu(160\micron)$ ratio.  Since the local volume lacks ``strong'' AGN, loosely defined here as AGN that dominate a galaxy's emission over substantial portions of the electromagnetic spectrum, it is unlikely that AGN contribute much to the scatter in Figure~\ref{fig:global_colors}.

The (42--122\m) far-infrared ({\it FIR}) and total infrared are frequently used as indications of the star formation rate in galaxies (Kennicutt 1998; Bell 2003).  However, in many instances the far-infrared continuum is unavailable or not detected, so monochromatic infrared proxies for {\it FIR} or {\it TIR} are occasionally employed (e.g., Papovich \& Bell 2002; Bavouzet et al.\ 2008).  Hence, the tightness (dispersion) in monochromatic-to-bolometric ratios are of general interest.  Five flavors of these ratios are displayed in the remaining panels of Figure~\ref{fig:global_colors}, and a tabulation of median LVL infrared colors and monochromatic-to-bolometric infrared ratios can be found in 
Table~4 along with their dispersions.  Panel b of Figure~\ref{fig:global_colors} shows the distribution of the 8.0\m\ emission with respect to the 3--1100\m\ total infrared, a distribution which exhibits a dispersion of 0.23~dex, similar to that for $f_\nu(70\micron)/f_\nu(160\micron)$ and $f_\nu(8.0\micron)/f_\nu(24\micron)$.  While it is evident that the LVL sample is distributed fairly evenly by morphology across $f_\nu(70\micron)/f_\nu(160\micron)$ ratios, the bulk of the systems exhibiting relatively low $f_\nu(8.0\micron)/f_\nu(24\micron)$ and $\nu f_\nu(8.0\micron$)/{\it TIR} ratios are from late-type spirals and irregulars.  Walter et al.\ (2007) also find somewhat unusual infrared colors for dwarf irregulars compared to normal spiral galaxies, and they attribute the difference to the lower dust content and higher dust temperatures in dwarf galaxies (see also Hirashita \& Ichikawa 2009 and Mu\~noz-Mateos et al.\ 2009).

The preponderance of late-type spirals and irregulars showing relatively low 8.0\m\ emission is amplified when ``dust-only'' 8.0\m\ emission is considered.  Panel e of Figure~\ref{fig:global_colors} shows a plot similar to that in Panel b but with the stellar emission removed using the expression presented in Helou et al.\ (2004): 
\begin{equation}
\label{eq:fnu8*}
  \nu f_\nu(8.0\mu {\rm m})_{\rm dust} = \nu f_\nu(8.0\mu {\rm m}) - \eta^{8*} \nu f_\nu(3.6\mu {\rm m})
\end{equation}
where $\eta^{8*}=0.232 \times 3.6/8.0$ (see also Engelbracht et al.\ 2008 for a similar scale factor).  The dispersion (0.40~dex) and overall range are significantly larger when the dust-only 8.0\m\ emission is normalized to the total infrared.  It is possible that a portion of these increases in dispersion and range is due to the inapplicability of Equation~\ref{eq:fnu8*} to late-type spirals, but it should be noted that Equation~\ref{eq:fnu8*} is based on a detailed analysis of NGC~300, a local system with an Sd morphological classification.  Another possibility is that the late-type spirals and irregulars within LVL are on average less abundant in heavy metals, and thus either the formation of PAH molecules is starved or the relatively fragile PAHs are photo-dissociated in the hard radiation fields typically associated with low-metallicity environments (Engelbracht et al.\ 2005; Madden et al.\ 2006; Wu et al.\ 2006; Jackson et al.\ 2006; Draine et al.\ 2007; Sloan et al.\ 2008; Gordon et al.\ 2008; Dale et al.\ 2009).  PAH emission from galaxies with normal metallicities, on the other hand, has been shown to be correlate strongly with far-infrared and submillimeter emission (e.g., Haas, Klaas, \& Bianchi 2002; Bendo et al.\ 2008).  Figure~\ref{fig:global_colorsb} is similar to Figure~\ref{fig:global_colors} but the flux ratios are displayed as a function of absolute $B$ magnitude.  Although additional data and detailed follow-up utilizing LVL metallicities would be required to address this issue, clearly the lowest luminosity galaxies in the LVL sample are driving most of the scatter in 8\m-to-TIR measures.  

In contrast to the 8.0\m-to-{\it TIR} measures, the $\nu f_\nu(24\micron$)/{\it TIR} ratio (Panel d of Figure~\ref{fig:global_colors}) shows a range less than an order of magnitude and a dispersion of 0.16~dex; the 70\m-to-{\it TIR} and 160\m-to-{\it TIR} ratios have even smaller dispersions (Panels c \& f and Table~4).  
The implication is that, compared to the 8.0\m\ PAH emission from galaxies, the infrared emission from large grains at 70 and 160\m\ is far more tightly coupled to the bolometric infrared emission.  The very small grain emission at 24\m\ shows an intermediate coupling to the total infrared, though these results may hinge on the relatively high percentage of low metallicity systems in the LVL sample.

The 70\m-to-{\it TIR} and 160\m-to-{\it TIR} ratios cling remarkably closely to the model predictions, with dispersions from the model of 0.039 and 0.032~dex, respectively.  These tight dispersions reflect the importance of the 70 and 160\m\ fluxes in determining the total infrared using just MIPS data.  However, there is evidence for a slightly increasing mismatch between model and data at the warmest far-infrared colors.  This inconsistency may reflect the differences between the LVL sample and the sample used to construct the models (Dale et al.\ 2000).  Galaxies in the LVL sample typically have lower star formation rates (per area) and thus more of the infrared emission stems from cold dust grains emitting at longer wavelengths (e.g., 70 and 160\m), leading to comparatively lower 24\m/TIR and higher 70\m/TIR and 160\m/TIR ratios.  

\subsection{The Infrared-to-Ultraviolet Ratio and Ultraviolet Spectral Slope}
\label{sec:beta}

The infrared-to-ultraviolet ratio is a measure of dust extinction in the ultraviolet for star-forming galaxies (e.g., Gordon et al.\ 2000), and thus should be related to the amount of reddening in their ultraviolet spectra.  Indeed, starbursting galaxies follow a tight correlation between the ratio of infrared-to-ultraviolet emission and the ultraviolet spectral slope (e.g., Calzetti 1997; Meurer et al.\ 1999).  Compared to the relation defined by starbursts, normal star-forming galaxies are offset to redder ultraviolet spectral slopes, exhibit lower infrared-to-ultraviolet ratios, and show significantly larger scatter (Buat et al.\ 2002, 2005; Bell 2002; Kong et al.\ 2004; Gordon et al.\ 2004; Burgarella et al.\ 2005; Calzetti et al.\ 2005; Seibert et al.\ 2005; Cortese et al.\ 2006; Boissier et al.\ 2007; Gil de Paz et al.\ 2007; Dale et al.\ 2007).  Conversely, extremely dusty galaxies with infrared luminosities above $\sim10^{11}~L_\odot$ are known to be offset {\it above} the canonical starbursting relation, to higher dust extinction levels (Goldader et al.\ 2002).  Offsets from the locus formed by starbursting 
galaxies can be particularly pronounced for systems lacking significant current star formation, such as elliptical galaxies, systems for which the luminosity is more dominated by a passively evolving older, redder stellar population.  The LVL survey provides a unique sample for exploring the relationship between the infrared-to-ultraviolet ratio and the ultraviolet slope, as it consists of a statistically complete set of star-forming galaxies, nearly two-thirds of which are dwarf/irregular systems.  

Figure~\ref{fig:beta} displays the LVL infrared-to-ultraviolet ratios as a function of the ultraviolet spectral slope.  As expected, the well-known starbursts in the LVL sample lie close to the starburst curve: NGC~0253, NGC~4631, NGC~4449, NGC~1705, and NGC~4736, with the latter formally known as a post-starburst galaxy (Walker, Lebofsky, \& Rieke 1988).  Overall, the LVL population is broadly segregated in the infrared-to-ultraviolet ratio according to optical morphology, with Sb and earlier-type galaxies showing relatively high values, Sc/Sd/Sm systems exhibiting intermediate values, and the bulk of the relatively optically thin irregulars appearing near the bottom of the diagram and significantly below the more dust-obscured starburst galaxies.  Interestingly, compared to the Dale et al.\ (2006) normal galaxy curve shown in Figure~\ref{fig:beta}, most LVL targets either have lower infrared-to-ultraviolet ratios for a given ultraviolet color, or are redder for a given infrared-to-ultraviolet ratio.  Lower infrared-to-ultraviolet ratios could arise from the typically less dusty nature of dwarf/irregulars, or the patchy distribution of dust allowing a higher fraction of ultraviolet photons to escape (e.g., Dale et al. 2007; Mu\~noz-Mateos et al.\ 2009).  Inspection of the imaging also shows that the (dust) infrared and \hal\ emission is frequently more centrally concentrated than the ultraviolet emission, and thus comparing the {\it global} infrared and ultraviolet fluxes in galaxies with spatially-extended ultraviolet emission will result in artificially lowered infrared-to-ultraviolet ratios.  Redder colors in LVL galaxies could be related to less efficient star formation capabilities in less massive galaxies (e.g., Kaufmann, Wheeler, \& Bullock 2007), as well as temporally-extended star formation histories suggested by the lack of widespread \hal\ emission in many cases.

Differences in the infrared-to-ultraviolet ratio can be quantified according to the amount of dust extinction (e.g., Meurer et al.\ 1999; Gordon et al.\ 2000; Buat et al.\ 2005).  The distribution of far-ultraviolet extinctions estimated by the infrared-to-ultraviolet-based prescription presented in Buat et al.\ (2005) is provided in Figure~\ref{fig:AFUV}.  Buat et al.\ derive their prescription by averaging over the results from many PEGASE-based model star formation histories (constant, burst, exponential decay) and dust attenuation configurations (foreground screen, clumpy mixture, etc.).  The median far-ultraviolet extinction in the LVL sample using this method is 0.54~mag with ninety percent of the sources having far-ultraviolet extinctions less than 1.7~mag, or equivalently with the aid of the Li \& Draine (2001) extinction curve, the median optical extinction is $A_V \sim 0.2$~mag and ninety percent have $A_V\lesssim0.64$~mag.  

To further explore possible correlations with the dominant stellar population and recent star formation history, Figure~\ref{fig:dstar} provides two observable tracers of the ``birthrate parameter'' as a function of the (perpendicular or closest) distance to the starburst curve in Figure~\ref{fig:dstar}.  The birthrate parameter is defined as the ratio of the current star formation rate to its overall lifetime average (Kennicutt et al.\ 1994), roughly the star formation rate per stellar mass, and thus provides a normalized measure of the star formation activity.  Both the ratio of far-ultraviolet-to-near-infrared luminosity and the H$\alpha$ equivalent width have been previously used as tracers of the birthrate (e.g., Kennicutt et al.\ 1994; Boselli et al.\ 2001; Cortese et al.\ 2006; Lee et al.\ 2009b; Mu\~noz-Mateos et al.\ 2009).  In the top panel of Figure~\ref{fig:dstar} the far-ultraviolet, which tracks star formation averaged over the most recent $\sim$100 Myr, is normalized by the near-infrared luminosity, which probes the total stellar mass built up over much longer timescales.  The bottom panel of Figure~\ref{fig:dstar} incorporates an observable indicator of the birthrate parameter that is much less affected by extinction: the \hal\ equivalent width (taken from Kennicutt et al.\ 2008).  This parameter is also a measure of the birthrate, since the \hal\ flux is a measure of the line emission in \HII\ regions primarily produced by massive ($>10~M_\odot$) stars on $\sim$3--20~Myr timescales (e.g., Kennicutt 1998; Meynet \& Maeder 2000) while the red continuum emission near 6563~\AA\ that provides the normalization for the equivalent width traces the total mass of stars built up over much longer timescales.  The H$\alpha$ equivalent widths shown here are measured over the entire extent of galaxies via narrowband and $R$ band imaging (as opposed to spectroscopic measurements), and thus are representative of global, galaxy-averaged values.

Both panels of Figure~\ref{fig:dstar} show a clear trend, with lower birthrate systems exhibiting larger distances from the starburst trend, consistent with the study of Kong et al.\ (2004).  The deviations from the starburst curve are presumably driven by the differential effects that young and old stellar populations and their local dust opacities have on the age-reddened and dust-reddened luminosities, suggesting that a galaxy's star formation history plays an important role in determining its location within Figure~\ref{fig:beta}.  To more directly interpret these deviations from the starburst curve as a function of the age of the stellar population, the righthand axis of the upper panel of Figure~\ref{fig:dstar} is quantified according to the age of a continuously star-forming system inferred from theoretical spectra.  This comparison is accomplished by convolving {\it GALEX} far-ultraviolet and {\it Spitzer} 3.6\m\ filter transmission profiles with stellar spectra similar to those described in \S~\ref{sec:seds} but for a wide range of ages (1~Myr to 10~Gyr).  The ages for the respective simulated spectra are shown along the righthand axis at levels corresponding to the computed theoretical far-ultraviolet-to-near-infrared ratios along the lefthand axis.  Assuming these theoretical spectra and a continuous star formation are broadly applicable to the LVL sample, the ages range from several million years up to a few billion years, with the longest-lasting star formation episodes corresponding to the largest deviations from the starburst curve.  It should be stressed that this comparison with theoretical stellar population ages is indicative---if the actual star formation rates have been declining with time instead of remaining constant, then the age axis in Figure~\ref{fig:dstar} overestimates the true ages.


\section {Summary}
\label{sec:summary}

In this contribution we describe the Local Volume Legacy, a {\it Spitzer Space Telescope} infrared imaging program built upon a foundation of {\it GALEX} ultraviolet and ground-based H$\alpha$ imaging of 258 galaxies within 11~Mpc, approximately two-thirds of which are irregulars or dwarf spheroidals.  The proximity and nearly volume-limited nature of the survey are key aspects to this program, enabling multi-wavelength analyses of star formation with high spatial resolution in a manner that is statistically representative of the nearby galaxy population.  First results based primarily on the near-, mid-, and far-infrared data are reported here.  Whereas monochromatic tracers of the far-infrared at 70 and 160\m\ closely track the 3-1100\m\ total-infrared emission, the mid-infrared-to-total-infrared ratios show large dispersions.  The large scatter in comparing dust emission at 8.0\m\ to the total dust emission is likely due to the notable deficiency of PAH emission from the low-metallicity, low-luminosity galaxies prevalent in the LVL survey.  The LVL sample shows a correlation between infrared-to-ultraviolet ratio and ultraviolet spectral slope, but it is shifted to redder colors and/or lower infrared-to-ultraviolet ratios than what is seen for starbursting galaxies and most star-forming galaxies.  In many instances the $\sim$100~Myr ultraviolet emission is more spatially extended than the $\sim$10~Myr (dust) infrared and \hal\ emission, suggesting that the outer parts of many LVL galaxies are older than their inner regions.  Thus, {\it global} flux ratios will naturally yield older (redder) and small infrared-to-ultraviolet ratios than is typically found in starbursts and normal star-forming galaxies.  Theoretical models are utilized to buttress the idea that deviations from the starburst relation correspond to the age of the stellar population that drives the bulk of the ultraviolet luminosity.

\acknowledgements 
Support for this work, part of the {\it Spitzer Space Telescope} Legacy Science Program, was provided by NASA 
and
issued by the Jet Propulsion Laboratory, California Institute of Technology under NASA contract 1407.  This research has made use of the NASA/IPAC Extragalactic Database which is operated by JPL/Caltech, under contract with NASA.  This publication makes use of data products from the Two Micron All Sky Survey, which is a joint project of the University of Massachusetts and the Infrared Processing and Analysis Center/California Institute of Technology, funded by the National Aeronautics and Space Administration and the National Science Foundation.  IRAF, the Image Reduction and Analysis Facility, has been developed by the National Optical Astronomy Observatories and the Space Telescope Science Institute.  We gratefully acknowledge NASA's support for construction, operation, and science analysis for the GALEX mission, developed in cooperation with the Centre National d'Etudes Spatiales of France and the Korean Ministry of Science and Technology.

\begin{deluxetable}{lrrrrccrrrr}
\def\a{\tablenotemark{a}}
\def\b{\tablenotemark{b}}
\def\c{\tablenotemark{c}}
\def\d{$^\dagger$}
\def\p{$\pm$}
\tabletypesize{\scriptsize}
\tablenum{1}
\label{tab:sample}
\tablecaption{Galaxy Sample and Infrared Photometry Apertures}
\tablewidth{0pc}
\tablehead{
\colhead{Galaxy} &
\colhead{$m_B$} &
\colhead{c$z$} &
\colhead{$D$} &
\colhead{$T$} &
\colhead{{\it E(B-V)}} &
\colhead{$\alpha_0$~\&~$\delta_0$} &
\colhead{2$a$} &
\colhead{2$b$} &
\colhead{PA}  &
\colhead{log(TIR)} 
\\
\colhead{} &
\colhead{(mag)} &
\colhead{(km~s$^{-1}$)} &
\colhead{(Mpc)} &
\colhead{} &
\colhead{(mag)} &
\colhead{(J2000)} &
\colhead{($\arcsec$)} &
\colhead{($\arcsec$)} &
\colhead{($\degr$)} &
\colhead{($L_\odot$)} 
}
\startdata
WLM            &11.03&   -116& 0.92&  10&0.04&000158.9$-$152655& 672& 340&  0&    6.7\\
NGC0024        &12.19&    554& 8.13&  5 &0.02&000955.9$-$245755& 301& 216&225&    8.8\\
NGC0045        &11.32&    471& 7.07&  8 &0.02&001404.6$-$231101& 577& 456&336&    8.9\\
NGC0055        & 8.42&    129& 2.17&  9 &0.01&001508.2$-$391256&2251& 714&106&    9.2\\
NGC0059        &13.12&    382& 5.30&$-$3&0.02&001525.8$-$212646& 256& 180&302&    7.5\\
ESO410-G005    &14.95&\nodata& 1.90&$-$1&0.01&001531.6$-$321052& 122&  90&308&$<$ 6.0\\
Sculptor-dE1   &17.78&\nodata& 4.20&  10&0.02&002351.6$-$244204& 160& 103&  0&$<$ 6.8\\
ESO294-G010    &15.52&    117& 1.90&$-$3&0.01&002633.7$-$415120& 165& 102&  0&$<$ 6.1\\
\enddata     
\tablecomments{\footnotesize Magnitudes, heliocentric redshifts, distances, and optical morphologies are from Kennicutt et al.\ (2008), and the Milky Way foreground extinctions are from Schlegel, Finkbeiner, \& Davis(1998).  The remaining columns define the elliptical apertures used for the photometry presented in Table~\ref{tab:data}; the position angles are measured east of north.}
\tablenotetext{a}{\footnotesize Infrared photometry for the Large and Small Magellanic Clouds are from Bernard et al.\ (2008) and Gordon et al.\ (2009), respectively.}
\tablenotetext{b}{\footnotesize NGC~1800 and MCG-05-13-004 spatially overlap so separate photometry for MCG-05-13-004 is not provided.}
\end{deluxetable}
\begin{deluxetable}{lrrrrrrrrrr}
\rotate
\def\a{\tablenotemark{a}}
\def\b{\tablenotemark{b}}
\def\c{\tablenotemark{c}}
\def\d{\tablenotemark{d}}
\def\e{\tablenotemark{e}}
\def\f{\tablenotemark{f}}
\def\p{$\pm$}
\tabletypesize{\scriptsize}
\tablenum{2}
\label{tab:data}
\tablecaption{Infrared Flux Densities}
\tablewidth{0pc}
\tablehead{
\colhead{Galaxy} &
\colhead{2MASS J} &
\colhead{2MASS H} &
\colhead{2MASS K$_{\rm s}$} &
\colhead{IRAC} &
\colhead{IRAC} &
\colhead{IRAC} &
\colhead{IRAC} &
\colhead{MIPS} &
\colhead{MIPS} &
\colhead{MIPS} 
\\
\colhead{} &
\colhead{1.25\m} &
\colhead{1.65\m} &
\colhead{2.17\m} &
\colhead{3.6\m} &
\colhead{4.5\m} &
\colhead{5.8\m} &
\colhead{8.0\m} &
\colhead{24\m} &
\colhead{70\m} &
\colhead{160\m} 
\\
\colhead{} & 
\colhead{(Jy)} &
\colhead{(Jy)} &
\colhead{(Jy)} &
\colhead{(Jy)} &
\colhead{(Jy)} &
\colhead{(Jy)} &
\colhead{(Jy)} &
\colhead{(Jy)} &
\colhead{(Jy)} &
\colhead{(Jy)} 
}
\startdata
WLM\e\f        &2.45\p0.15E-1&3.13\p0.20E-1&1.17\p0.16E-1&8.81\p1.19E-2&6.29\p0.86E-2&4.11\p0.52E-2&5.10\p0.64E-2&7.57\p0.83E-2&2.01\p0.25E+0&3.88\p0.61E+0\\      
NGC0024\e\f    &2.32\p0.12E-1&2.49\p0.14E-1&1.90\p0.12E-1&1.02\p0.14E-1&6.84\p1.01E-2&8.22\p1.16E-2&1.18\p0.16E-1&1.20\p0.13E-1&2.23\p0.27E+0&7.12\p1.11E+0\\      
NGC0045        &2.08\p0.13E-1&2.29\p0.17E-1&1.67\p0.18E-1&2.00\p0.27E-1&1.27\p0.17E-1&6.56\p0.83E-2&1.69\p0.21E-1&1.88\p0.20E-1&3.86\p0.47E+0&1.23\p0.19E+1\\      
NGC0055\e\f    &3.31\p0.17E+0&3.08\p0.16E+0&2.63\p0.14E+0&2.02\p0.27E+0&1.39\p0.19E+0&1.52\p0.19E+0&2.41\p0.30E+0&6.29\p0.68E+0&1.26\p0.15E+2&2.53\p0.40E+2\\      
NGC0059        &8.48\p0.55E-2&1.13\p0.08E-1&6.36\p0.72E-2&3.29\p0.45E-2&2.23\p0.31E-2&1.57\p0.20E-2&1.18\p0.15E-2&4.40\p0.48E-2&5.17\p0.66E-1&4.03\p0.73E-1\\      
ESO410-G005    &$<$8.18E-3   &$<$1.22E-2   &$<$1.49E-2   &1.09\p0.15E-2&4.52\p0.62E-3&6.13\p0.82E-3&4.52\p0.58E-3&$<$4.61E-3   &$<$6.23E-2   &$<$2.35E-1   \\      
Sculptor-dE1   &$<$5.04E-3   &$<$7.50E-3   &$<$9.14E-3   &$<$2.80E-4   &$<$4.30E-4   &$<$1.56E-3   &$<$1.70E-3   &$<$5.68E-3   &$<$7.66E-2   &$<$2.89E-1   \\      
ESO294-G010    &1.73\p0.23E-2&2.12\p0.34E-2&1.35\p0.39E-2&4.79\p0.65E-3&3.12\p0.43E-3&$<$1.56E-3   &$<$1.71E-3   &$<$5.69E-3   &$<$7.68E-2   &$<$2.90E-1   \\
\enddata     
\vskip-0.3in
\tablecomments{\footnotesize The compact table entry format T.UV$\pm$W.XYEZ implies (T.UV$\pm$W.XY)$\times10^{\rm Z}$.  See \S~\ref{sec:data} for corrections that have been applied to the data.  The uncertainties include both statistical and systematic effects ($\lesssim10$\% for the near-infrared data).  5$\sigma$ upper limits are provided for non-detections.}
\tablenotetext{a}{\footnotesize Infrared photometry for the Large and Small Magellanic Clouds are from Bernard et al.\ (2008) and Gordon et al.\ (2009), respectively.}
\tablenotetext{b}{\footnotesize NGC~1800 and MCG-05-13-004 spatially overlap so separate photometry for MCG-05-13-004 is not provided.}
\tablenotetext{c}{\footnotesize The bright cores of NGC~3034 (M82) and NGC~0253 produce particularly challenging {\it Spitzer} data, including the effects of saturation at 24\m.  The 24\m\ flux densities are approximated from {\it IRAS} data via $\nu f_\nu(24\micron) = \nu f_\nu(25\micron)$, and the data at other wavelengths should be used with caution.}
\tablenotetext{d}{\footnotesize Only 4.5 and 8.0\m\ data were obtained for IC~5152.}
\tablenotetext{e}{\footnotesize IRAC imaging taken from the {\it Spitzer} archives.}
\tablenotetext{f}{\footnotesize MIPS imaging taken from the {\it Spitzer} archives.}
\end{deluxetable}
\begin{deluxetable}{cccc}
\tabletypesize{\scriptsize}
\tablenum{3}
\label{tab:IRAC_ap}
\tablecaption{IRAC Aperture Correction Parameters}
\tablewidth{0pc}
\tablehead{
\colhead{$\lambda$} &
\colhead{A} &
\colhead{B} &
\colhead{C} 
}
\startdata
3.6\m&   0.82&    0.370&   0.910\\
4.5\m&   1.16&    0.443&   0.940\\
5.8\m&   1.49&    0.207&   0.710\\
8.0\m&   1.37&    0.330&   0.740\\
\enddata
\tablecomments{\footnotesize See Equation~\ref{eqn:irac} and http://ssc.spitzer.caltech.edu/irac/calib/extcal/}
\end{deluxetable}
\begin{deluxetable}{llc}
\tabletypesize{\scriptsize}
\tablenum{4}
\label{tab:ratios}
\tablecaption{LVL Infrared Ratios and Dispersions}
\tablewidth{0pc}
\tablehead{
\colhead{Ratio} &
\colhead{Median} &
\colhead{Dispersion} 
\\
\colhead{} &
\colhead{} &
\colhead{(dex)} 
}
\startdata
$f_\nu(70\micron)/f_\nu(160\micron$)        &   0.53~&0.21\\
$f_\nu(8.0\micron)/f_\nu(24\micron$)        &   0.53~&0.27\\
$\nu f_\nu(8.0\micron$)/{\it TIR}           &   0.12~&0.23\\
$\nu f_\nu(8.0\micron)_{\rm dust}$/{\it TIR}&   0.083&0.39\\
$\nu f_\nu(24\micron$)/{\it TIR}            &   0.074&0.16\\
$\nu f_\nu(70\micron$)/{\it TIR}            &   0.46~&0.11\\
$\nu f_\nu(160\micron$)/{\it TIR}           &   0.39~&0.12\\
\enddata
\end{deluxetable}

\clearpage
\newpage
\begin{figure}
 \caption{Schematic representation of the tiered volume coverage for the Local Volume Legacy survey.  The 258 galaxies in the LVL sample include {\it i}) 69 early- and late-type galaxies primarily within the inner 3.5~Mpc for which {\it HST} observations exist from the ANGST program, and {\it ii}) a magnitude-limited sample of spiral and irregular galaxies from 11HUGS out to 11~Mpc.  The {\it Spitzer} infrared imaging data that are being being collected by LVL complement the ground-based H$\alpha$ and {\it GALEX} ultraviolet imaging already available for the sample.  Image credit: Pete Marenfeld (NOAO/AURA/NSF).}
 \label{fig:volume_coverage}
\end{figure}

\begin{figure}
 \plotone{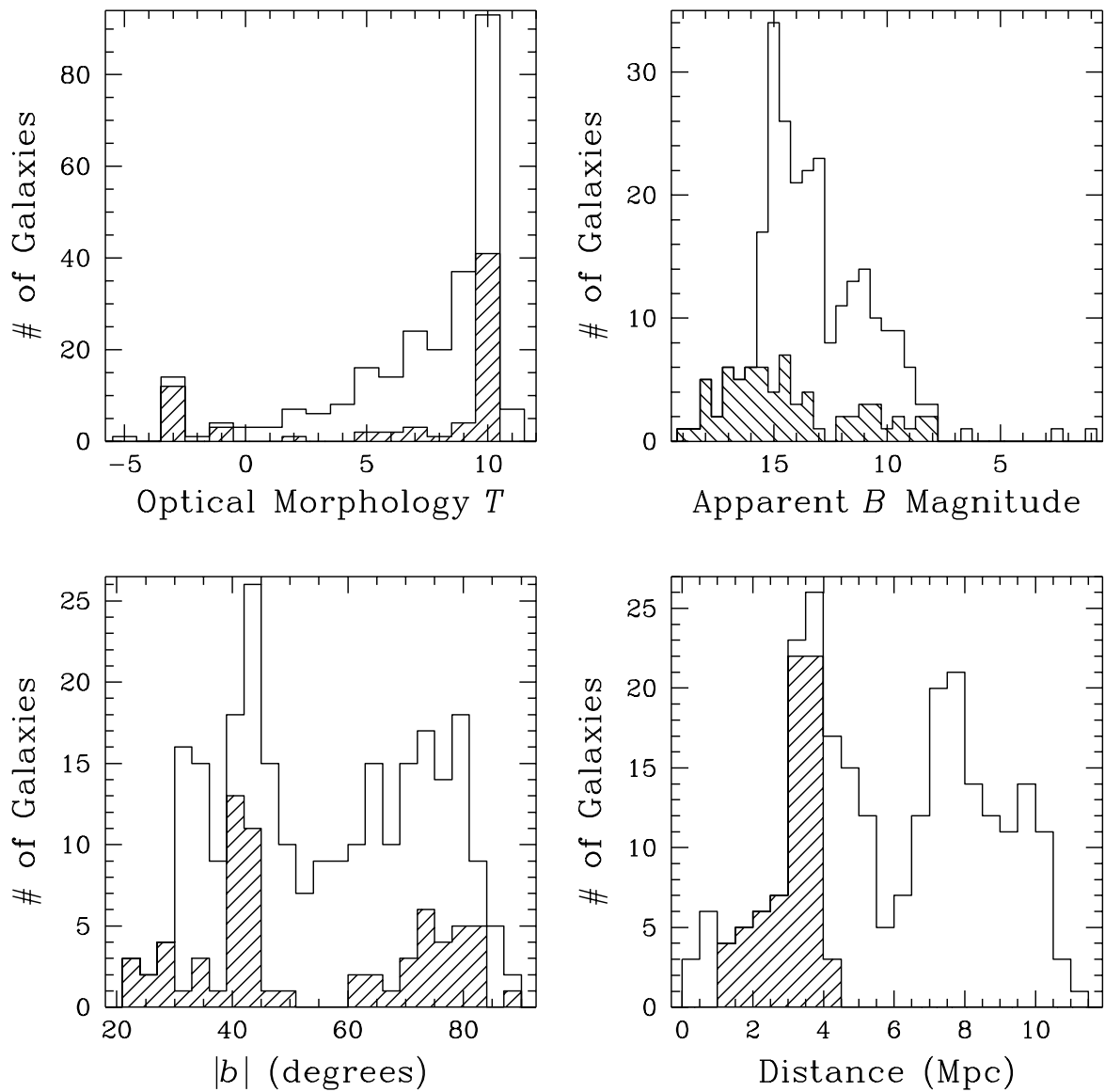}
 \caption{The distributions of RC3 morphological type (top left), apparent B magnitude (top right), Galactic latitude (bottom left) and distance (bottom right) for the LVL sample.  The histograms outlined by the solid lines show the entire sample of 258 galaxies, whereas the shaded portions indicate the ANGST sub-sample for which {\it HST} resolved stellar populations observations are available.}
 \label{fig:sample}
\end{figure}

\begin{figure}
 \plotone{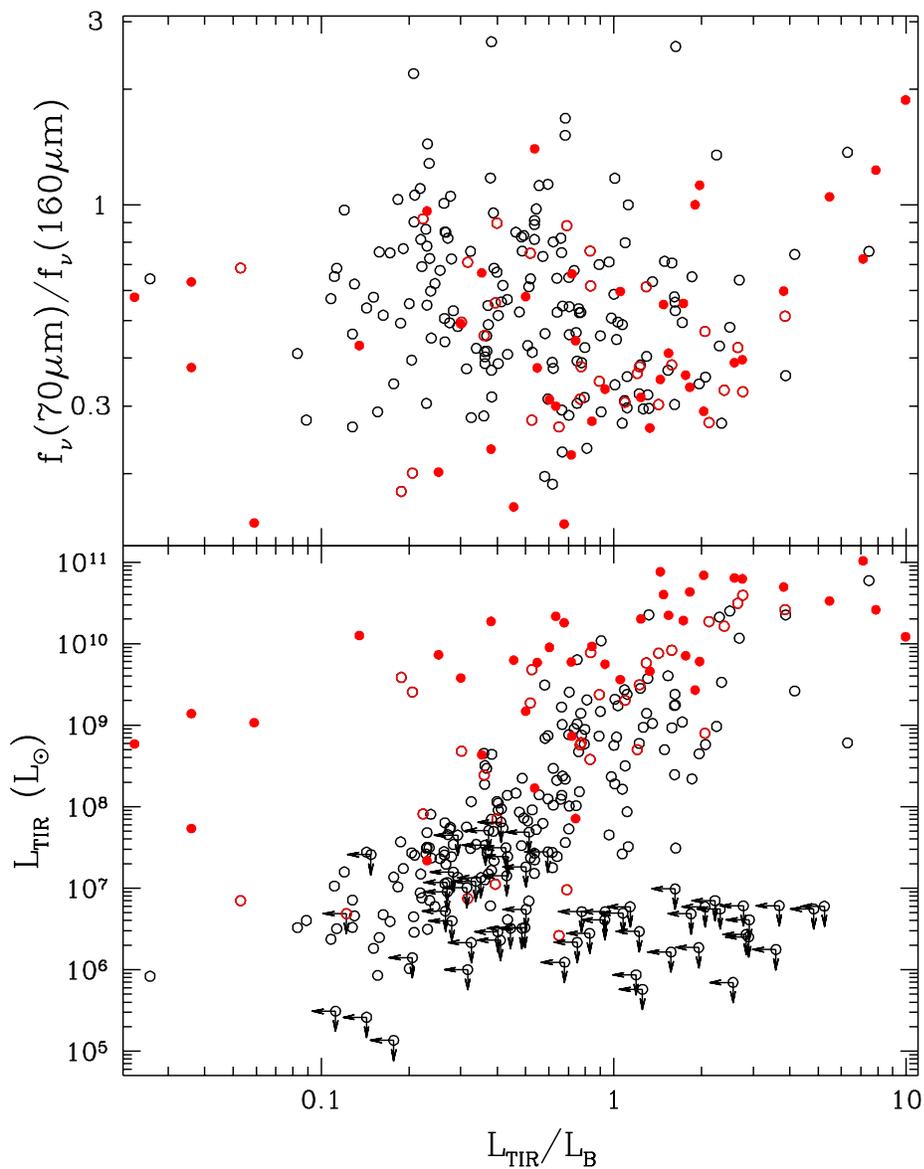}
 \caption{The range of 3--1100\m\ total infrared and total infrared-to-optical properties in the LVL (open black circles) and SINGS (filled red circles) samples.  Open red circles denote galaxies in both samples.  SINGS data are taken from Dale et al.\ (2007).  Arrows indicate 3$\sigma$ upper limits.}
 \label{fig:sample_phase_space}
\end{figure}

\begin{figure}
 \plotone{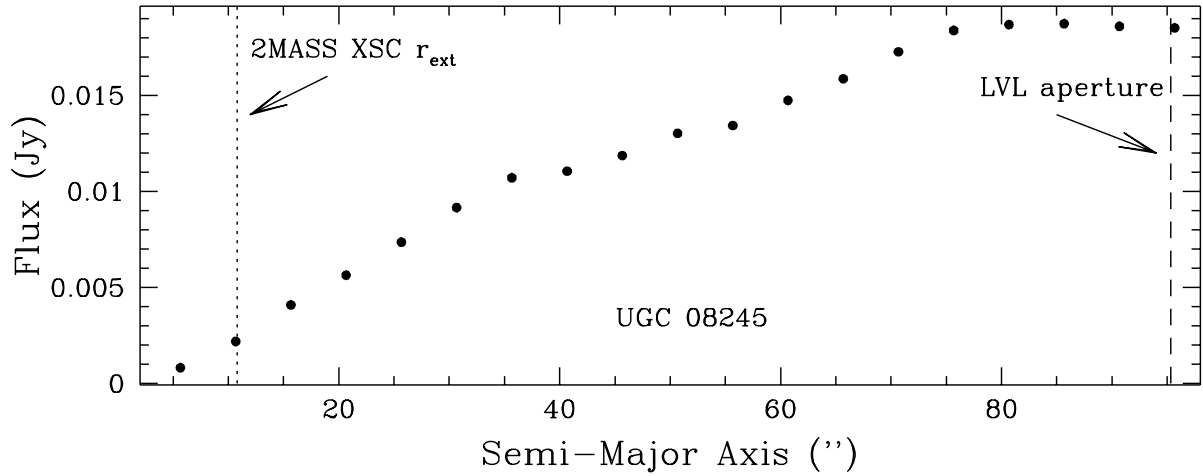}
 \caption{A $K$ band curve-of-growth plot that includes a comparison of extraction apertures for UGC~08245 from our work (dashed line) and from the 2MASS `XSC' Extended Source Catalog (dotted line).  The larger aperture captures more of the faint diffuse emission from the galaxy's disk.}
 \label{fig:u8245}
\end{figure}

\begin{figure}
 \plotone{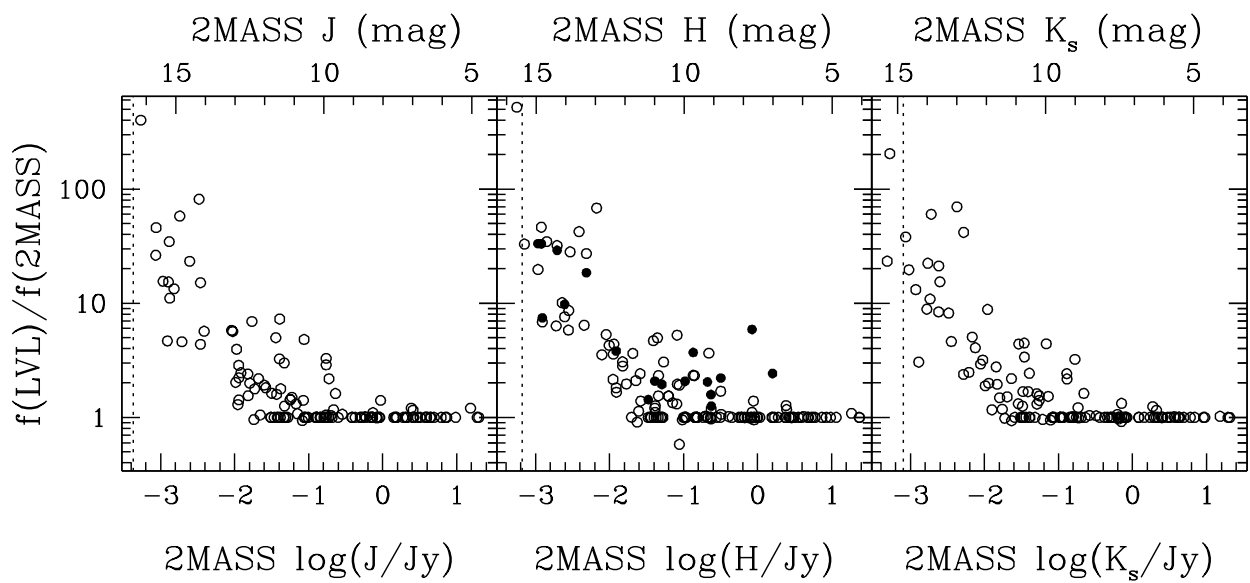}
 \caption{A comparison of near-infrared fluxes independently extracted from the 2MASS image archives using the apertures described in \S~\ref{sec:aperture} with those from the 2MASS Extended Source Catalog.  The filled circles are based on deep $H$ band imaging of nearby galaxies with the 3.9~m Anglo-Australian Telescope (Kirby et al.\ 2008).  The vertical dotted lines indicate the 2MASS 10$\sigma$ point source detection limits.}
 \label{fig:2mass}
\end{figure}

\newpage
\begin{figure}
 \plotone{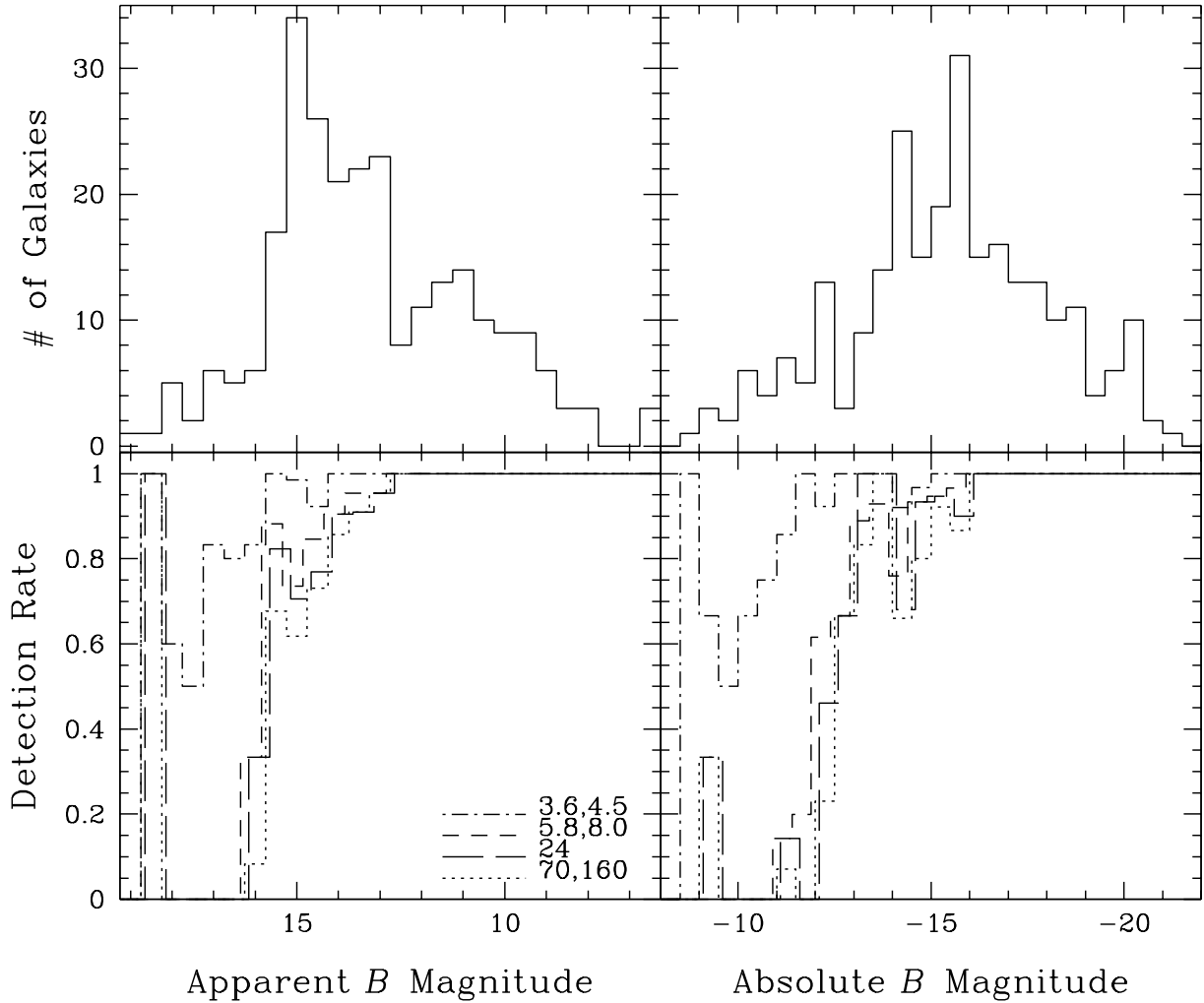}
 \caption{Top: The distribution of LVL galaxies as a function of apparent and absolute $B$-band magnitudes.  Bottom: The imaging detection rates for different \Spitzer\ wavelengths.  Two histograms are slightly offset in magnitude for purposes of clarity.  Note that the {\it average} detection rate is displayed at 3.6 and 4.5\m\ (dot dashed line) and at 5.8 and 8.0\m\ (short dashed line).} 
 \label{fig:detection_rate}
\end{figure}

\begin{figure}
 \plotone{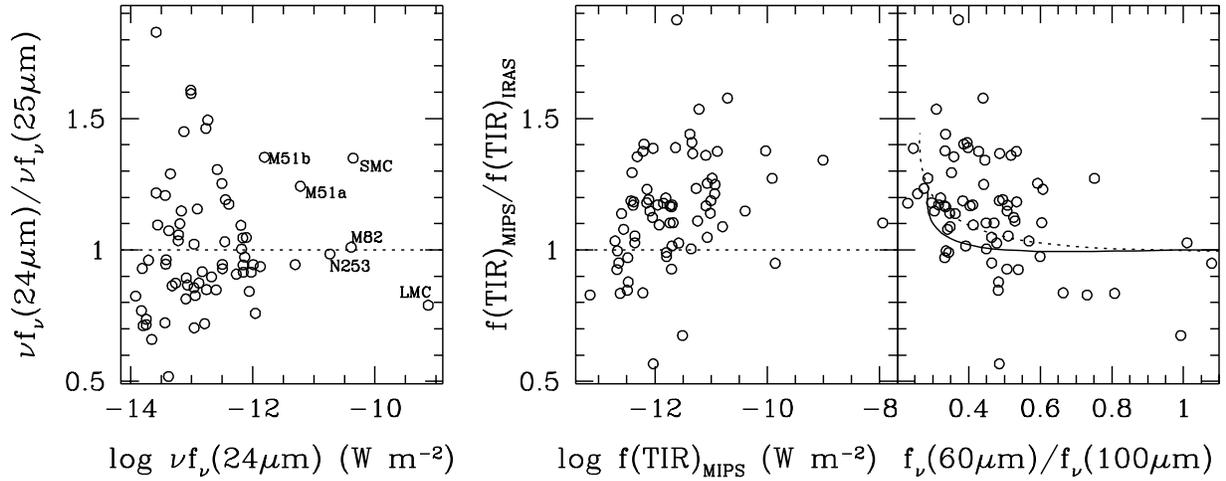}
 \caption{A comparison of {\it Spitzer} and {\it IRAS} data for the LVL sample.  The left-most panel shows the ratio of {\it Spitzer} 24\m\ and {\it IRAS} 25\m\ data, and the right two panels compare the total infrared as measured from {\it Spitzer} and {\it IRAS}.  The dotted line in the left two panels indicate a ratio of unity, whereas the solid and dotted lines in the right-most panel indicate model predictions from Dale \& Helou (2002) and Dale et al.\ (2001), respectively.}
 \label{fig:iras}
\end{figure}

\newpage
\begin{figure}
 \caption{A multi-wavelength mosaic of spiral galaxy NGC~5236, displaying the range of imaging available for the LVL survey.  The \hal\ image is continuum-subtracted.  The images are all 9\farcm9$\times$10\farcm7, or 13~kpc~$\times$~14~kpc at a distance of 4.47~Mpc.}
 \label{fig:mosaicA}
\end{figure}

\begin{figure}
 \caption{The same as Figure~\ref{fig:mosaicA} but for the irregular galaxy UGC~05829.  The images are all 5\farcm0$\times$5\farcm4, or 11~kpc~$\times$~12~kpc at a distance of 7.88~Mpc.}
 \label{fig:mosaicB}
\end{figure}

\begin{figure}
 \plotone{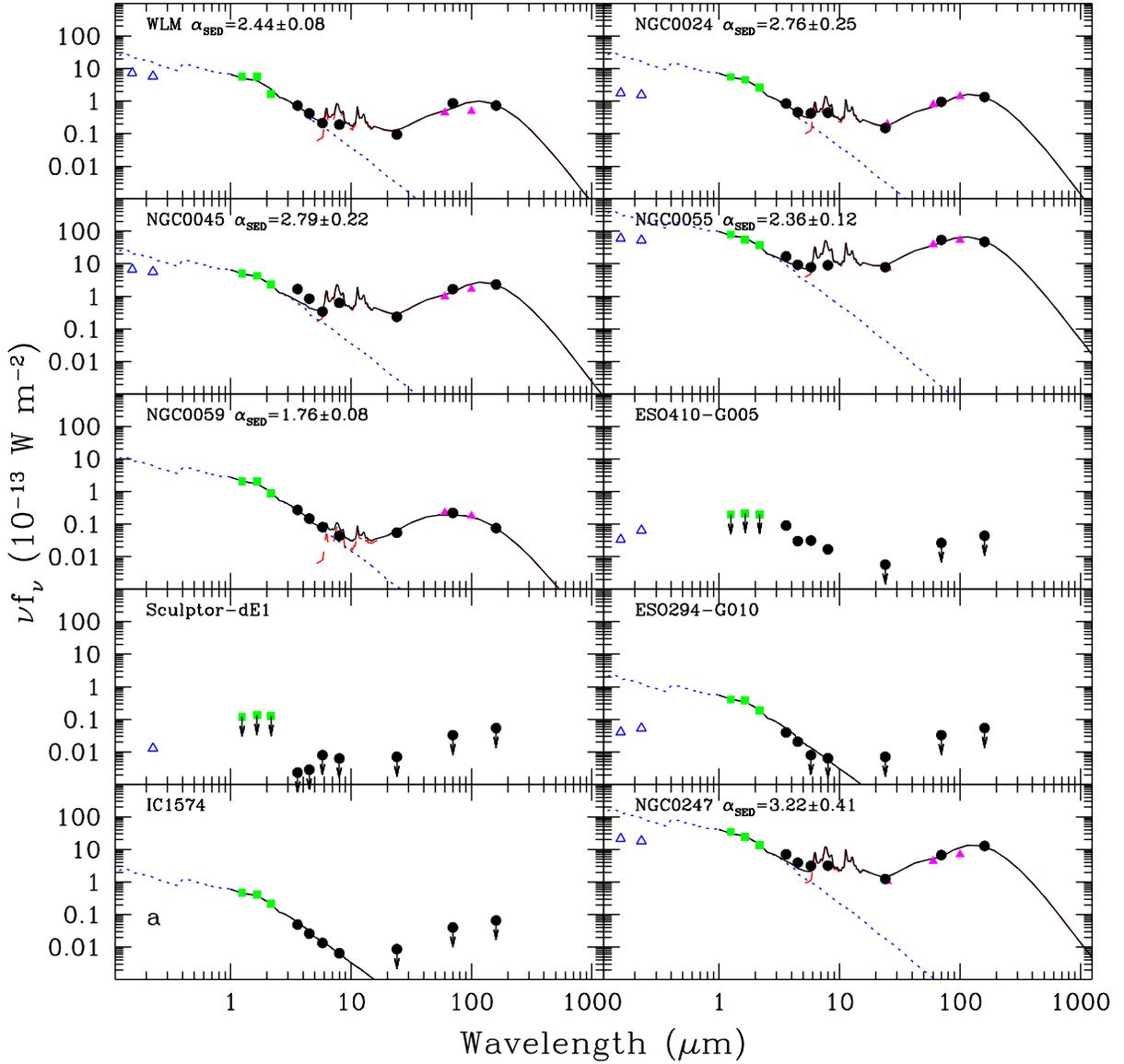}
 \caption{Globally-integrated 0.15-160\m\ spectral energy distributions for the LVL sample.  {\it GALEX}, 2MASS, {\it IRAS}, and {\it Spitzer} data are represented by open triangles, filled squares, filled triangles, and filled circles, respectively.  Downward-pointing arrows, if present, indicate 5$\sigma$ upper limits.  The solid curve is the sum of a dust (dashed) and a stellar (dotted) model.  The dust curve is a Dale \& Helou (2002) model fitted to the amplitude and ratios of the observed 24, 70, and 160\m\ fluxes; the $\alpha_{\rm SED}$ listed within each panel parameterizes the distribution of dust mass as a function of heating intensity, as described in Dale \& Helou (2002).  The stellar curve, serving merely as a fiducial visual aid, is a 1~Gyr continuous star formation, solar metallicity curve from Vazquez \& Leitherer (2005) fitted to the 2MASS data (see \S~\ref{sec:seds} for details).}
 \label{fig:seds01}
\end{figure}

\addtocounter{figure}{-1}
\begin{figure} 
 \plotone{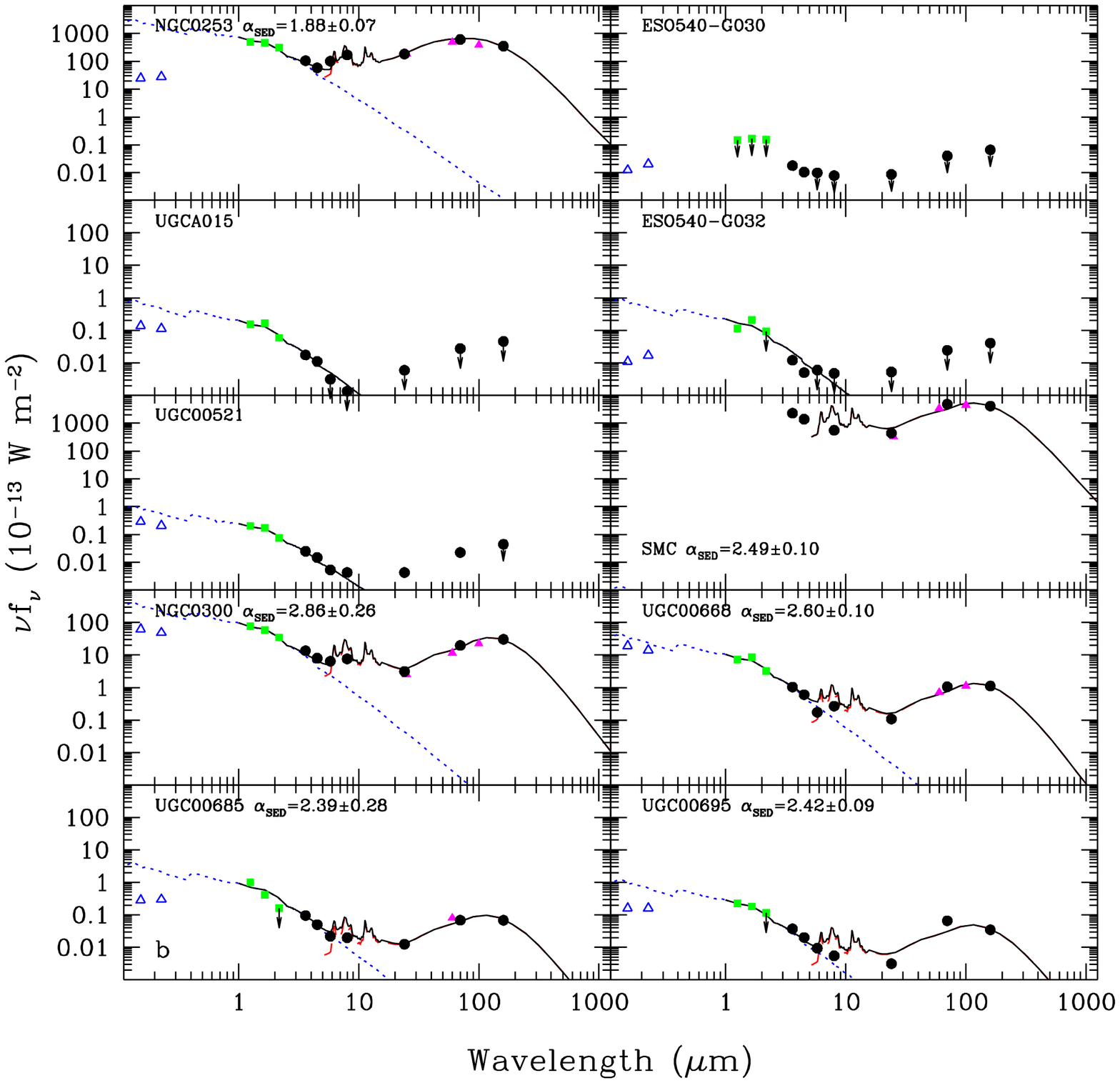} 
 \caption{Globally-integrated 0.15-160\m\ spectral energy distributions for the LVL sample (continued).}
\end{figure}

\addtocounter{figure}{-1}
\begin{figure}
 \plotone{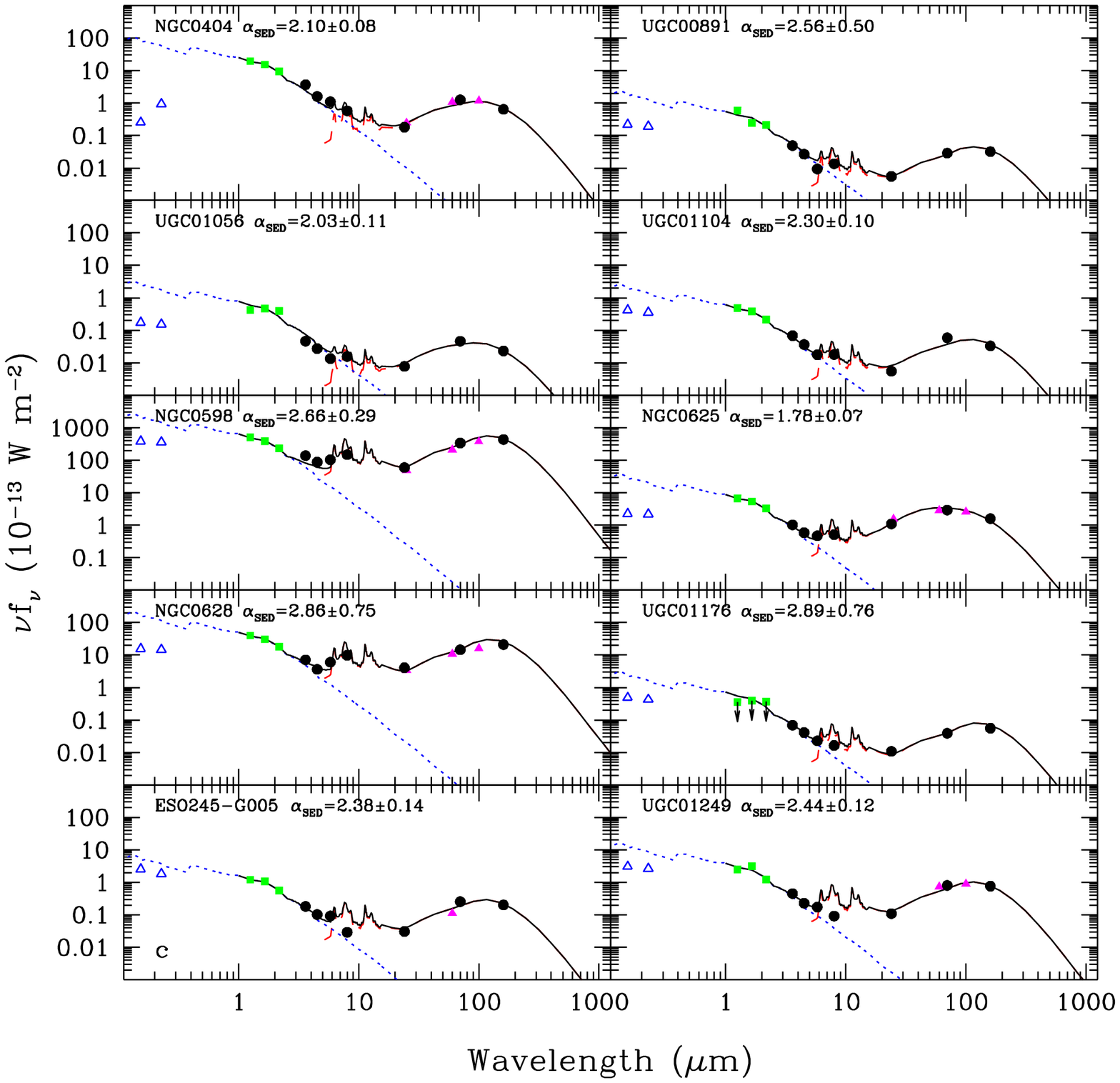}
 \caption{Globally-integrated 0.15-160\m\ spectral energy distributions for the LVL sample (continued).}
\end{figure}

\addtocounter{figure}{-1}
\begin{figure}
 \plotone{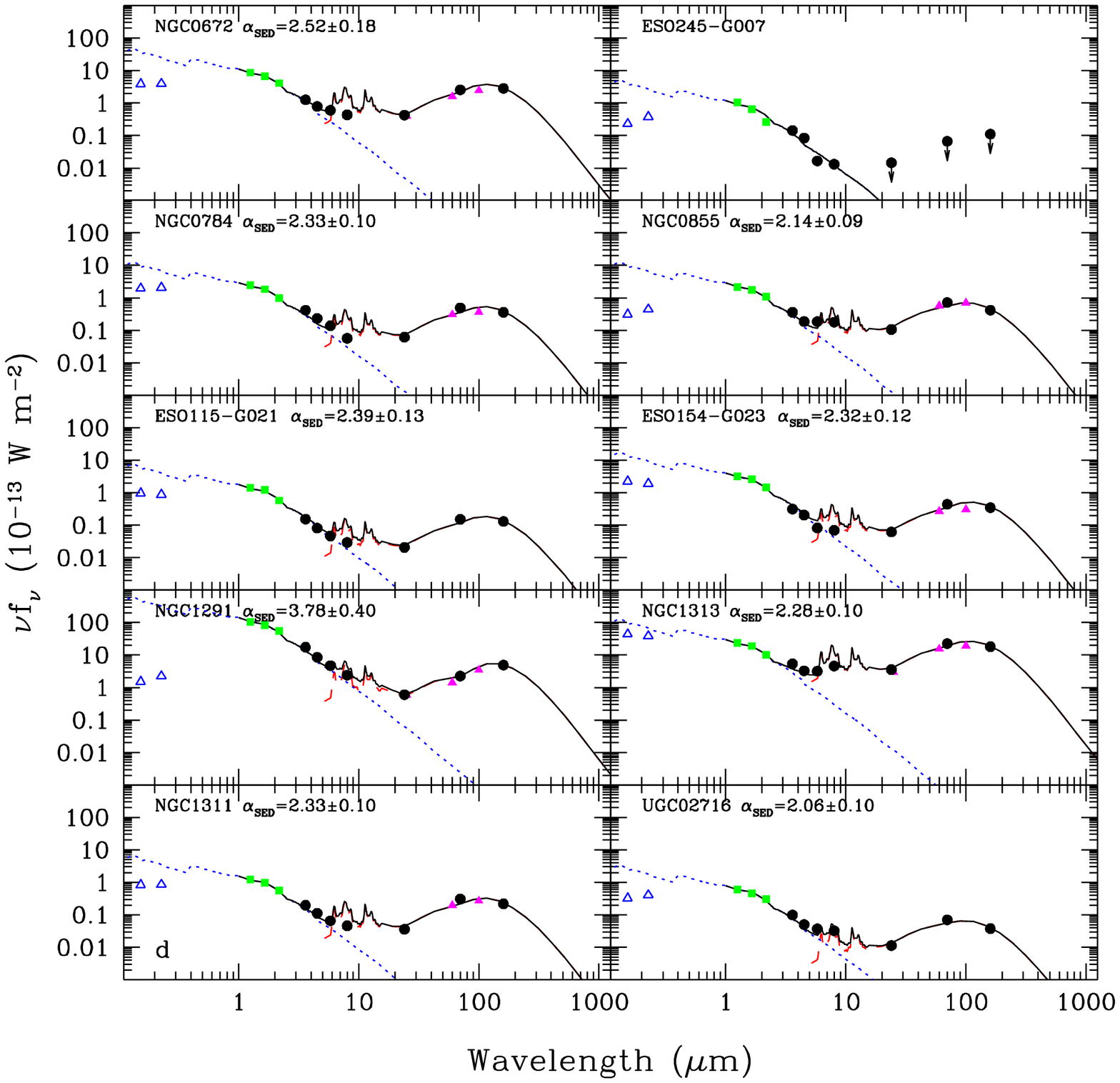}
 \caption{Globally-integrated 0.15-160\m\ spectral energy distributions for the LVL sample (continued).}
\end{figure}

\addtocounter{figure}{-1}
\begin{figure}
 \plotone{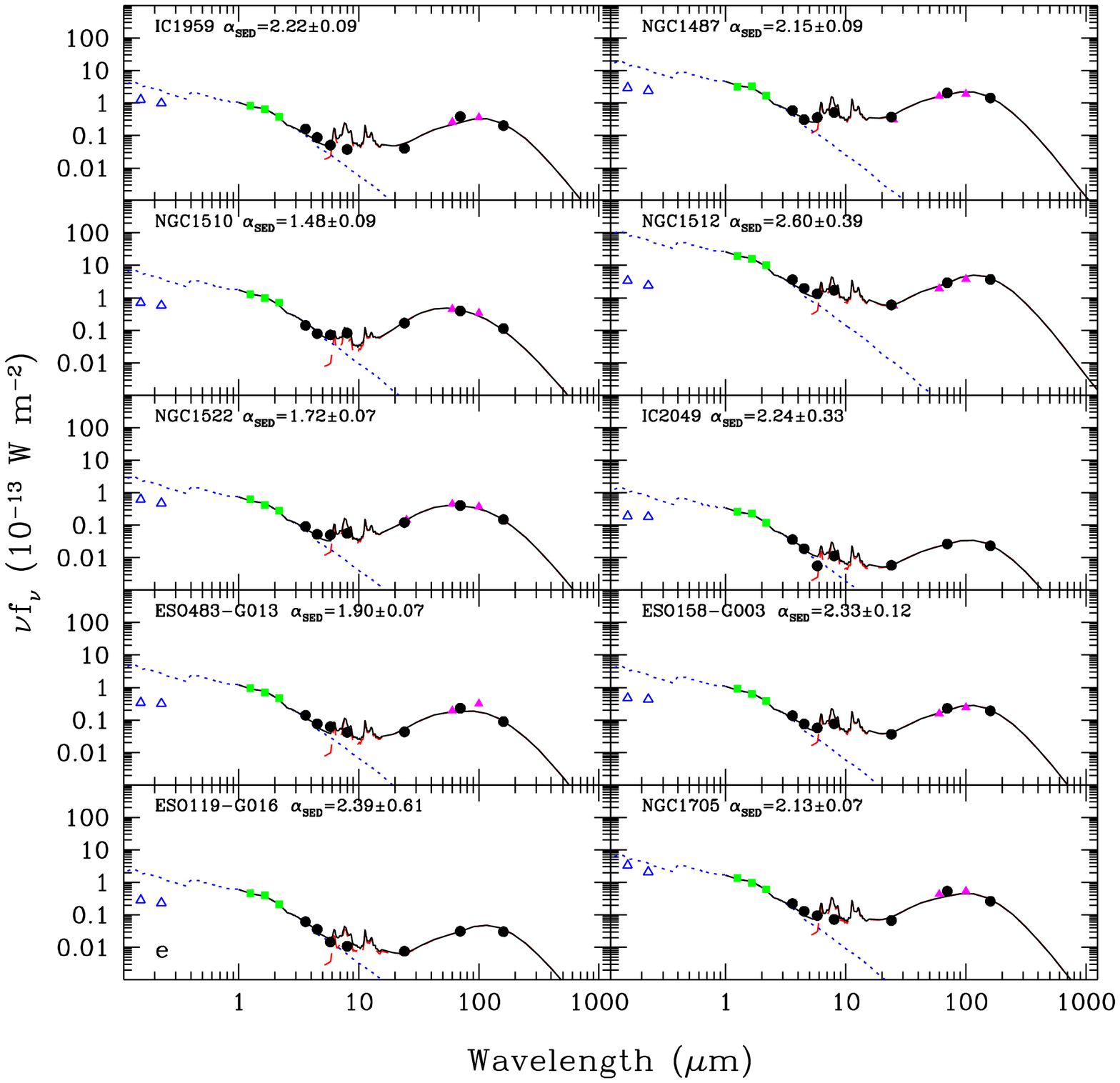}
 \caption{Globally-integrated 0.15-160\m\ spectral energy distributions for the LVL sample (continued).}
\end{figure}

\addtocounter{figure}{-1}
\begin{figure}
 \plotone{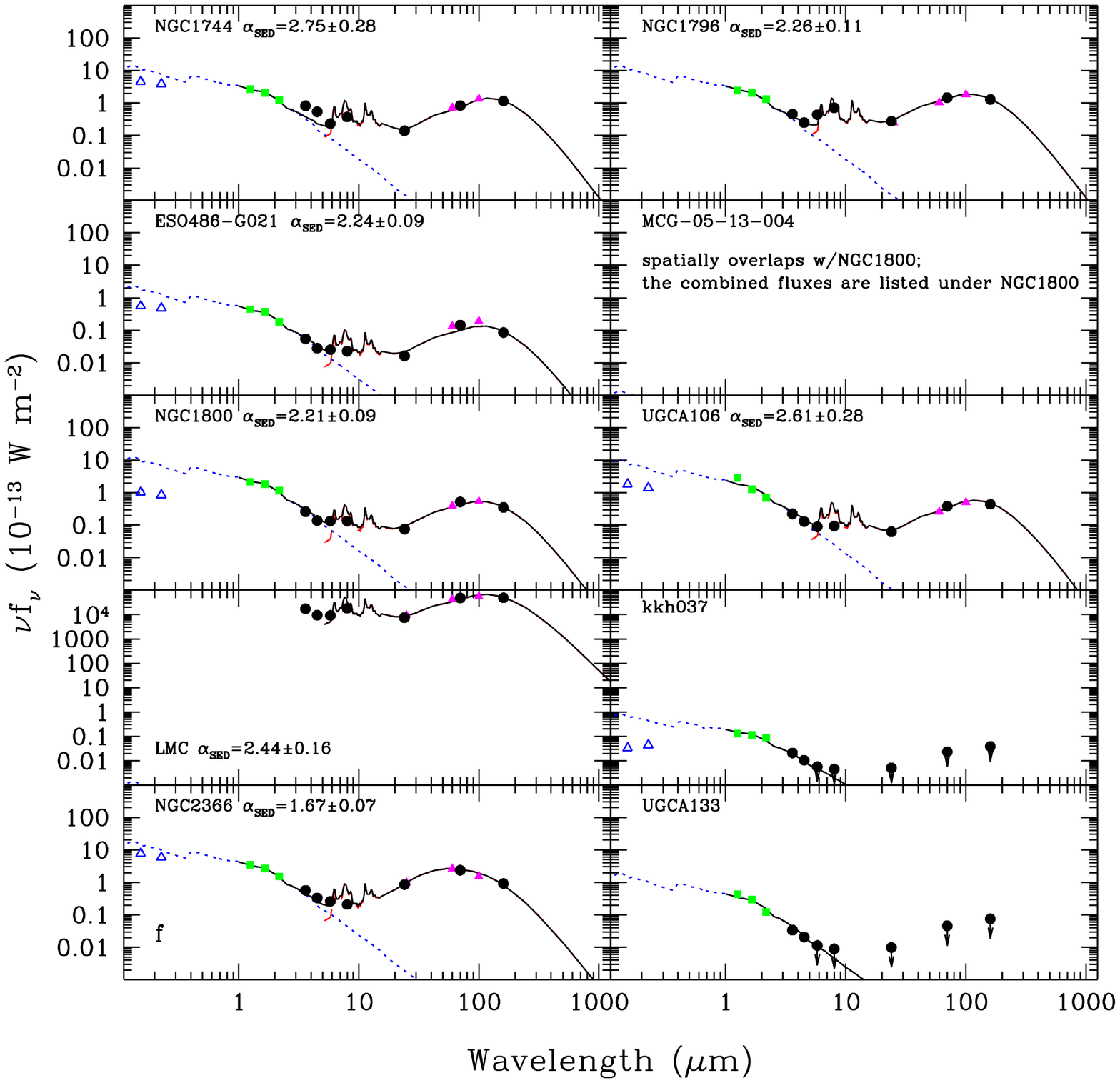}
 \caption{Globally-integrated 0.15-160\m\ spectral energy distributions for the LVL sample (continued).}
\end{figure}

\addtocounter{figure}{-1}
\begin{figure}
 \plotone{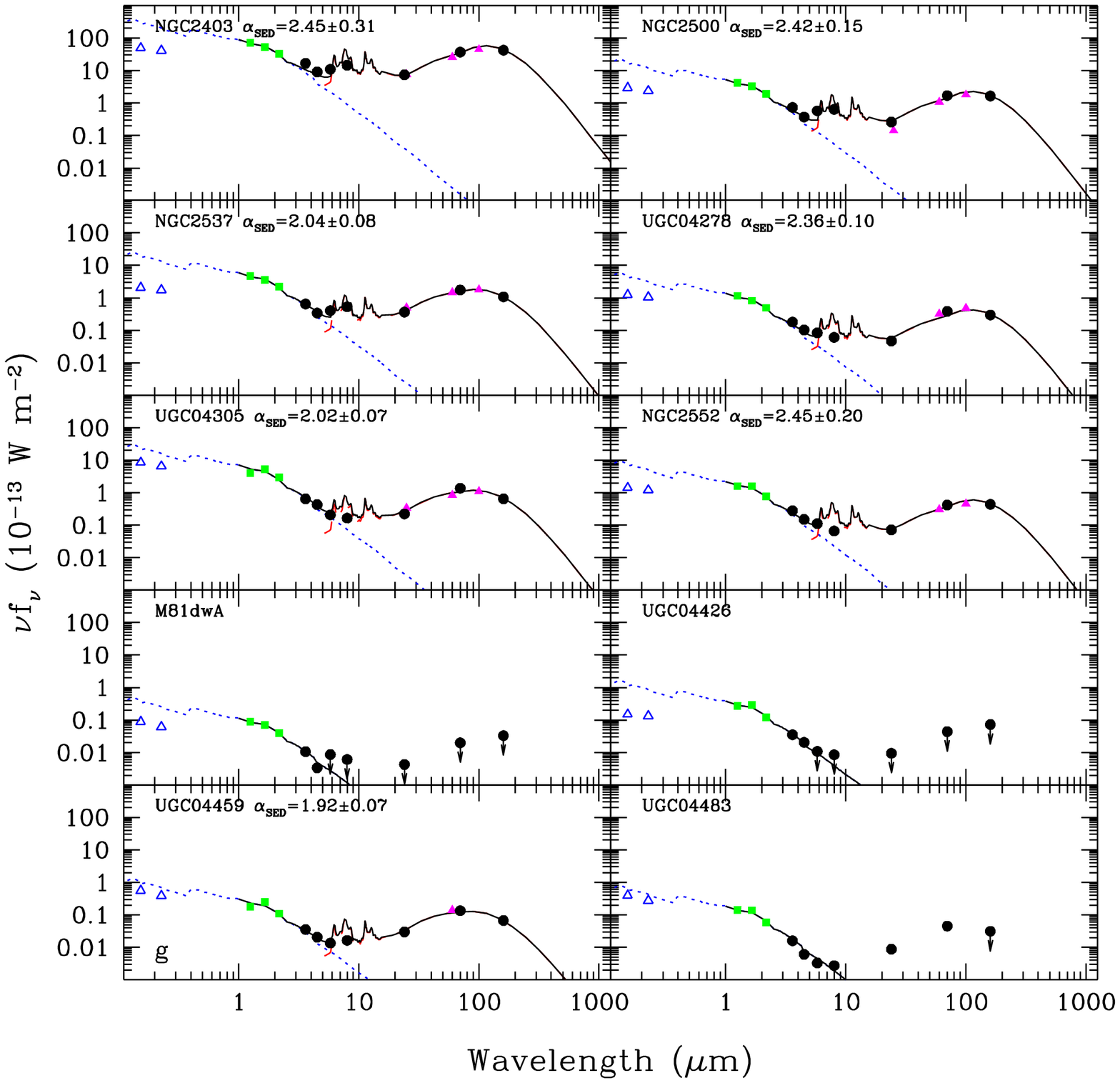}
 \caption{Globally-integrated 0.15-160\m\ spectral energy distributions for the LVL sample (continued).}
\end{figure}

\addtocounter{figure}{-1}
\begin{figure}
 \plotone{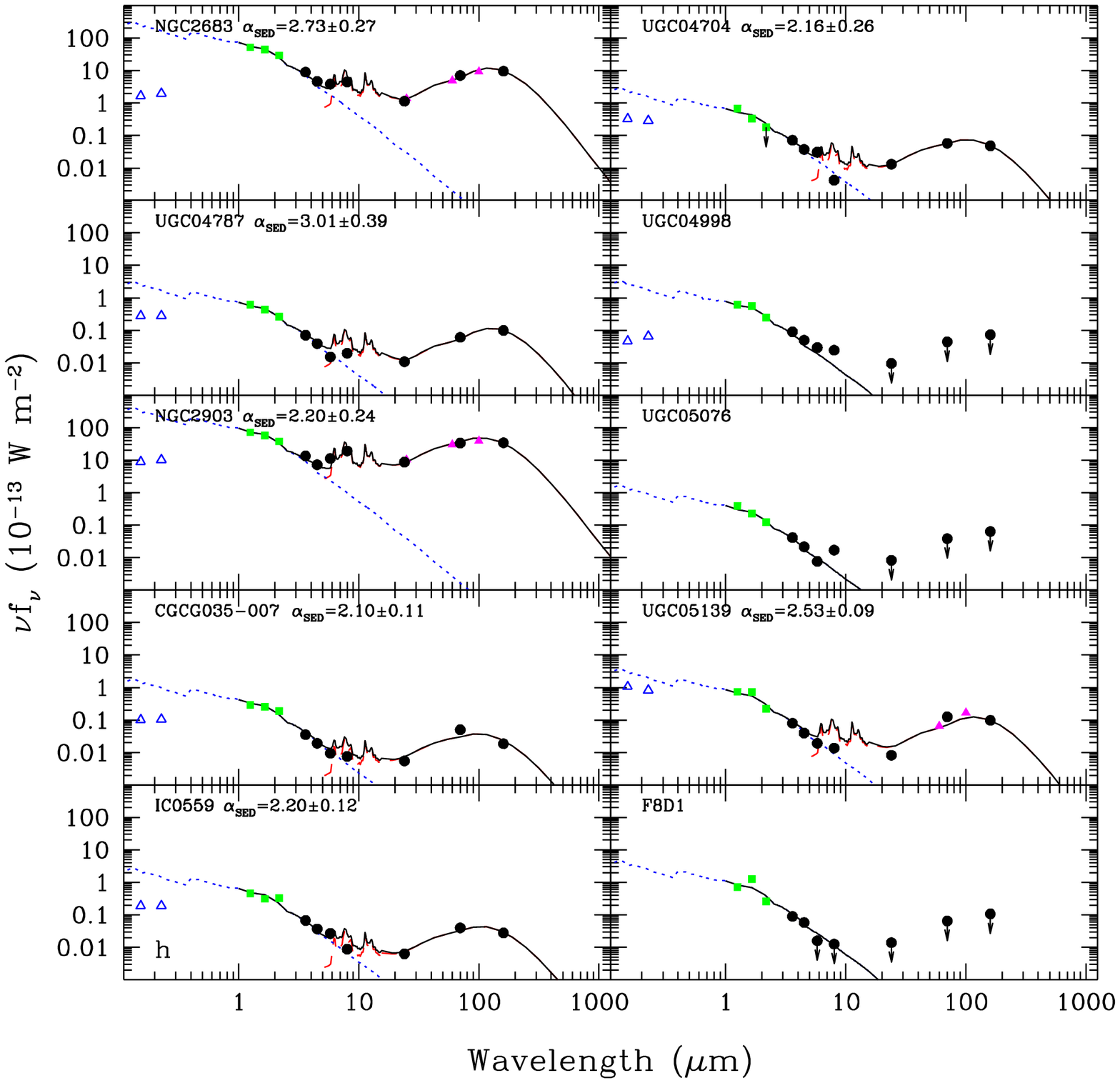}
 \caption{Globally-integrated 0.15-160\m\ spectral energy distributions for the LVL sample (continued).}
\end{figure}

\addtocounter{figure}{-1}
\begin{figure}
 \plotone{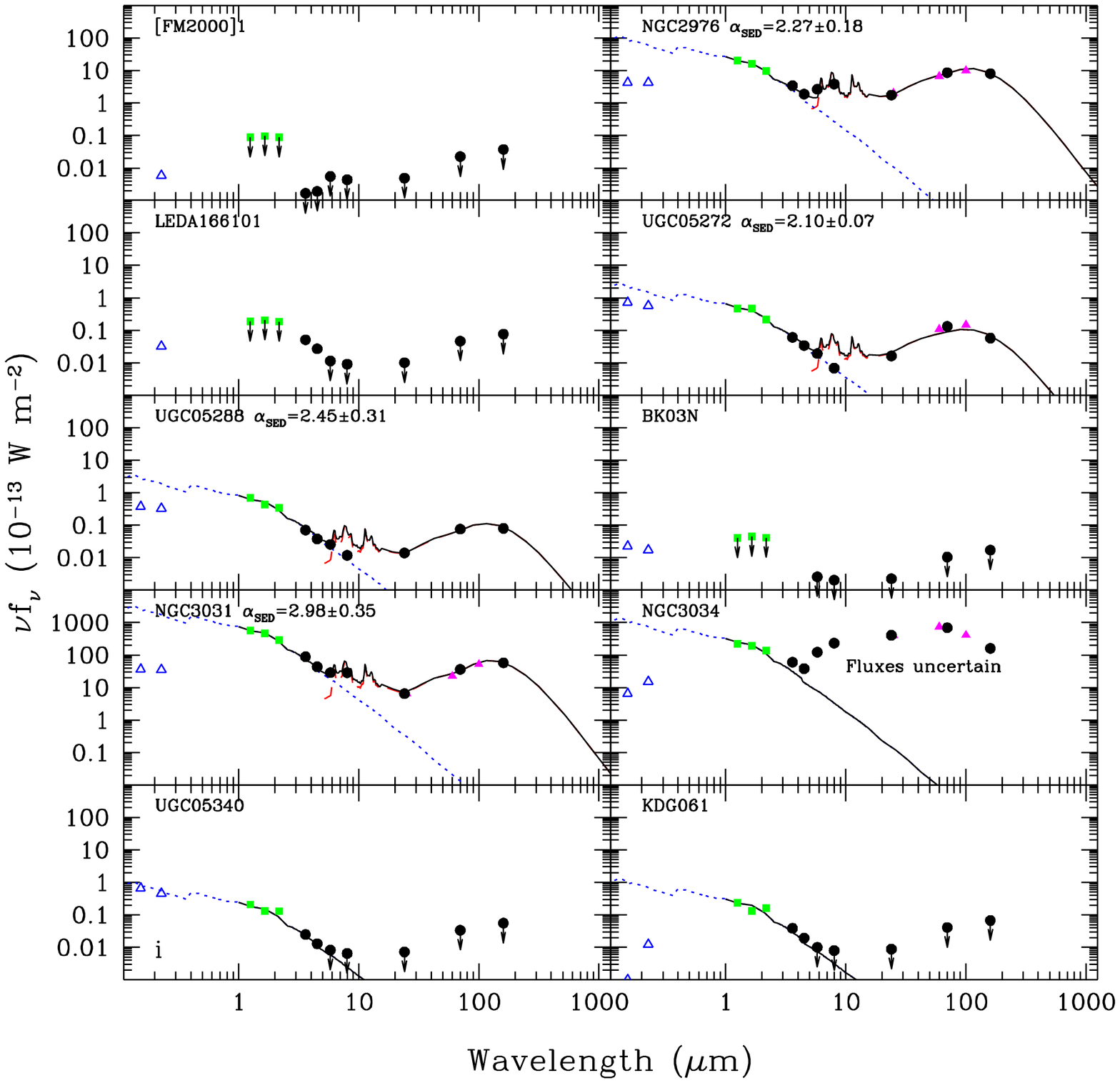}
 \caption{Globally-integrated 0.15-160\m\ spectral energy distributions for the LVL sample (continued).}
\end{figure}

\addtocounter{figure}{-1}
\clearpage
\begin{figure}
 \plotone{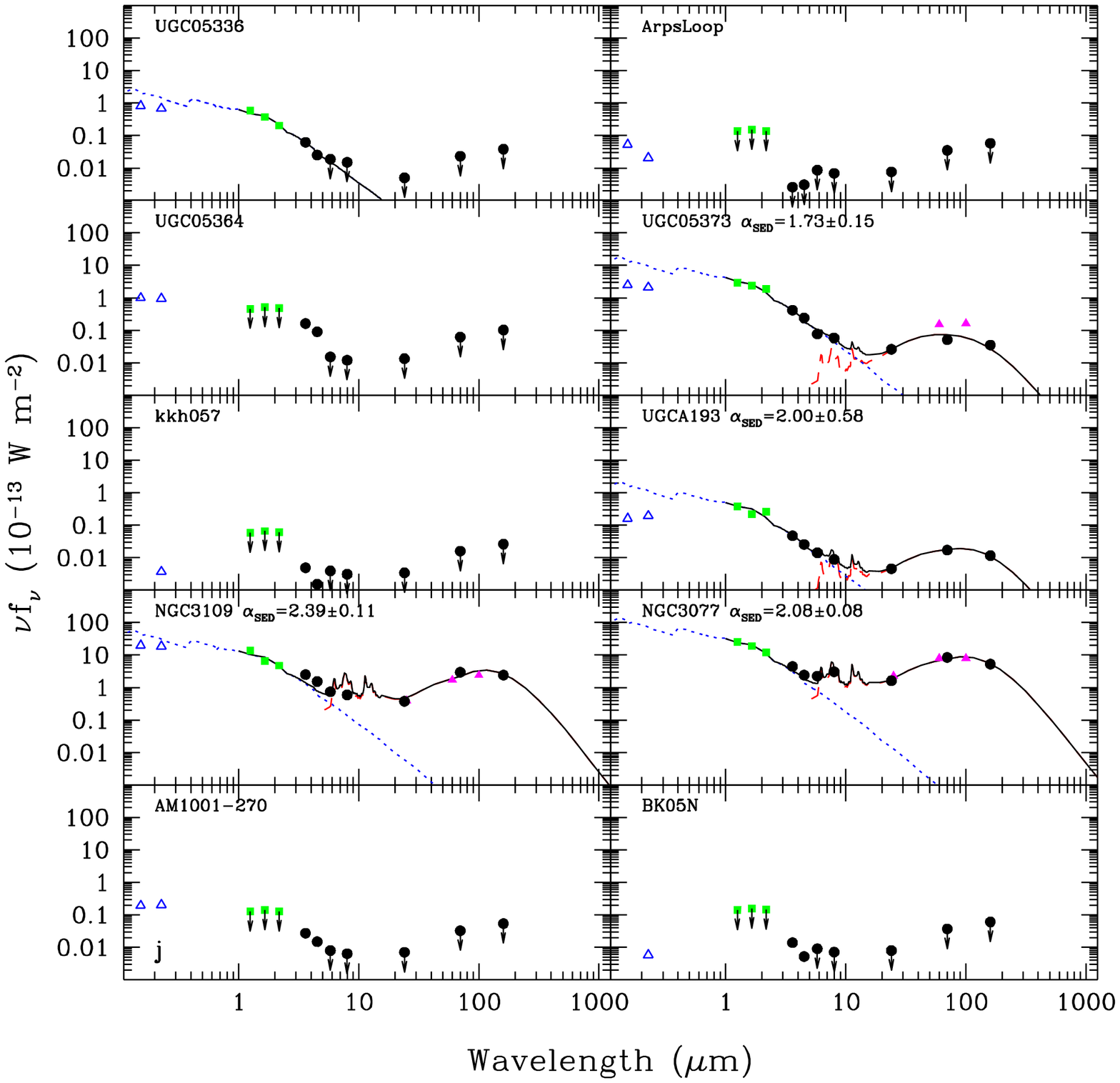}
 \caption{Globally-integrated 0.15-160\m\ spectral energy distributions for the LVL sample (continued).}
\end{figure}

\addtocounter{figure}{-1}
\begin{figure}
 \plotone{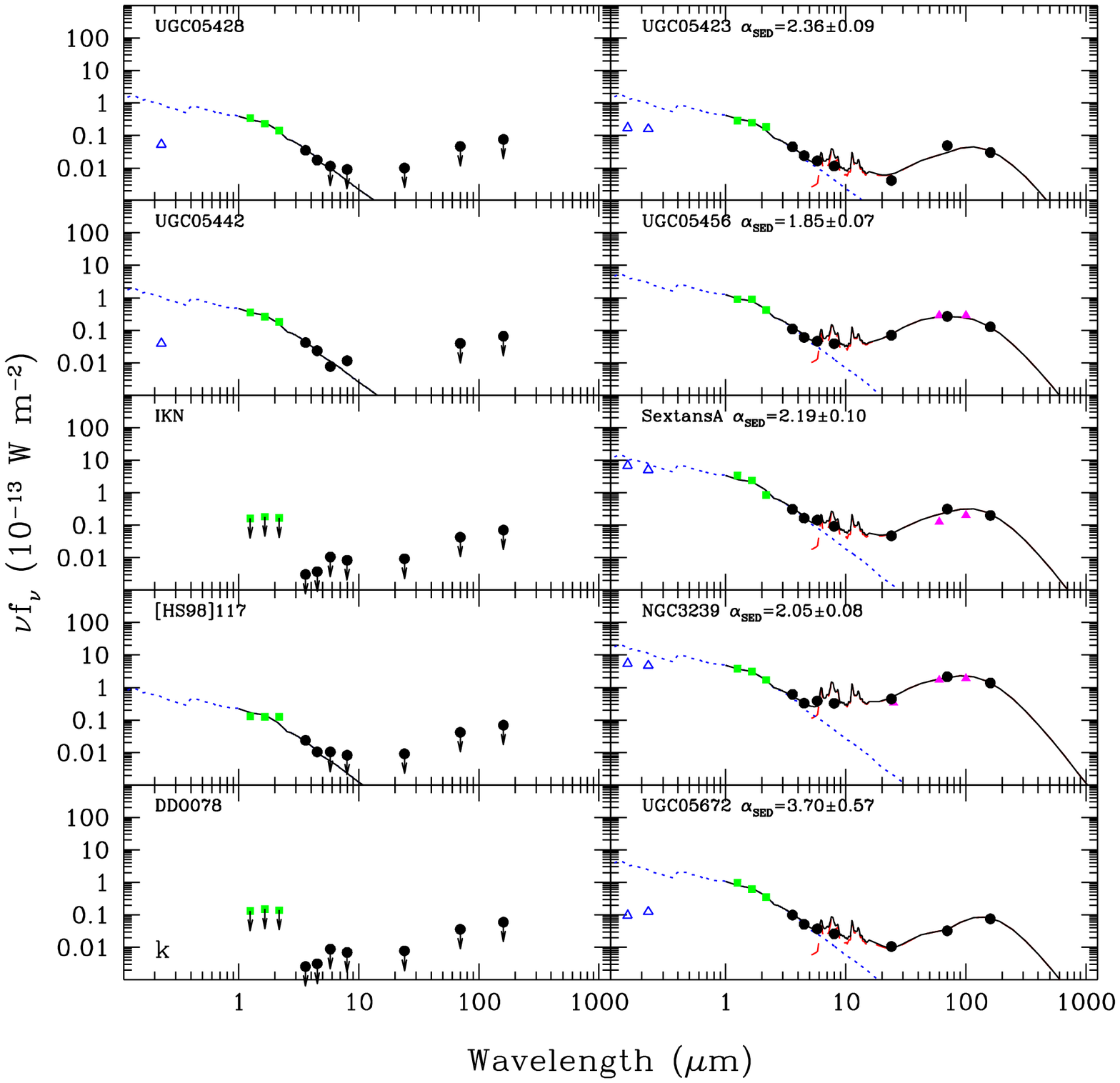}
 \caption{Globally-integrated 0.15-160\m\ spectral energy distributions for the LVL sample (continued).}
\end{figure}

\addtocounter{figure}{-1}
\begin{figure}
 \plotone{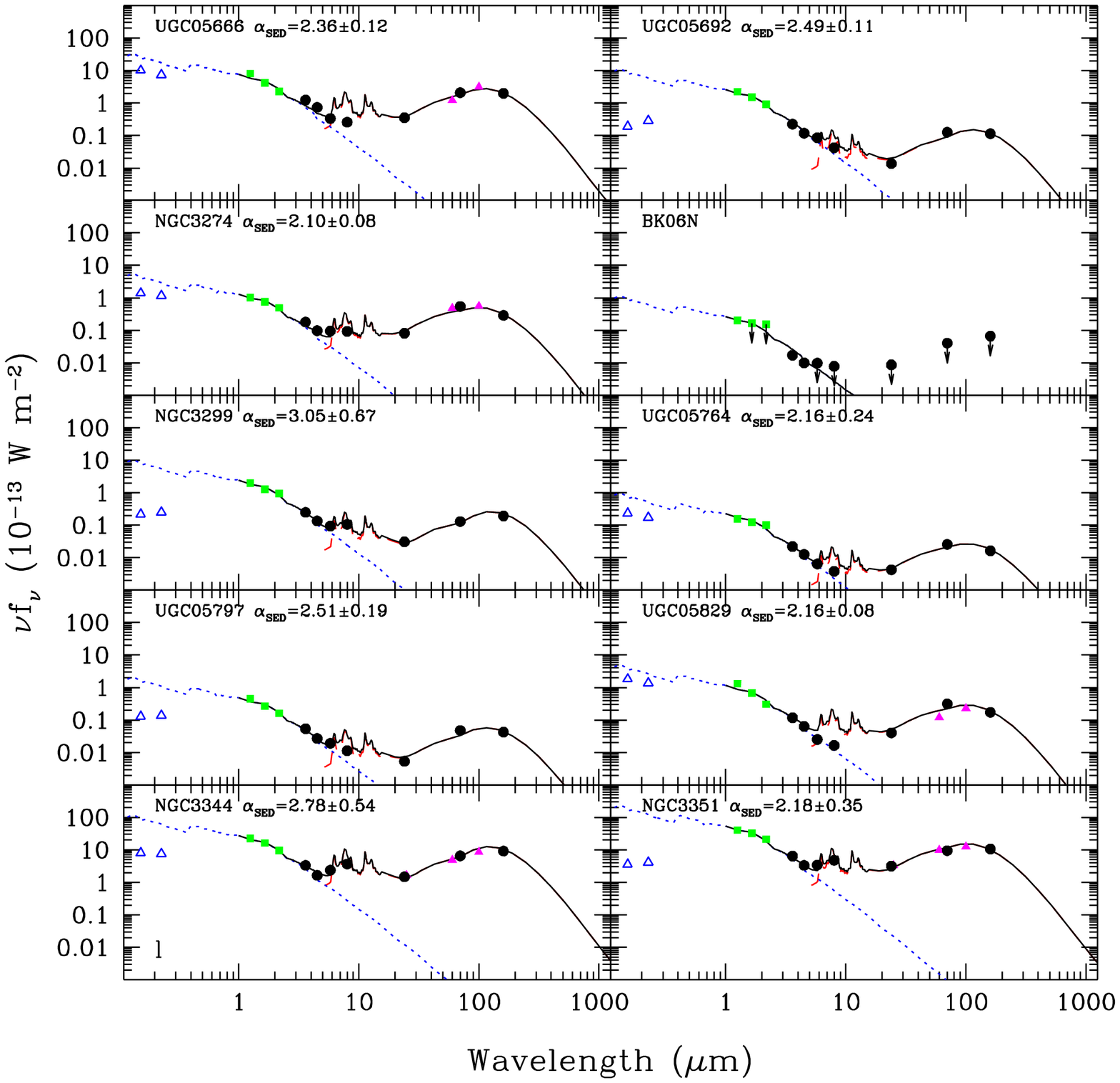}
 \caption{Globally-integrated 0.15-160\m\ spectral energy distributions for the LVL sample (continued).}
\end{figure}

\addtocounter{figure}{-1}
\begin{figure}
 \plotone{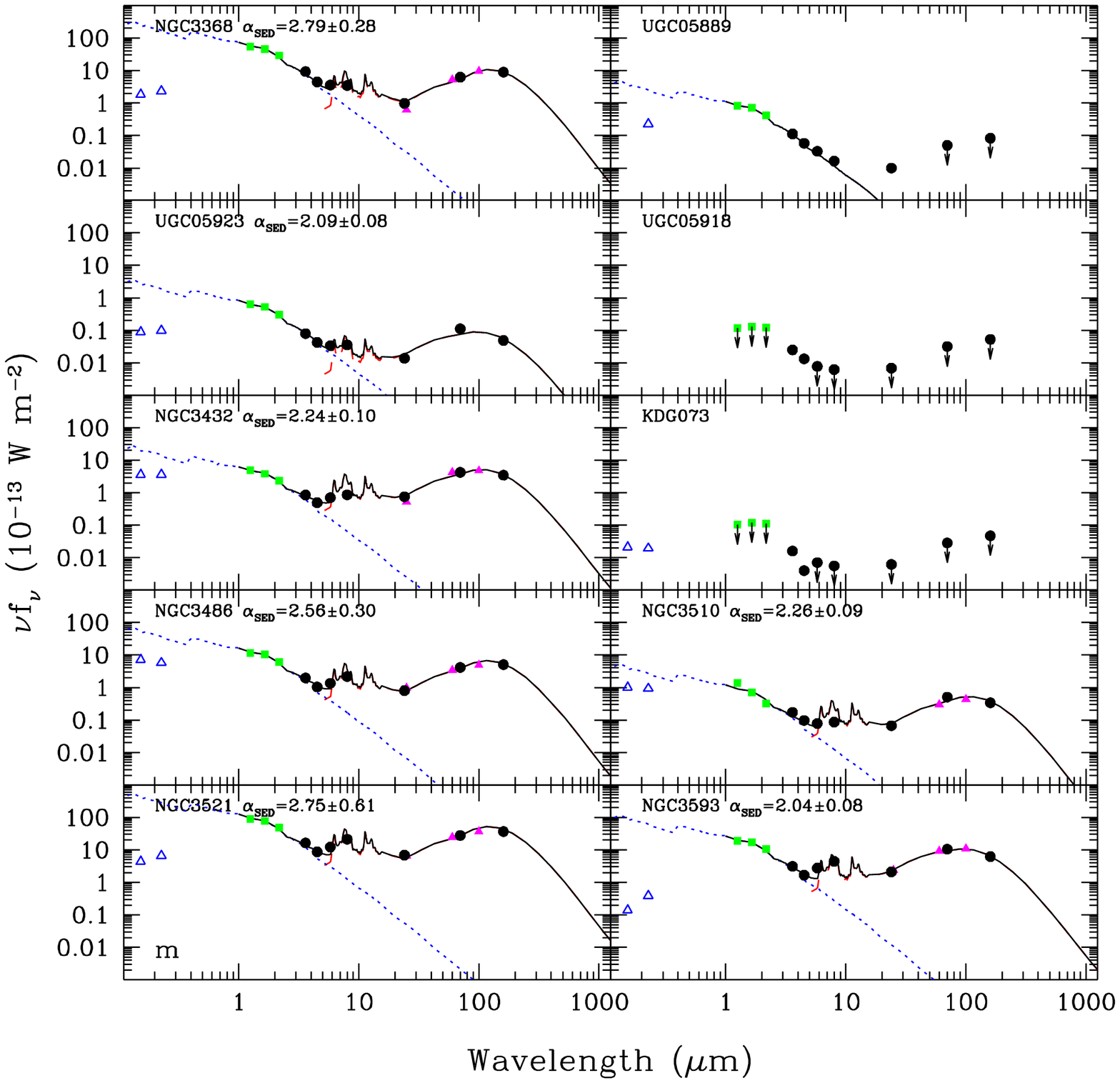}
 \caption{Globally-integrated 0.15-160\m\ spectral energy distributions for the LVL sample (continued).}
\end{figure}

\addtocounter{figure}{-1}
\begin{figure}
 \plotone{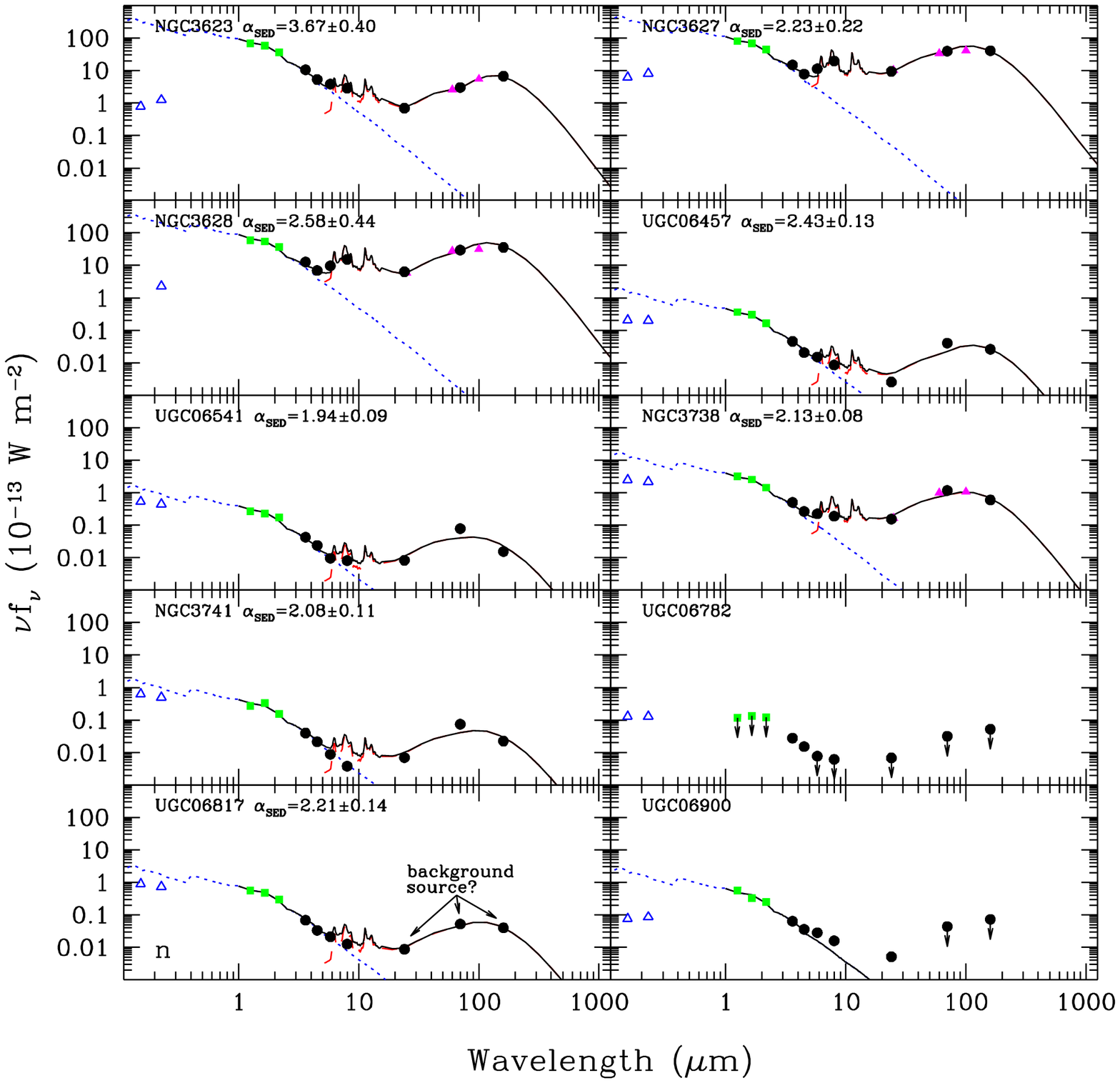}
 \caption{Globally-integrated 0.15-160\m\ spectral energy distributions for the LVL sample (continued).}
\end{figure}

\addtocounter{figure}{-1}
\begin{figure}
 \plotone{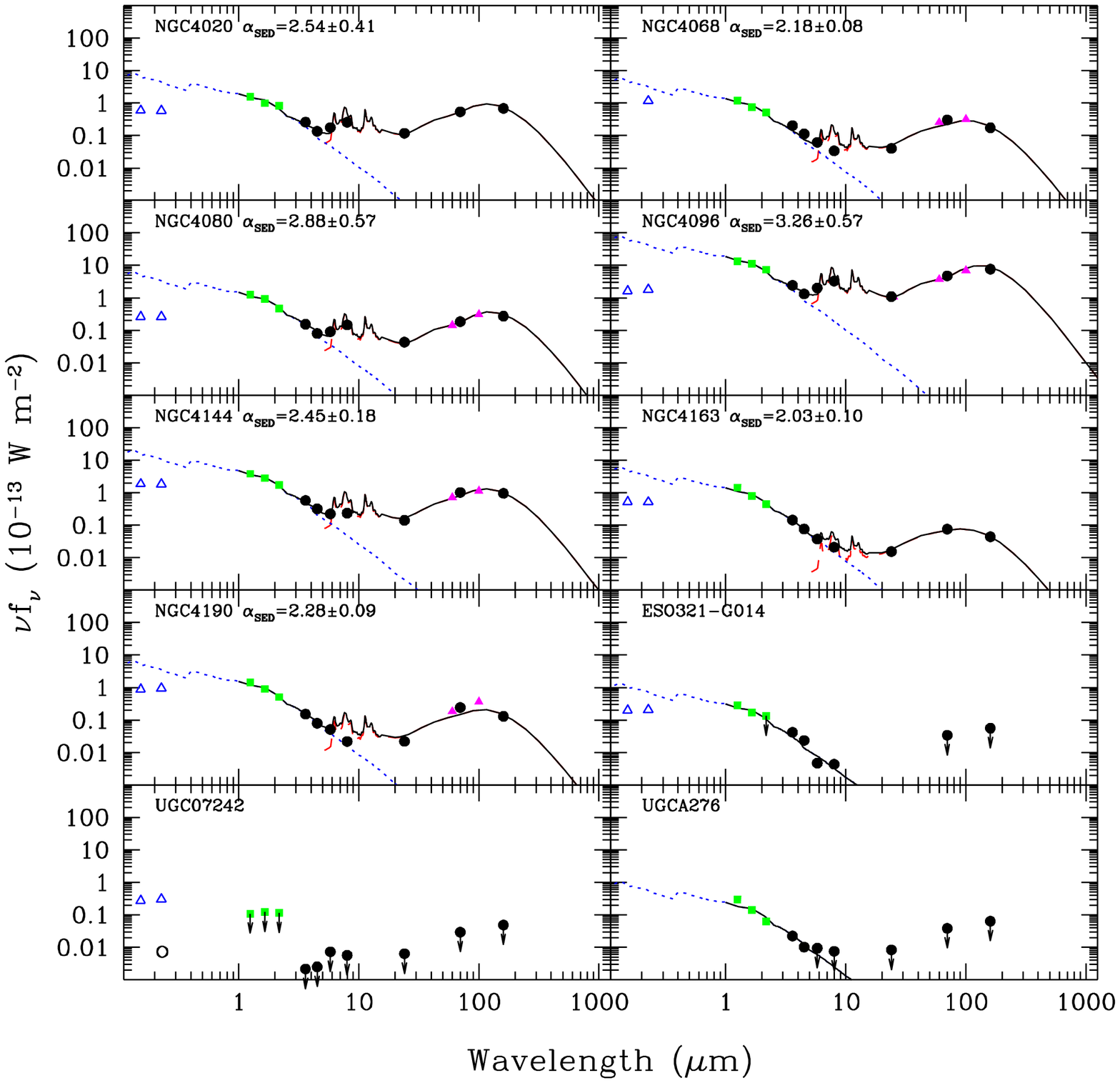}
 \caption{Globally-integrated 0.15-160\m\ spectral energy distributions for the LVL sample (continued).}
\end{figure}

\addtocounter{figure}{-1}
\begin{figure}
 \plotone{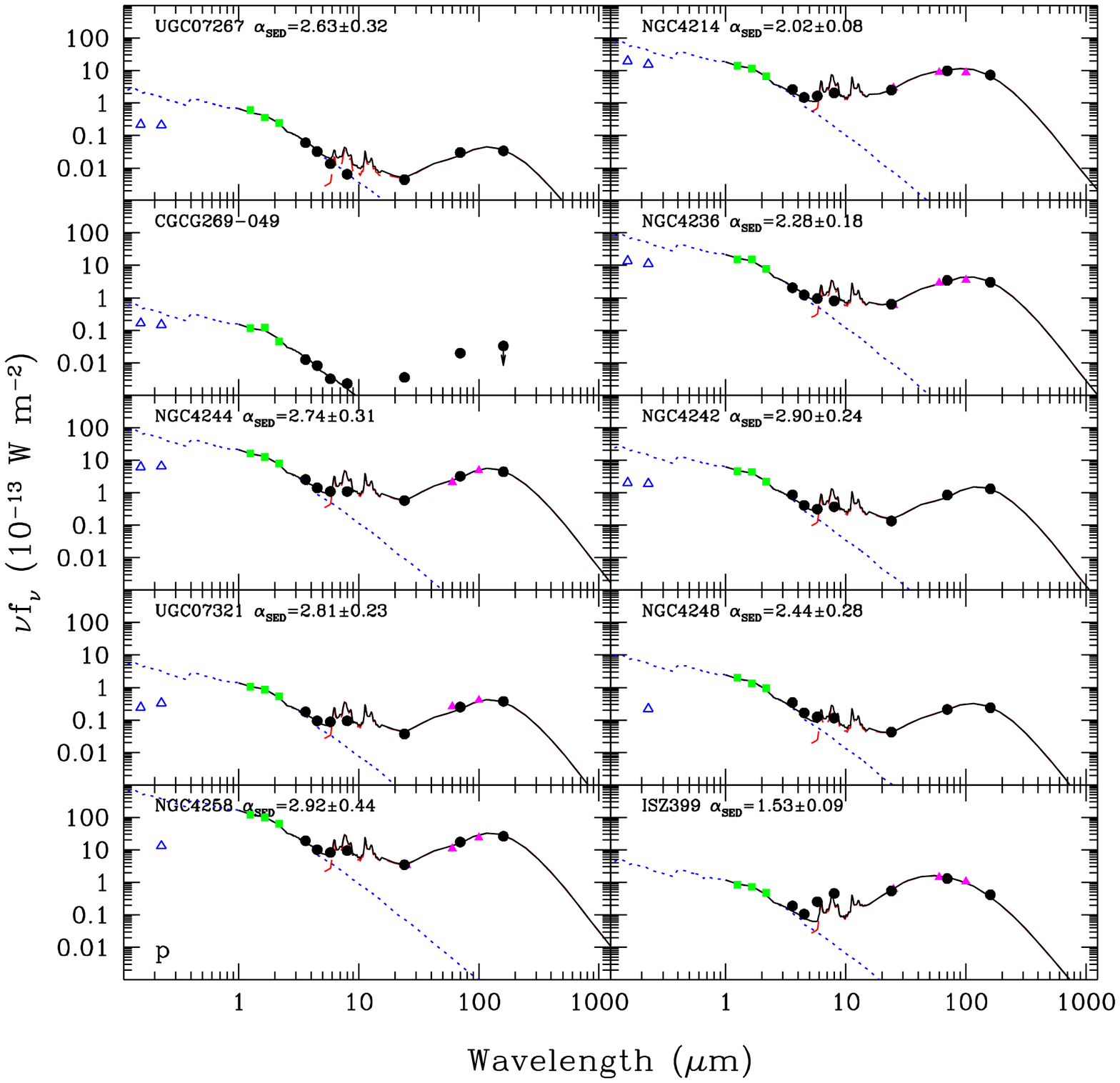}
 \caption{Globally-integrated 0.15-160\m\ spectral energy distributions for the LVL sample (continued).}
\end{figure}

\addtocounter{figure}{-1}
\begin{figure}
 \plotone{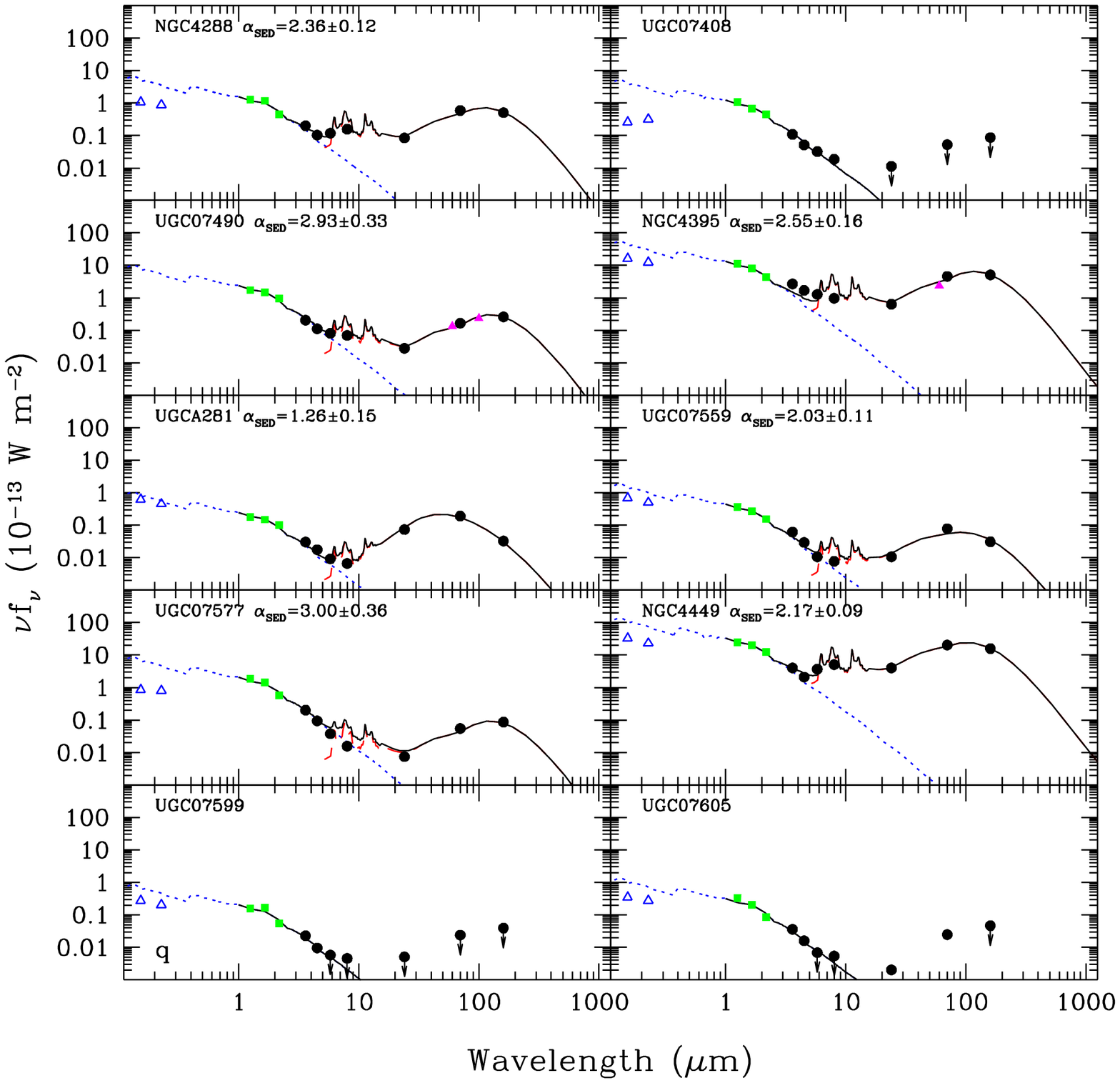}
 \caption{Globally-integrated 0.15-160\m\ spectral energy distributions for the LVL sample (continued).}
\end{figure}

\addtocounter{figure}{-1}
\begin{figure}
 \plotone{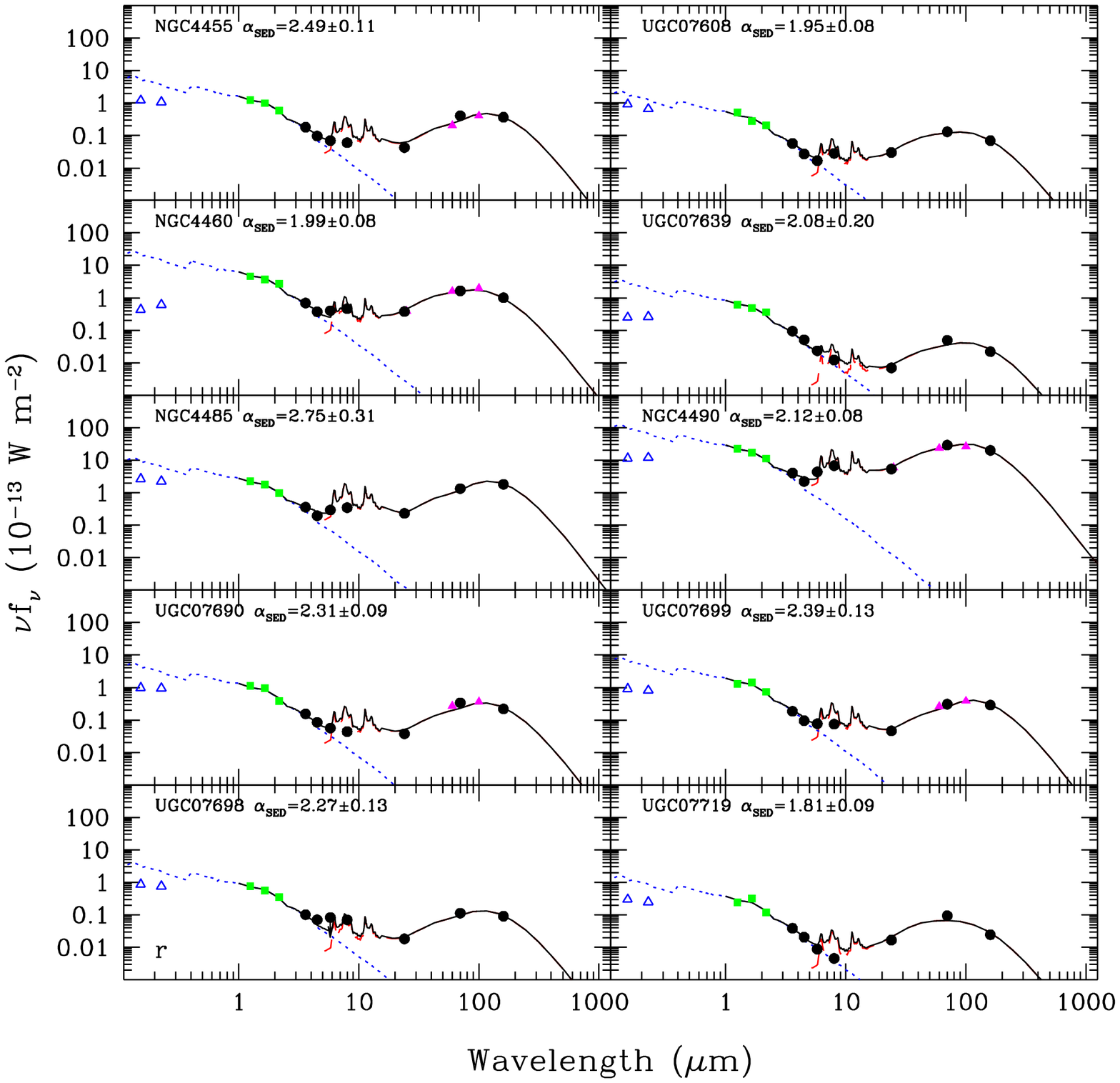}
 \caption{Globally-integrated 0.15-160\m\ spectral energy distributions for the LVL sample (continued).}
\end{figure}

\addtocounter{figure}{-1}
\begin{figure}
 \plotone{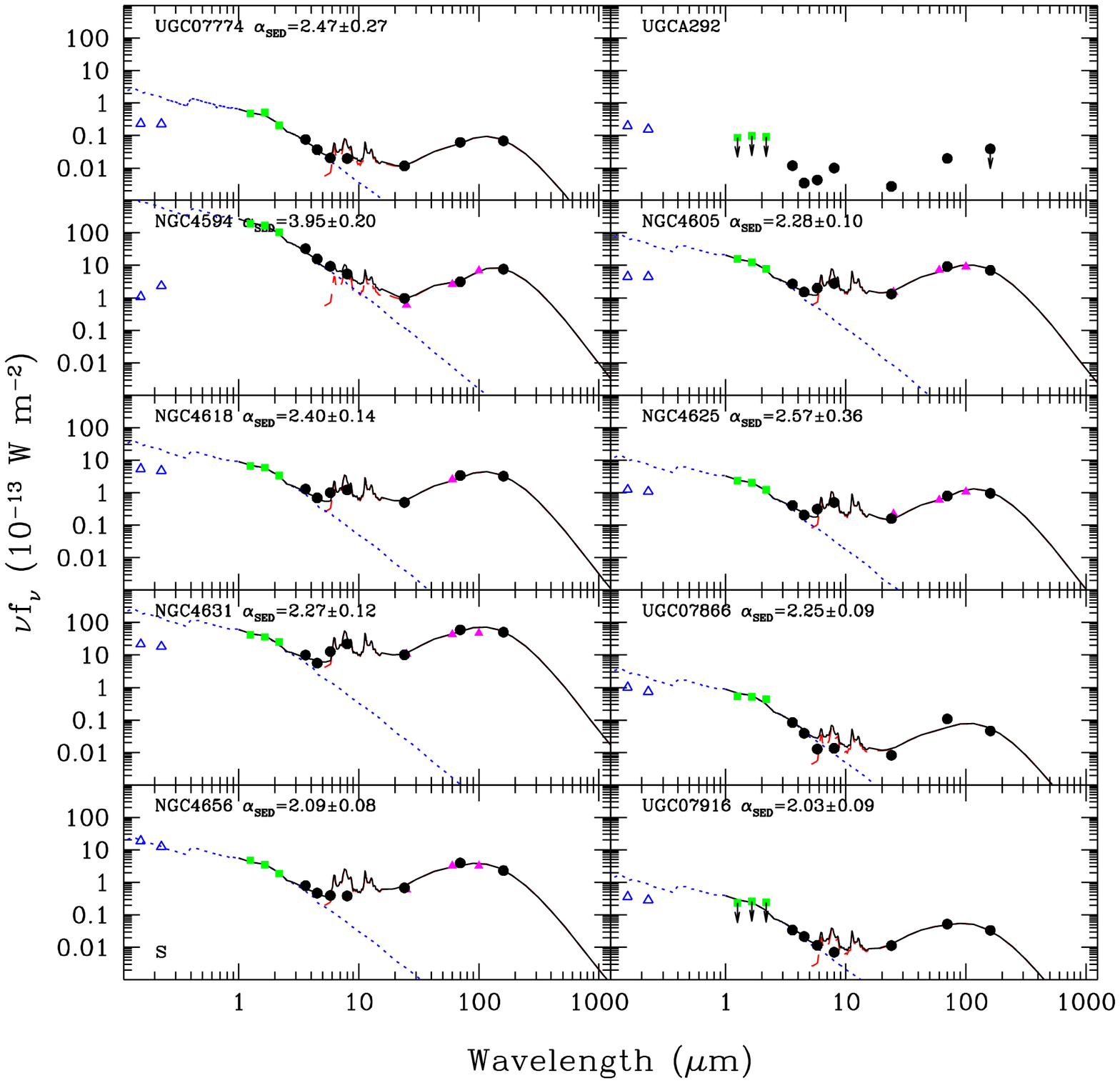}
 \caption{Globally-integrated 0.15-160\m\ spectral energy distributions for the LVL sample (continued).}
\end{figure}

\addtocounter{figure}{-1}
\clearpage
\begin{figure}
 \plotone{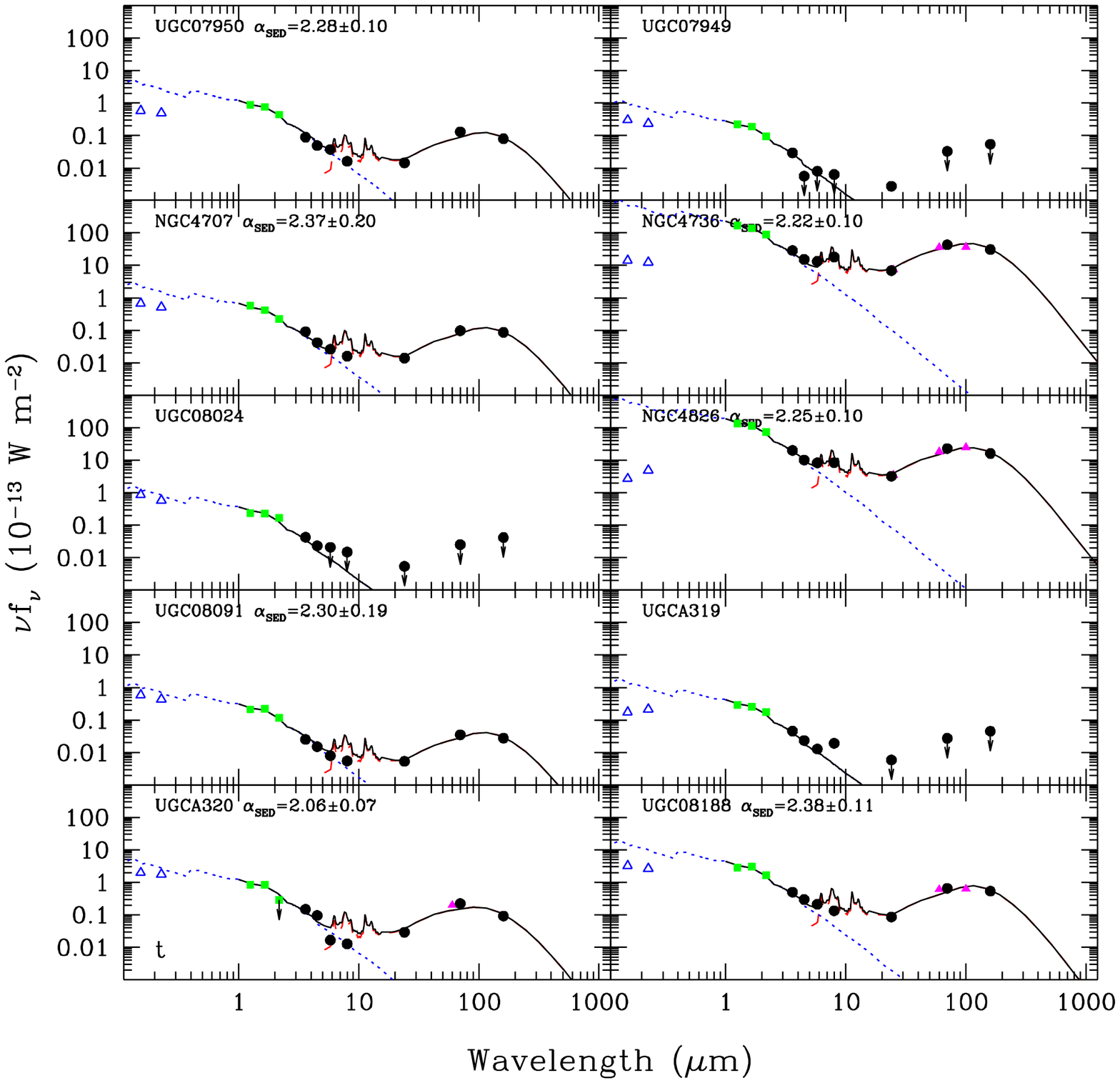}
 \caption{Globally-integrated 0.15-160\m\ spectral energy distributions for the LVL sample (continued).}
\end{figure}

\addtocounter{figure}{-1}
\begin{figure}
 \plotone{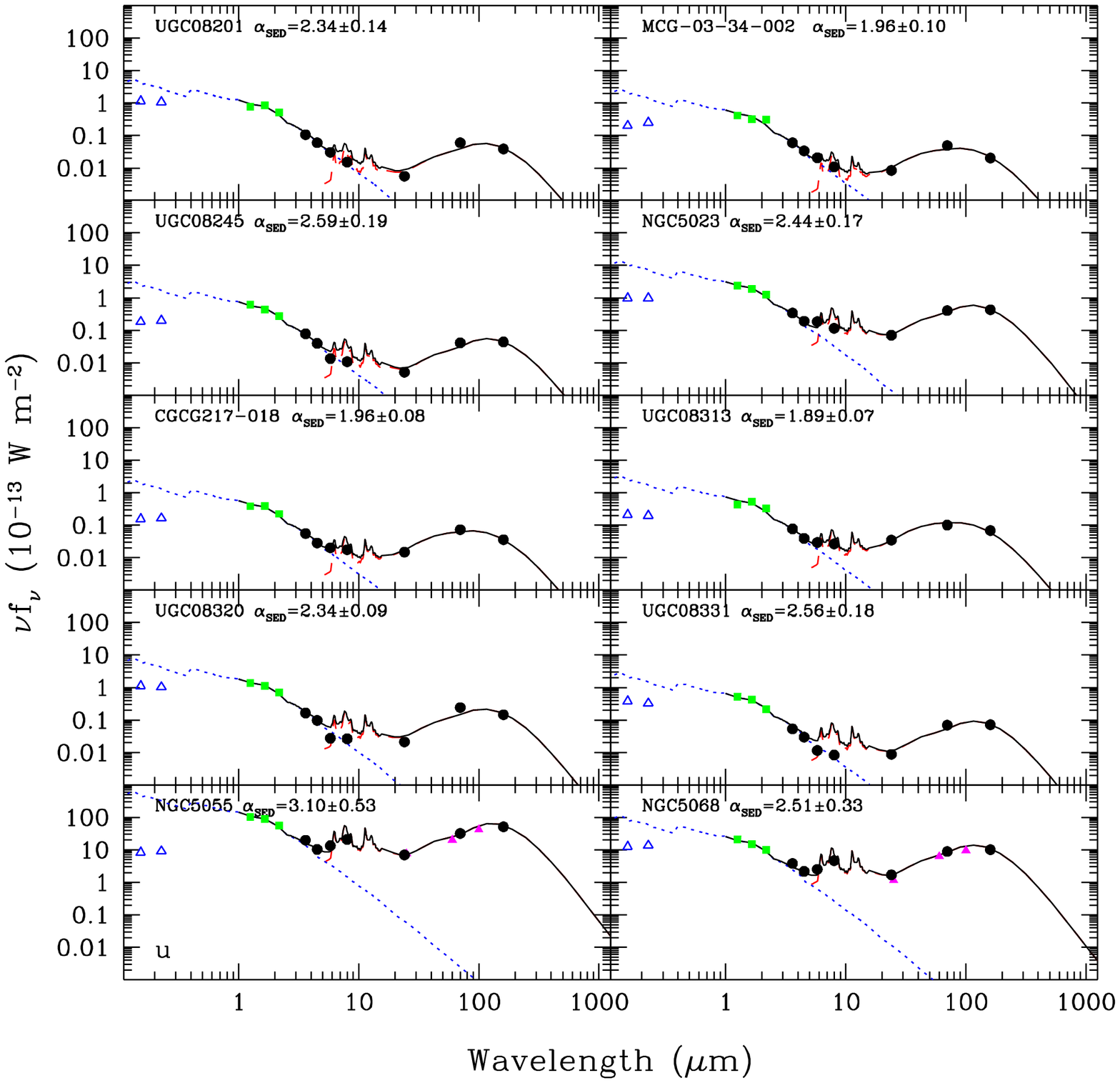}
 \caption{Globally-integrated 0.15-160\m\ spectral energy distributions for the LVL sample (continued).}
\end{figure}

\addtocounter{figure}{-1}
\begin{figure}
 \plotone{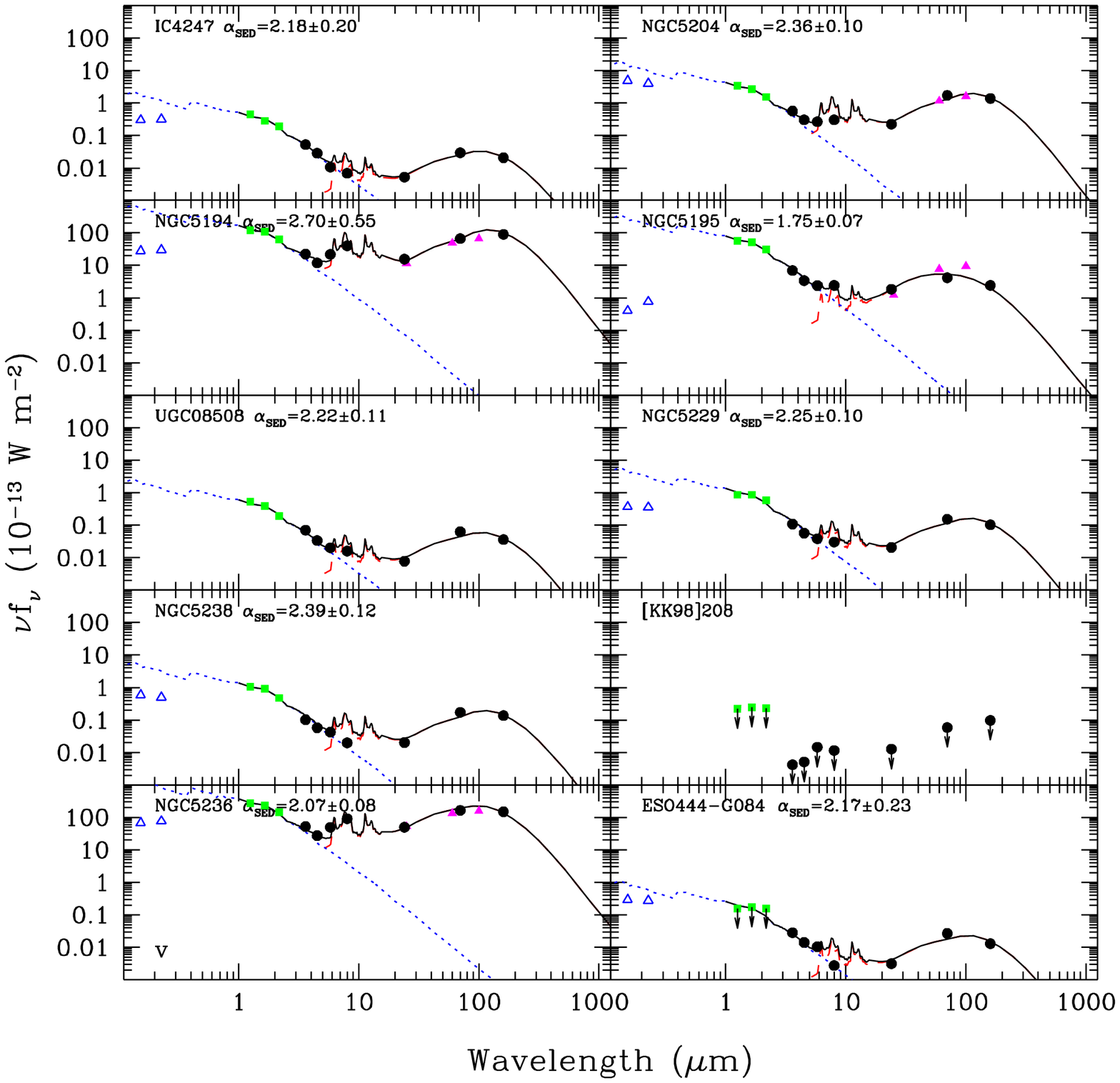}
 \caption{Globally-integrated 0.15-160\m\ spectral energy distributions for the LVL sample (continued).}
\end{figure}

\addtocounter{figure}{-1}
\begin{figure}
 \plotone{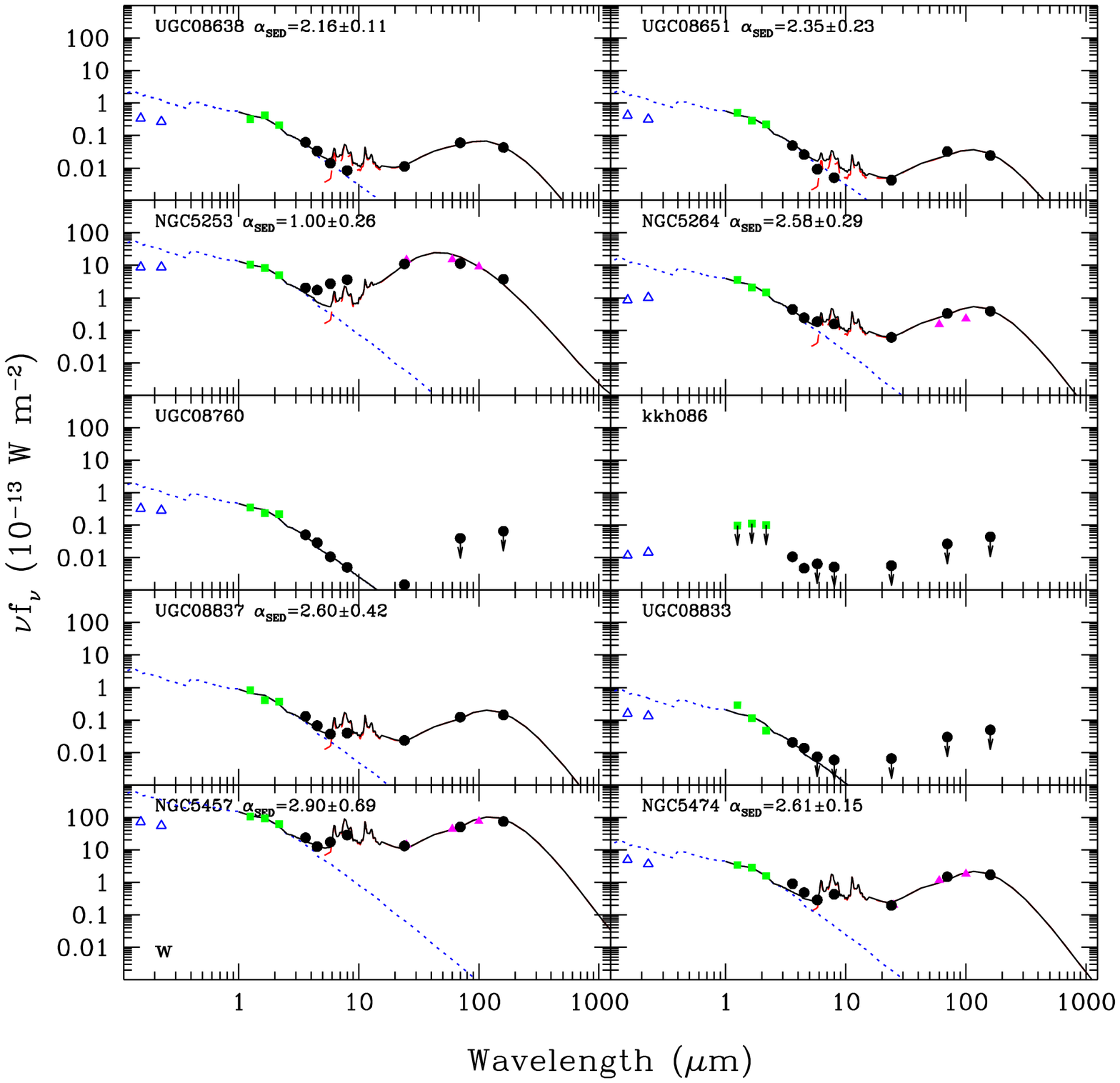}
 \caption{Globally-integrated 0.15-160\m\ spectral energy distributions for the LVL sample (continued).}
\end{figure}

\addtocounter{figure}{-1}
\begin{figure}
 \plotone{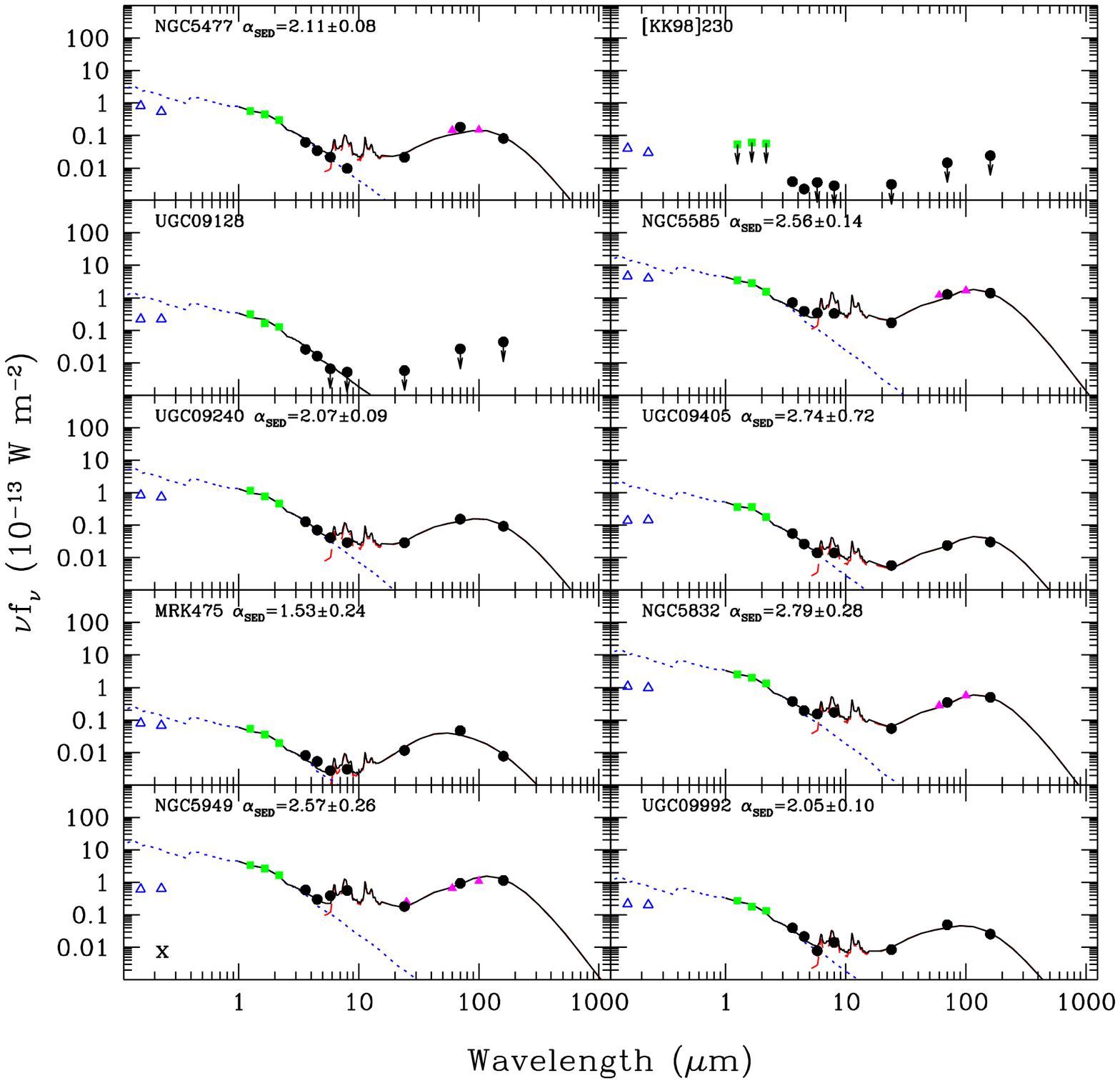}
 \caption{Globally-integrated 0.15-160\m\ spectral energy distributions for the LVL sample (continued).}
\end{figure}

\addtocounter{figure}{-1}
\begin{figure}
 \plotone{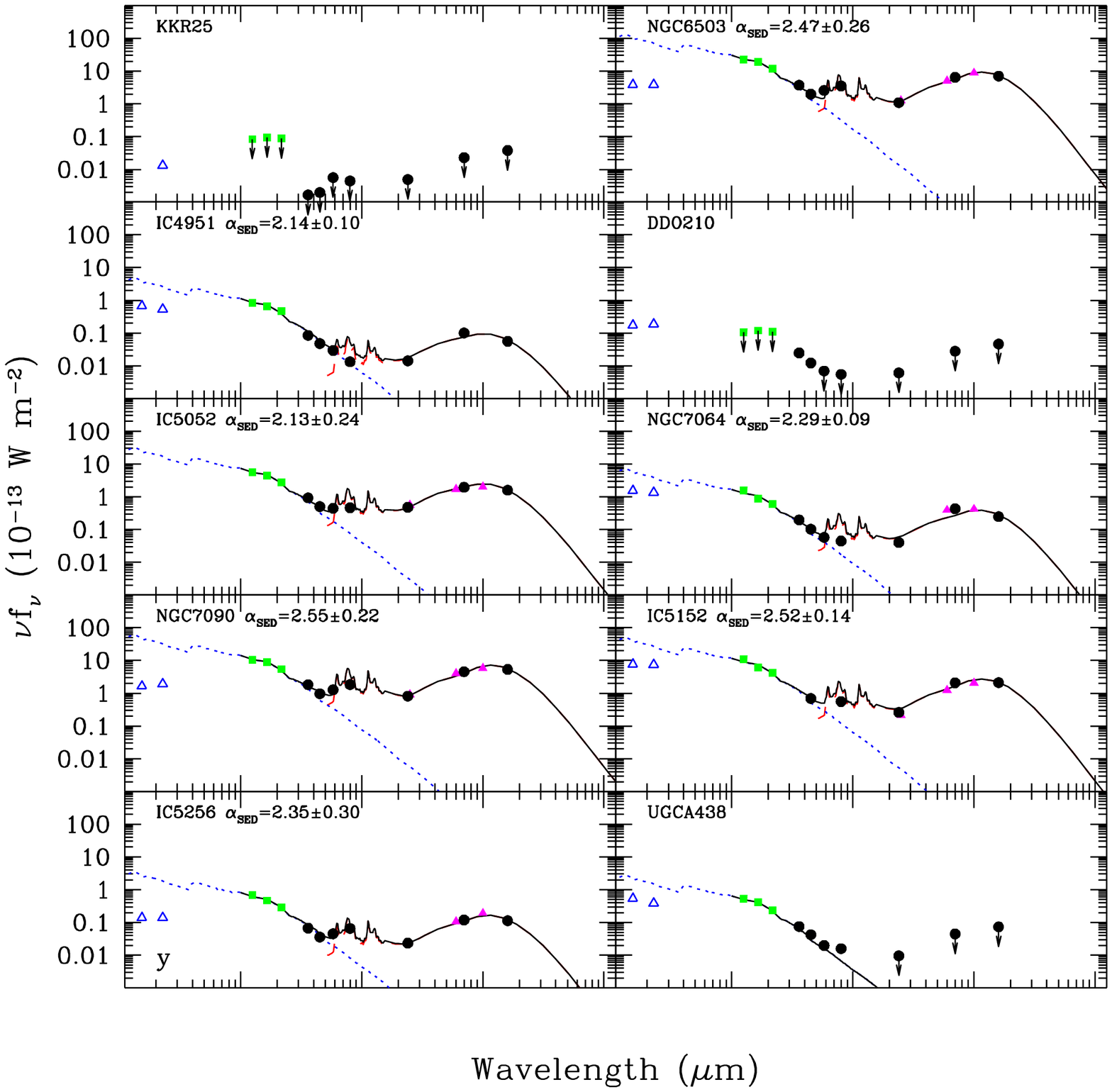}
 \caption{Globally-integrated 0.15-160\m\ spectral energy distributions for the LVL sample (continued).}
\end{figure}

\addtocounter{figure}{-1}
\begin{figure}
 \plotone{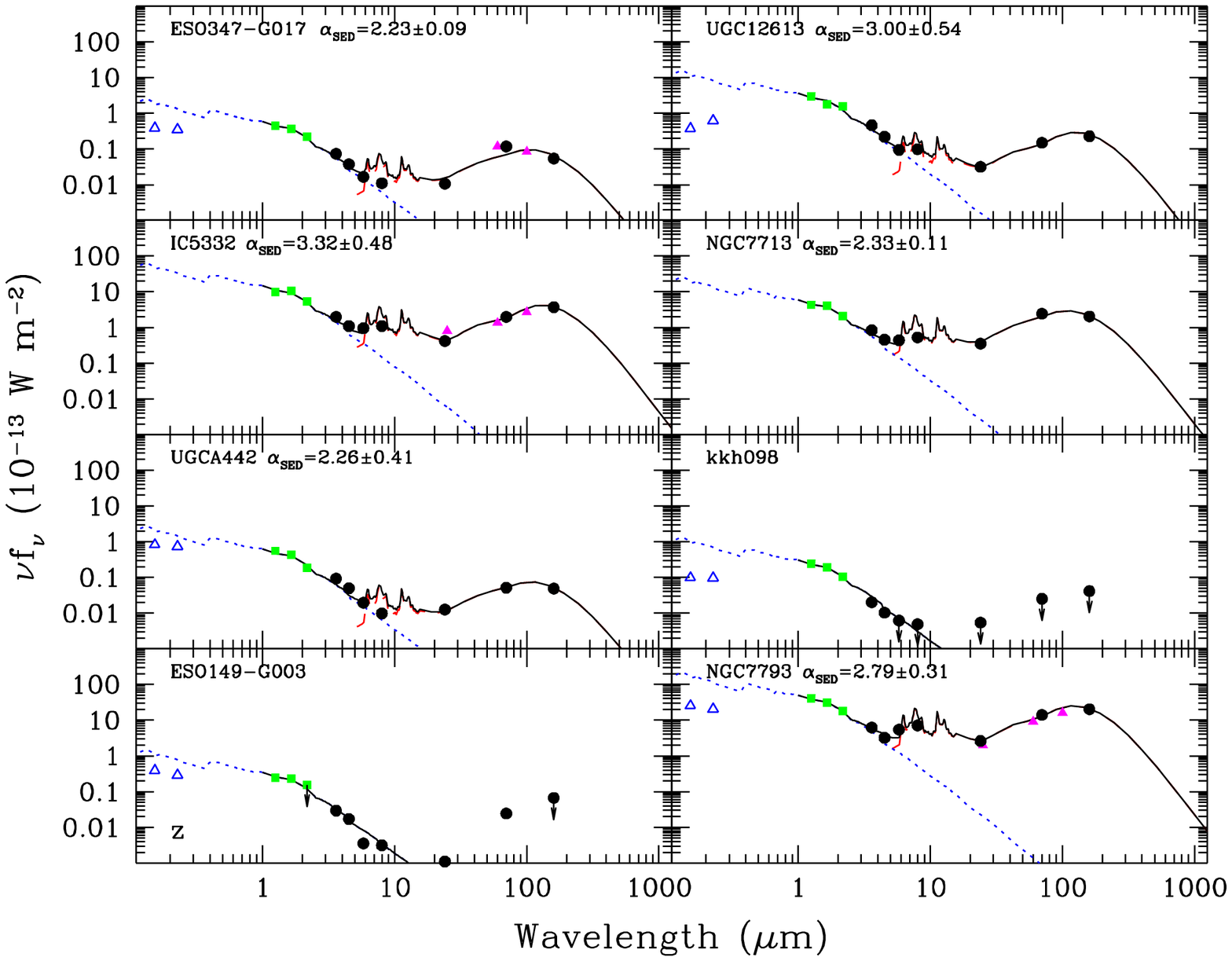}
 \caption{Globally-integrated 0.15-160\m\ spectral energy distributions for the LVL sample (continued).}
\end{figure}

\begin{figure}
 \plotone{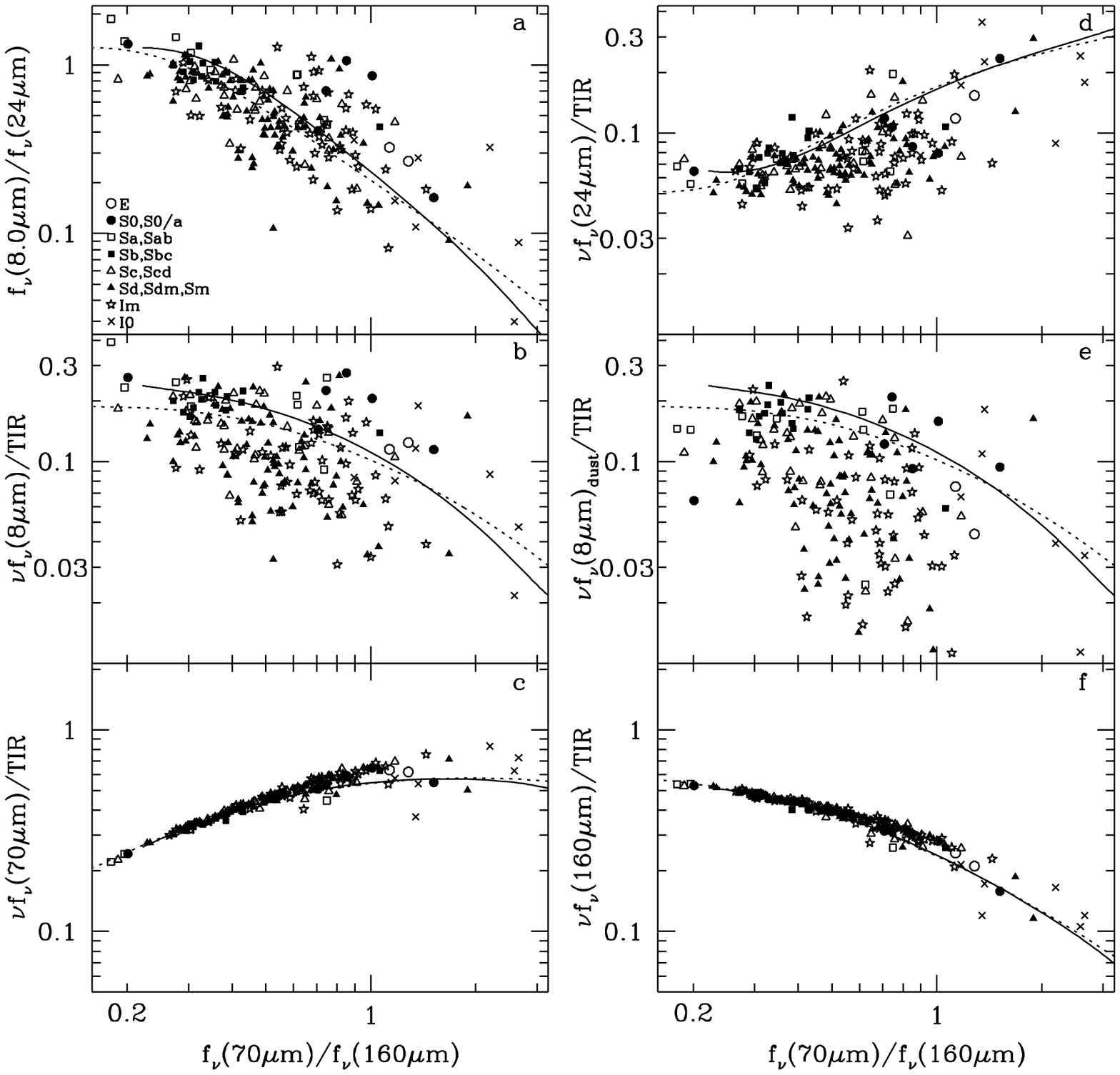}
 \caption{The {\it Spitzer} infrared colors and monochromatic-to-bolometric infrared ratios for globally-integrated LVL data.  The solid and dotted lines indicate the dust-only SED models of Dale \& Helou (2002) and Dale et al.\ (2001), respectively, derived from the average global trends for a sample of normal star-forming galaxies observed by \ISO\ and \IRAS.  A $y$-axis logarithmic range of $\sim$1.6~dex is the same for panels b--f for ease of comparison.}
 \label{fig:global_colors}
\end{figure}

\begin{figure}
 \plotone{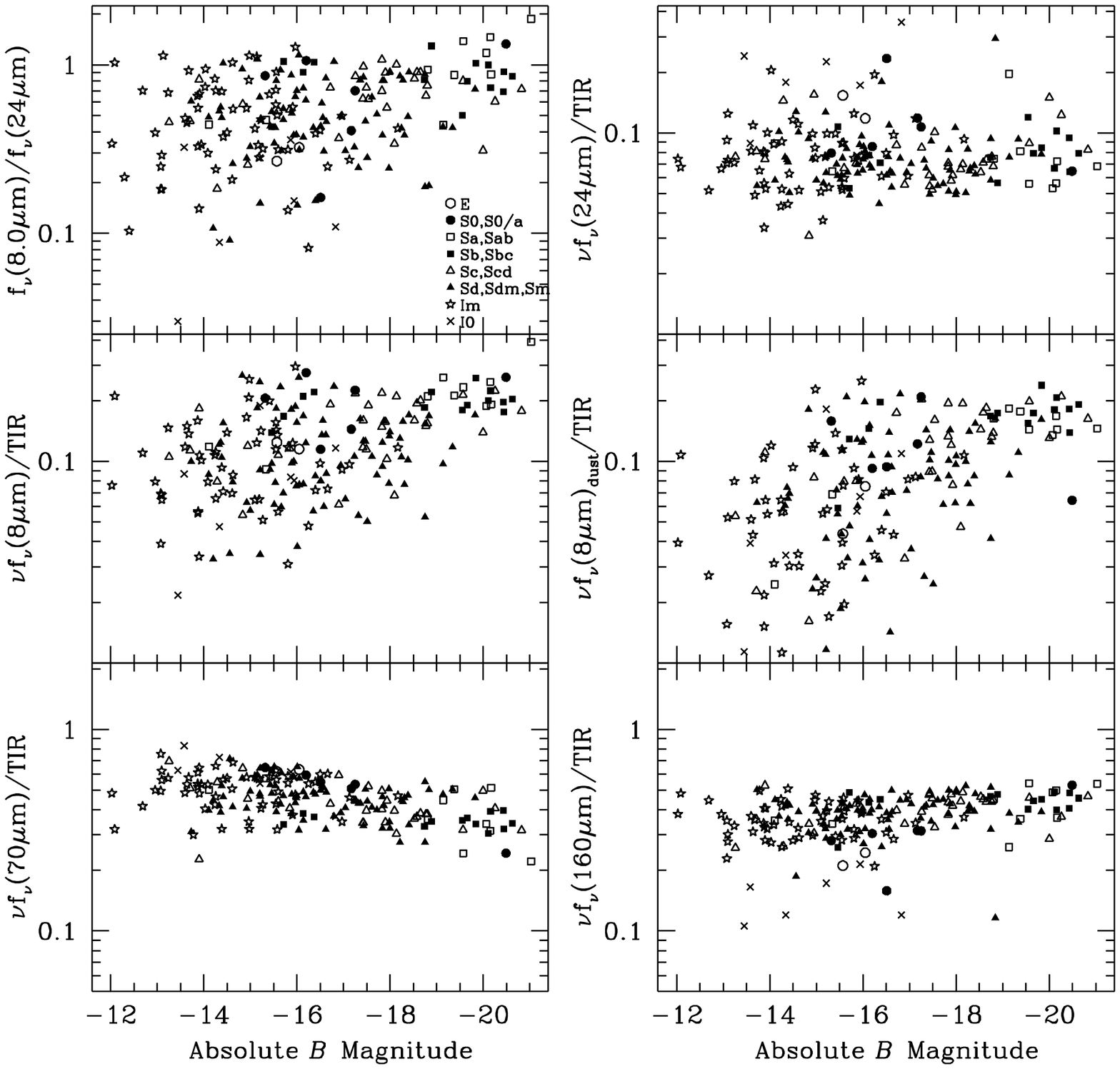}
 \caption{Similar to Figure~\ref{fig:global_colors} except as a function of absolute $B$ magnitude.}
 \label{fig:global_colorsb}
\end{figure}

\begin{figure}
 \plotone{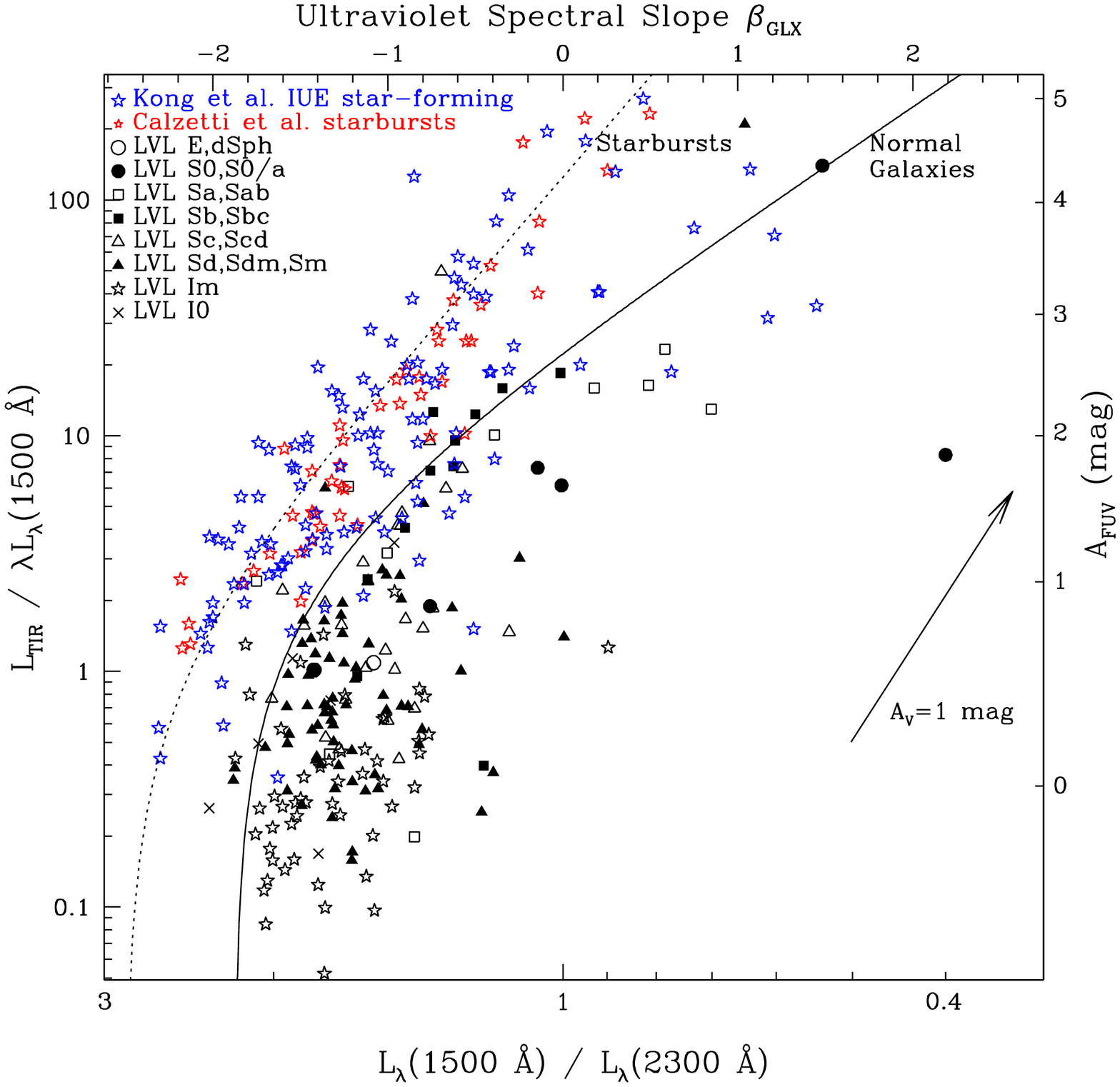}
 \caption{The infrared-to-far-ultraviolet ratio as a function of ultraviolet spectral slope.  Normal star-forming and starbursting galaxies from Kong et al.\ (2004) and Calzetti et al.\ (1995) are plotted in addition to the LVL data points.  The dotted curve is that for starbursting galaxies from Kong et al.\ (2004) and the solid curve is applicable to normal star-forming galaxies (Dale et al.\ 2007).  The reddening vector assumes the reddening curve of Li \& Draine (2001) and the far-ultraviolet extinction prescription used for the righthand axis is from Buat et al. (2005).}
 \label{fig:beta}
\end{figure}

\begin{figure}
 \plotone{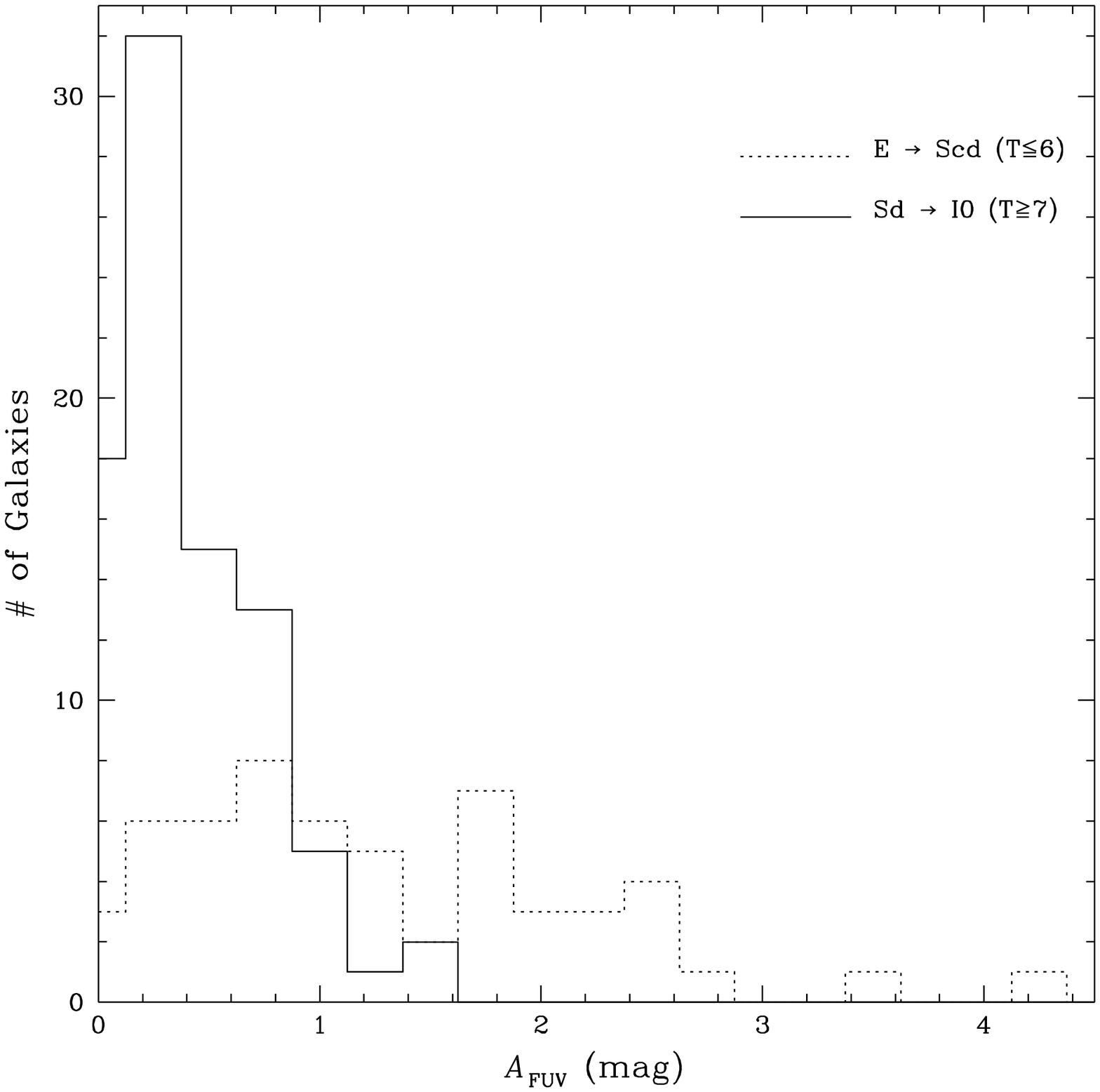}
 \caption{The distribution of far-ultraviolet extinctions, computed using the infrared-to-ultraviolet ratio and Equation~2 of Buat et al.\ (2005).}
 \label{fig:AFUV}
\end{figure}

\begin{figure}
 \plotone{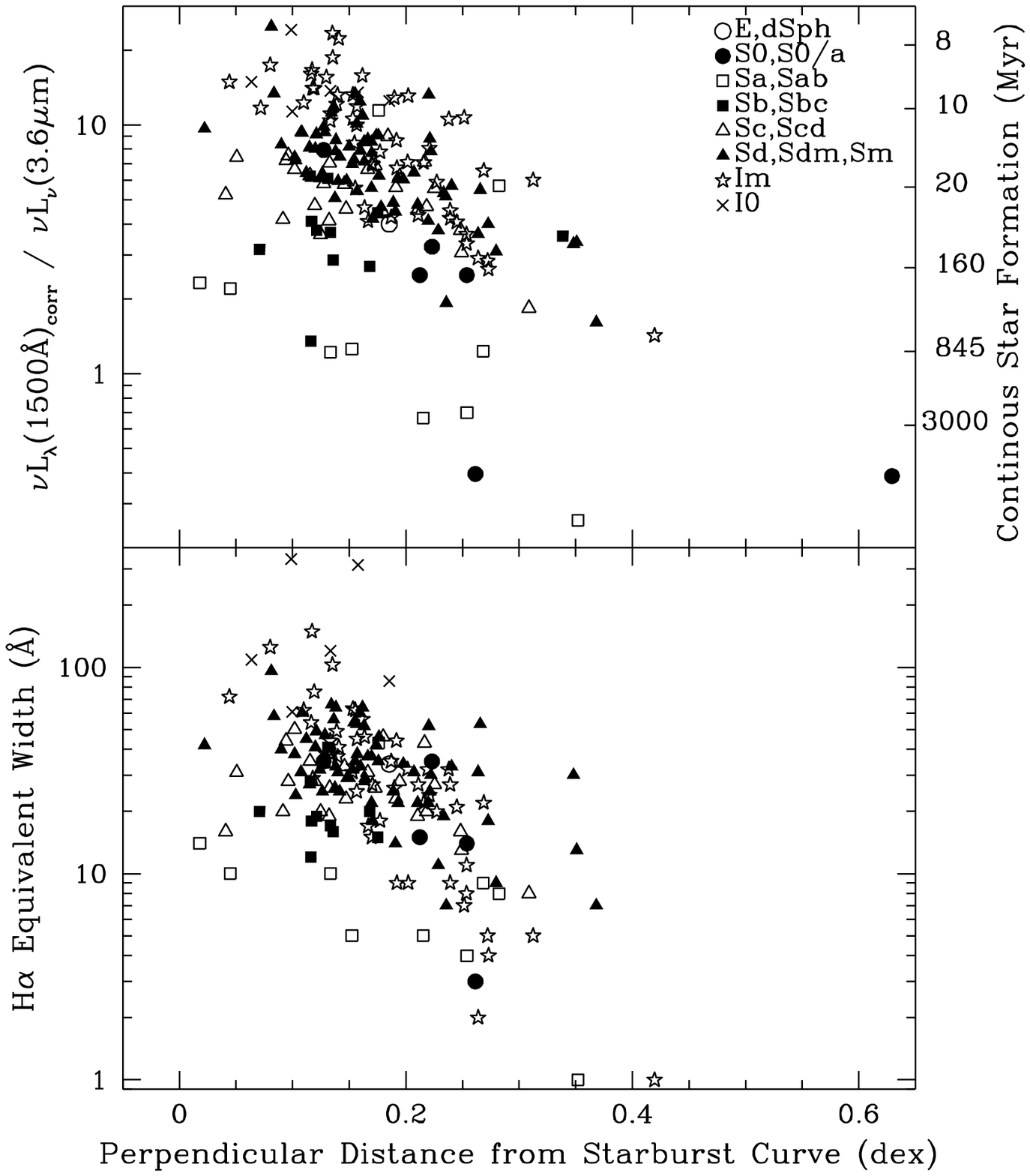}
 \caption{The dependence of galaxy star formation history as a function of distance from the starburst relation infrared-to-ultraviolet versus ultraviolet slope, as shown in Figure~\ref{fig:beta}.  The lefthand axes are observable diagnostics of the birthrate parameter, the current star formation rate normalized to the average star formation rate.  Top: The far-ultraviolet-to-near-infrared ratio, with the righthand axis showing the number of years (continuous) star formation has been occurring, as measured from theoretical spectra.  The theoretical spectra utilized are solar metallicity, 1~$M_\odot~{\rm yr}^{-1}$ continuous star formation curves assuming a double power law initial mass function, with $\alpha_{\rm 1,IMF}=1.3$ for $0.1<m/M_\odot<0.5$ and $\alpha_{\rm 2,IMF}=2.3$ for $0.5<m/M_\odot<100$ (Vazquez \& Leitherer 2005).  The far-ultraviolet emission is corrected for extinction using the recipe formulated in Buat al. (2005) and described in \S~\ref{sec:beta}.  Bottom: The global \hal\ equivalent width measured from narrowband and $R$ band imaging.}
 \label{fig:dstar}
\end{figure}
%

\end{document}